\documentclass[12pt]{article}
\usepackage{geometry}                
\usepackage{graphicx}

\begin{document}


%
%

\title{Recent Observation of Short Range  Nucleon Correlations in Nuclei  
and their Implications for  the Structure of Nuclei and Neutron Stars}

\author{Leonid Frankfurt\\
School of Physics and Astronomy, Tel Aviv University Ramat Aviv\\ Tel Aviv, 69978, Israel\\ \\
Misak Sargsian \\
Florida International University, University Park\\ Miami, FL 33199,   USA\\ \\
Mark Strikman\\
104 Davey Lab, The Pennsylvania State University\\  University Park, PA 16803, USA}

\maketitle


\begin{abstract}
Novel processes probing  the  decay  of nucleus after removal of a nucleon with momentum larger than Fermi momentum by hard 
probes finally proved unambiguously the evidence for long sought presence of  short-range correlations~(SRCs) in nuclei. In 
combination with the analysis of large $Q^2$, A(e,e')X processes 
at $x>1$ they allow us to conclude that (i) practically all nucleons with  momenta $\ge$ 300 MeV/c belong to SRCs, 
consisting mostly of 
two nucleons, ii) probability of such SRCs in medium and heavy nuclei is $\sim 25\%$,  iii) a fast removal of such nucleon 
practically always leads to emission of 
correlated nucleon with approximately opposite momentum, 
iv) proton  removal  from two-nucleon SRCs in 90\% of cases is accompanied by a removal of a neutron 
and only in 10\% by a removal of another proton. We  explain that observed absolute probabilities and the isospin structure of 
two nucleon  SRCs  confirm  the  important role that 
tensor forces play  in  internucleon interactions. 
We find also that the presence  of SRCs requires modifications of the Landau Fermi liquid approach to highly asymmetric 
nuclear matter and 
leads to 
 a significantly faster Êcooling of cold neutron stars with neutrino cooling operational even for
$N_p/N_n \le 0.1$. The effect is even stronger for the hyperon stars.
  Theoretical challenges raised by  the discovered 
dominance of nucleon degrees of freedom in SRCs and important role of the spontaneously 
broken chiral symmetry in quantum chromodynamics~(QCD) in resolving them 
are considered.   We also outline directions for future theoretical and experimental 
studies of the physics relevant for SRCs.


\end{abstract}

\section{Introduction}

\subsection{Definition of short range correlations and  key questions in their studies}
  
It was understood many decades ago that stability of heavy nuclei and saturation of the nuclear density requires interplay between 
nucleon-nucleon attraction at intermediate distances $\le 1.5~$fm and 
significantly stronger repulsion which sets in at distances $\le 0.5$~fm.  
A strong compensation between attractive and repulsive potentials leads to the 
binding energy  per nucleon which is much smaller than both average kinetic and potential energies 
(see e.g. Ref. \cite{Bethe}). 

Presence of the strong short range repulsion and intermediate range attraction between nearby 
nucleons generates nucleons with momenta substantially larger than Fermi momentum characteristic 
for the given nucleus.   Therefore the first implication of the presence of 
short-range
interaction in nuclei is  the presence of high momentum component in
the nuclear ground state wave  function in momentum space.

Since the main contribution to the high momentum component comes from spatial configurations 
where distances between two nucleons are significantly smaller  than the average internucleon 
distances it is natural to refer to {\it all } these configurations as 
short-range correlations (SRCs). Often in the literature one separates these correlations 
into medium and short -distance correlations. We will not follow this tradition since 
such correlations are manifested in a similar way in the high momentum component of the 
nuclear wave function.   Presence of high momentum component in the nuclear ground state wave function 
has been demonstrated in theoretical calculations of wave functions of light nuclei  and infinite nuclear 
matter based on  the nonrelativistic nuclear theory~(see e.g. Refs. \cite{Zabolitsky} and \cite{Pandharipande}).  
In these and similar calculations SRCs play an important role in the microscopic 
structure of nuclei with more than  50\% of kinetic energy 
originating from SRC\footnote{In a number of approaches such as
Landau-Migdal Fermi liquid approach, mean field shell models and  effective chiral theory approach, 
which aimed at describing low energy effects
in nuclei the SRC are  hidden in the parameters of the effective potential  describing quasiparticles, 
with  very little energy carried by nucleons with momenta above the Fermi momentum.}. 
Understanding the dynamics relevant to SRC is important also for building a realistic equation of 
state for dense nuclear matter such as neutron stars where typical 
internucleon distances in the core are close to those 
encountered in SRCs $\sim 0.5-1.5$~fm.

In this respect the key questions in studying the dynamics of SRC's are
\begin{itemize}
\item How large are the probabilities of SRCs in nuclei ? 
\item What is the isotopic structure of SRCs?
\item Are there significant three nucleon SRCs?
\item How significant are non-nucleonic degrees of freedom in the SRC?
\item What is kinematical range of applicability of the concept of SRC in QCD?
\item What is the impact of SRCs on the dynamics of compact stars: neutron stars, hyperon stars etc?
\end{itemize}

\subsection{SRC observables and strategy of their studies}

Identifying the processes in which one can unambiguously 
probe SRCs including  their {\it microscopic} properties 
is one of the goals of the present  review.  Nucleon momentum distributions in nuclei as well as  nuclear spectral and decay 
functions represent the set of observables which elucidate the different aspects of 
the dynamics of SRCs, in particular: (i) the nature of SRCs as high density fluctuations of 
nuclear matter, (ii) dynamical correlation between 
initial momentum of struck nucleon  and energy of residual nuclear system 
associated with a removal of a nucleon from two and three nucleon SRCs in the nucleus, and 
(iii) the isospin content of SRCs.
These questions are discussed in details in Sec.2.

First we discuss the conditions under which 
it is possible to probe SRCs 
by
suppressing contributions associated with long range (low momentum) 
processes in nuclei.

For many years SRCs were considered as an important but elusive 
feature of the nuclear structure. Referring to SRCs as elusive was 
due to lack of low energy processes which are dominated by the 
high momentum component of nuclear wave 
function\footnote{It was argued in Ref. \cite{Amado} that 
high momentum component of nuclear wave function could not be observed  even in principle due 
to inability to separate the influence of measuring process from the measured quantity. 
The argument was based on implicit assumption of the proximity of scales characterizing 
measuring process and structure of SRCs. Such arguments were valid for  intermediate 
energy processes in which  cases  energy and momentum  scales characteristic 
for the measuring process  and for the SRCs are comparable.}. 
It was  shown in Refs. \cite{FS77} and \cite{FS81} that the fundamental problem was the 
use of the processes with energy-momentum  scales comparable to that of the SRCs and 
that situation should drastically improve for high energy processes 
in which one can select kinematics corresponding to an energy and momentum transfer 
scale much larger than  the scale characteristic  for SRCs:
\begin{equation}
q_0 \gg V_{NN}, \ \ \ \ \ \ |\vec{q}| \gg 2k_F.
\label{minenergy}
\end{equation}
This condition is well satisfied in high energy  projectile-nucleus quasielastic and inelastic interactions. 
As a result, generic lepton/hadron-nucleus processes  could be treated as  instantaneous  
as compared to the nucleon motion within SRC\cite{FS77}---\cite{FS76}.
In such processes energy and momentum  transfered
to one of the nucleons in SRC  significantly exceed
relevant energies and momenta in SRC, leading to an effective  release of  nucleon spectators from the SRC. 
Formally this process is described by the decay function of the nucleus which will be defined below. 
Processes associated with a release of spectators produce 
significant correlation properties of the nuclear decay function, which 
can be used for identification of SRCs.  

Possibility of instantaneous removal of nucleon from nucleus 
in  high energy processes greatly simplifies  identification of the observables that are sensitive to 
the high momentum component of nuclear ground state wave function.  Therefore unambiguous identification 
of short  range nucleon correlations in a 
nucleus  requires an  effective use  of the  resolution power of high momentum transfer processes. 

\begin{figure}[ht]
\centering\includegraphics[height=6cm,width=12cm]{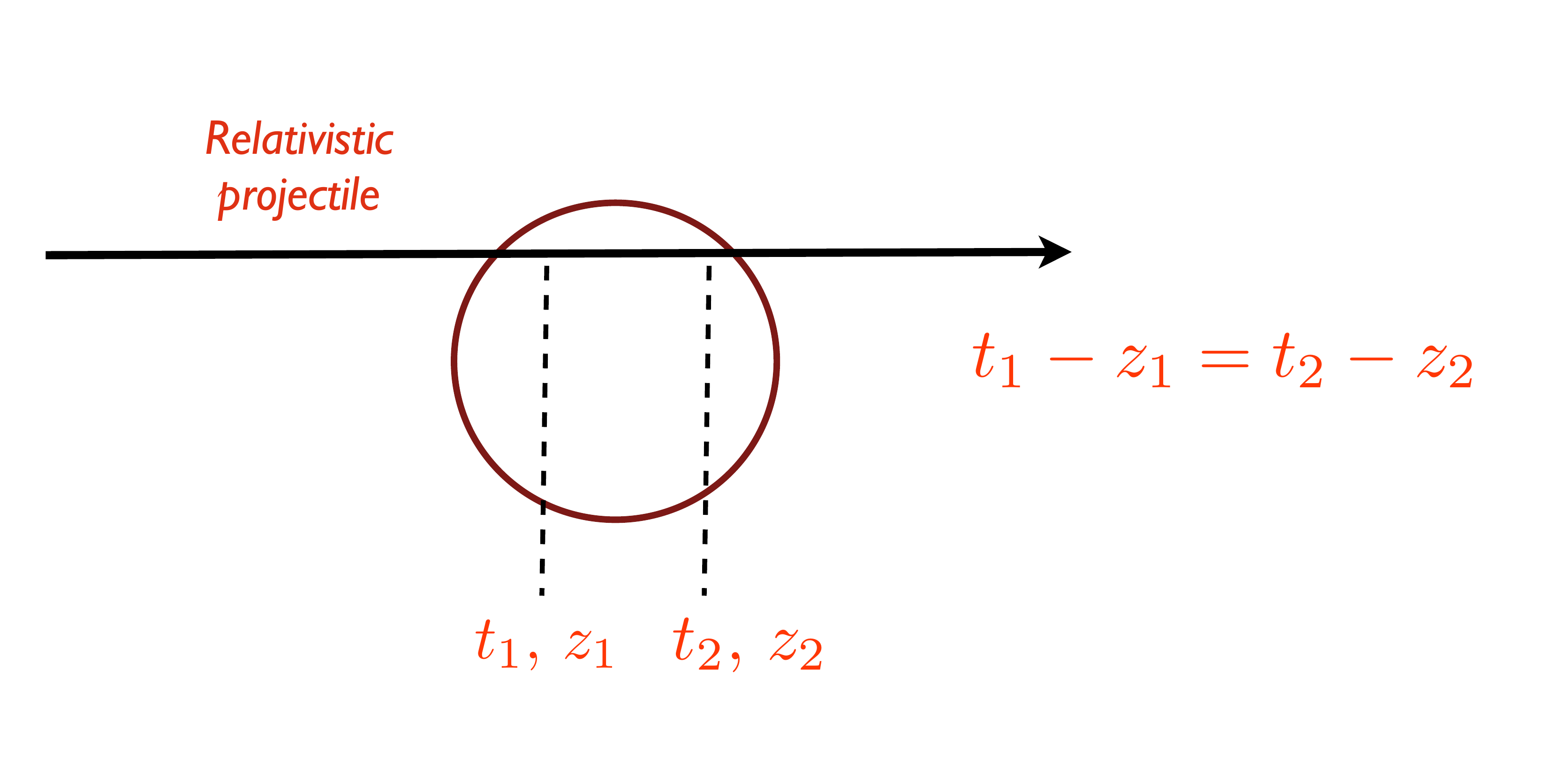}
\caption{Fast projectile interacting with nucleus selects a light-cone slice of the wave function.}
\label{lcsk}
\end{figure}

The theoretical challenge of using high energy and momentum 
transfer reactions is that projectile and some of the final state 
particles move with relativistic velocities making it impossible to 
apply directly non-relativistic approaches  for description of the nuclear wave functions  
as well as the scattering process itself. 
High energy projectile moving along the z-direction probes 
the light-cone (LC) slice of nuclear wave function near the hyperplane 
$t-z=const$~(Fig.\ref{lcsk}) - the LC 
wave function of the nucleus: $\psi_A(\alpha_1,k_{1,t},...\alpha_i,k_{i, t}, ...\alpha_A,k_{A,t})$, 
where 
\begin{equation}
\alpha_i= A\left( {E_{i}-p_{i, z}\over E_{A}-p_{A, z}}\right),
\label{alphai}
 \end{equation}
are the light-cone fractions (scaled by A) of the nucleus momentum  carried by 
constituent nucleons ($\sum\limits_{i=1}^A \alpha_i=A $). Here, ($E_i$, $p_{iz}$) and ($E_A,p_{Az})$ 
are the energy and longitudinal momentum of constituent nucleons and target nucleus respectively.
Due to the invariance of $\alpha_i$ with respect to the Lorentz boosts in $z$ direction,
in the nucleus rest frame $\alpha_i= A\left( {E_{i}-p_{i, z}\over M_A}\right)$,
where $E_i$, and $p_{i,z}$ now are the lab energy and $z$ component of bound nucleon in the nucleus 
with mass $M_A$.

The important feature of nuclear light cone wave function is that  there exists a simple  
connection between  LC and nonrelativistic wave functions of a nucleus:
  \begin{eqnarray}
& &  \psi_{nr}(\vec k_1,...,\vec{k}_i,...\vec k_A) = \nonumber \\
& & \ \ \ \ (m_N)^{-{A\over 2}}\psi_{LC}
\left(\alpha_1=1+{k_{1,z}\over m_N},k_{1,t},..,\alpha_i=1+{k_{i,z}\over m_N}, k_{i,t},..,
\alpha_A = 1+ {k_{A,z}\over m_N},k_{A,t}\right), \nonumber \\
\label{nr-lc}
 \end{eqnarray}
at $k_i\ll m_N$ for $i=1,...,A$.
Therefore the  knowledge of the nuclear LC wave function  allows us to study the rest frame 
nonrelativistic nuclear wave function as well. However for  large nucleon momenta  in the nucleus 
the correspondence between nonrelativistic and light cone wave functions  becomes more 
complex especially for the case of SRC of more than two nucleons.

\subsection{Short survey of recent progress in SRC studies}

For a long time the only class of high  energy processes which was systematically studied 
experimentally and which appeared to be dominated  by a projectile scattering off the SRC was 
production of fast backward nucleons and pions from nuclei in reactions 
(see Ref. \cite{FS81} and references therein):
\begin{equation} 
\gamma(\nu, \pi, p) +A\to \mbox{fast backward} \, p \,  (\pi) +X.
\label{fbn}
\end{equation}
It was demonstrated back in 1977~\cite{FS77}   that the data for the reaction (\ref{fbn}) 
for $\gamma+^{12}C$  scattering at high energies ($E_{\gamma} \ge 2$~GeV)  can be described 
as due to  the  
{\it decay of SRC after the inelastic interaction  of photon with  one of the nucleons of two-nucleon  SRC}. 
The  wave function of the SRC was found to be proportional to the deuteron wave function for 
$300 \le k\le 800 \ MeV/c$ with a proportionality coefficient $a_2(^{12}C)=4\div 5$.
The similarity of  emission spectra for interaction of different projectiles with 
lightest~($^2H,^4He$) and heavy~($Pb$) nuclei in kinematics in which 
scattering off the low momentum nucleons could not contribute, as well as  several other 
regularities have been naturally  explained based on the few-nucleon correlation model~\cite{FS81}.  
However these inclusive processes did not allow to reconstruct a complete final state of the reaction 
and therefore to perform quantitative investigation of the SRC structure of nuclei.  Rather direct 
confirmation of the significant role of SRC has been obtained in the processes of neutrino~(antineutrino)  
scattering off nuclei in the  observation of correlation between momenta of backward nucleon  and forward  
muon~\cite{nuA1,nuA2} which was predicted in Ref. \cite{FS77}.
There was also an evidence for universality of SRCs coming from the comparison~\cite{FS88,FSDS} 
of $A(e,e')X$  cross sections for different nuclei at $x\ge 1$ and  $Q^2 \ge 1$~GeV$^2$  measured 
in several different experiments performed in the 1980's.

A new qualitative and quantitative progress in determination of   the structure of  SRC have been  
achieved recently on the basis  of two new  theoretical ideas:
(i) that the presence of two energy-momentum scales:  high energy scale for the probe 
and lower energy scale for nuclear phenomena,  justifies the 
application of the closure approximation. For $A(e,e')X$  reaction at $x > 1$ and 
$Q^2 > 1.5$~GeV$^2$  closure application can be applied 
up to the final state interaction effects within  SRC which are  practically
independent  of $A$. The latter observation  leads to the scaling of the ratios of 
cross sections of different nuclei (see Sec.3); 
(ii)~observation that hard exclusive processes, in which  a nucleon from SRC is removed 
instantaneously, probe a new observable: the nuclear decay function $D_A(k_1,k_2,E_R)$  
\cite{FS77,FS81,FS88,SASF2}, which represents a  probability of emission from the nucleus 
of a nucleon with momentum $k_2$  after the removal of a fast nucleon with momentum $k_1$ and leading 
to a residual nuclear state with recoil energy $E_R$~(see Sec.2 for details).  Although in general 
$D_A$ is a very complicated function of its variables, in the case of  the removal of a nucleon 
from two-nucleon SRC its form is rather simple:  the second nucleon of the SRC is  released with 
momentum $\vec k_2\sim  - \vec k_1$. 
In this case emission of nucleons with $\vec k_2$ which are spatially far from the SRC 
from which the nucleon with $\vec k_1$ is  removed
does not contribute to $D_A$.
Decay functions can be evaluated  based on
nuclear models which include SRCs
in a rather wide  kinematical region.

For more than a decade there were practically no new experimental data in this field. 
In the last three years a qualitative progress was reached as the  new $A(e,e')X$ experiments 
have been performed at Jefferson Lab\cite{Kim1,Kim2} while studies of processes sensitive to the 
properties of the decay function in kinematics dominated by SRC were performed first at 
BNL\cite{eip1}---\cite{eip3} 
and then at JLab \cite{eip4,eip5}. 

Combined analysis of these data indicates that a) probability of short range correlations in carbon 
is $\sim 20\%$, b) these  correlations are predominantly consist of two nucleons,  
c)  probability for two protons to belong to a SRC as compared to that of  a  proton and neutron is very 
small $\sim 1/20$. These observations are in line with the concept of tensor forces 
dominating at intermediate to short distances in $I=0$, and $S=1$ channels of $NN$ interaction 
(see e.g. discussion in Refs. \cite{Paris}--\cite{Alvioli07}.
All these observations together demonstrate strong potential of high energy processes for addressing 
long standing issues of the short-range  nuclear structure.

\medskip
\medskip

In this review  we outline theoretical expectations of 
realistic nuclear models that describe SRCs, summarize new 
information about properties of SRCs obtained recently in high energy processes,   
discuss compatibility of the ideas of nuclear theory with basics of QCD, 
outline  implications of the discovery of SRCs for the theory of neutron stars and outline 
directions for the future studies both in theory and in high energy nuclear experiments.
In particular we emphasize studies of isotopic structure of SRCs relevant to the 
study of internucleon forces in the region of the nuclear core,
three-nucleon correlations, $\Delta$-isobar admixture, isospin effects etc.

\section{The status of high momentum component of nuclear wave function in nonrelativistic theory}

In this section we discuss the manifestations
of the high momentum component of nuclear wave 
function in the properties of nucleon momentum distribution, $n(k)$, spectral function,  $S(k,E)$ 
as well as decay function $D(k_1,k_2,E_r)$ (to be defined below).
In particular,
we  will demonstrate how one can identify the signatures of 
two- and  three- nucleon SRCs  in these quantities.
We will also discuss nuclear reactions in which the above functions can  be measured.  
Our consideration in this section is restricted to the  nonrelativistic theory, 
though a number of arguments indicate that many SRC related properties we 
discuss will reveal themselves in a similar way in  the relativistic theory.

\subsection{Momentum Distribution}

The nucleon 
momentum distribution $n_{A}(k)$ is given by 
the modulus square of the ground state nuclear wave function
integrated over all nucleon momenta except one, 
\begin{equation}
n(k) =\sum\limits_{i=1}^A\int \psi_A^2(k_1,k_2,k_i,...k_A)\delta^3(k-k_i)
\delta^3(\sum\limits_{j=1}^{A} k_j) \prod\limits_{l=1}^Ad^3k_l.
\label{nkdef}
\end{equation}

Properties of $n(k)$ at high momentum, $k\gg k_{Fermi}$ follow
directly from  the Schr\"{o}dinger equation in the momentum space.  In general for 
given two-nucleon interaction potential, $V$, the ground state wave function, $\psi_{A}$ 
satisfies the equation:
\begin{eqnarray}
(E_B - {k^2\over 2 m} -\sum_{i=2,..A} T_i) \psi_{A} & = &  
\sum_{i=2,...A} \int V(k-k'_i)\psi_{A}(k, k'_i,...k_j,...k_A) {d^3k'_i\over (2\pi)^3} 
\nonumber \\
& + & \sum_{i=2,...A}
\int V(k_i-k'_i)\psi_{A}(k, k'_i,...k_j,...,k_A) {d^3k'_i\over (2\pi)^3},
\label{Shreq}
\end{eqnarray}
where  $E_B$ is nuclear binding energy and $T_i$ are kinetic energies
of nucleon-spectators and $V(k) = \int V(r) e^{-i(kr)}d^3r$  is the 
$NN$ potential  in the momentum space. 

\subsubsection{Theorem on high momentum tail of nuclear ground state wave function}
Based on Eq.(\ref{Shreq}) it can be  proven that if the 
potential decreases at large $k$, like $V(k)\sim {1\over k^n}$  
and $n>1$  
then the $k$ dependence of the wave function for 
$k^2/2m_N\gg |E_B|$ is calculable in terms of the potential $V$ as follows:
\begin{equation}
\psi_{A} \sim {V_{NN}(k)\over k^2} f(k_3,...k_A),
\label{psi_asym}
\end{equation}
where $f(k_3,...k_A)$ is a smooth function of spectator nucleon's momenta with $k_2 \sim -k$.
To prove  the theorem it is sufficient to show that 
all higher order iterations will decrease faster with $k$ and therefore
preserve the form of Eq.(\ref{psi_asym}).  In the case of nucleon-nucleon potentials 
which contain both repulsive core and medium range attraction 
the accuracy of this equation maybe worse, however numerical studies described below are consistent with Eq.(\ref{psi_asym}).

\subsubsection{Nuclear wave function and short range correlations}

In nonrelativistic nuclear theory interaction potential  is constructed 
as a  sum of potentials involving 
two, three, and higher number of  nucleons. 
All realistic $NN$ potentials  deduced from fitting 
the NN scattering phase shifts at 
large $k$ have the property that 
the potential of two-nucleon interaction,
$V_{NN}$, decreases significantly slower with an increase of  $k$ 
than triple  and higher order nucleon potentials. 
(Such a behavior arises naturally if many body interaction  
results from the  iteration of 
%
two- 
nucleon interactions).
As a result, at large $k$ limit 
the contribution of the pair nucleon  potential $V_{NN}(k)$ in which
two nucleons have large relative momentum $k$ will dominate.
This will justify the use of Eq.(\ref{psi_asym}) for calculation of 
asymptotic form of the ground state nuclear wave function at large $k$. 
Consequently, using Eq.(\ref{nkdef}) one arrives at the 
asymptotic form of momentum distribution function:
\begin{equation}
n(k) \sim \left( {V_{NN}(k)\over k^2}\right)^2.
\label{momlargek}
\end{equation}
The above relation can be improved by taking into account the center of mass motion of the 
NN pair. However  the latter  effect decreases with increase of $k$.

\begin{figure}[ht]
\centering\includegraphics[height=8cm,width=12cm]{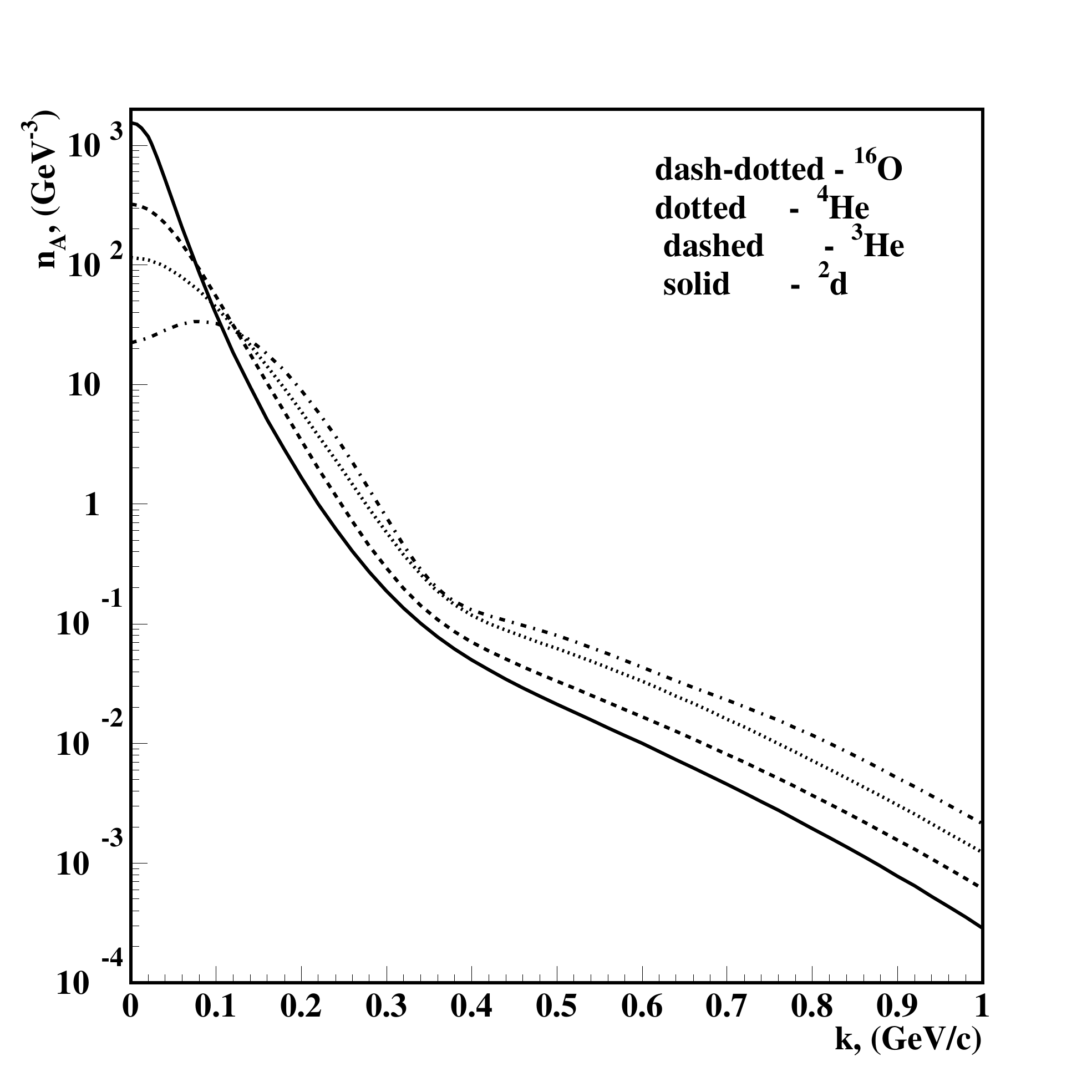}
\caption{Comparison of the momentum distributions calculated for different nuclei in Ref.24.}
\label{momdis}
\end{figure}

The direct consequence of Eq.(\ref{momlargek}) is the dominance of two nucleons SRCs in the 
high momentum part of  $n_A(k)$.
This results in the similarity of  the shapes of $n_A(k)$ for different nuclei at 
$k> 300$~MeV/c. This similarity is clearly seen in Fig.\ref{momdis} 
where momentum distributions for $^2H$, $^3He$, $^4He$ 
as well as $^{16}O$ from Ref. \cite{Pieper} are compared.

\begin{figure}[ht]
\centering\includegraphics[height=8cm,width=12cm]{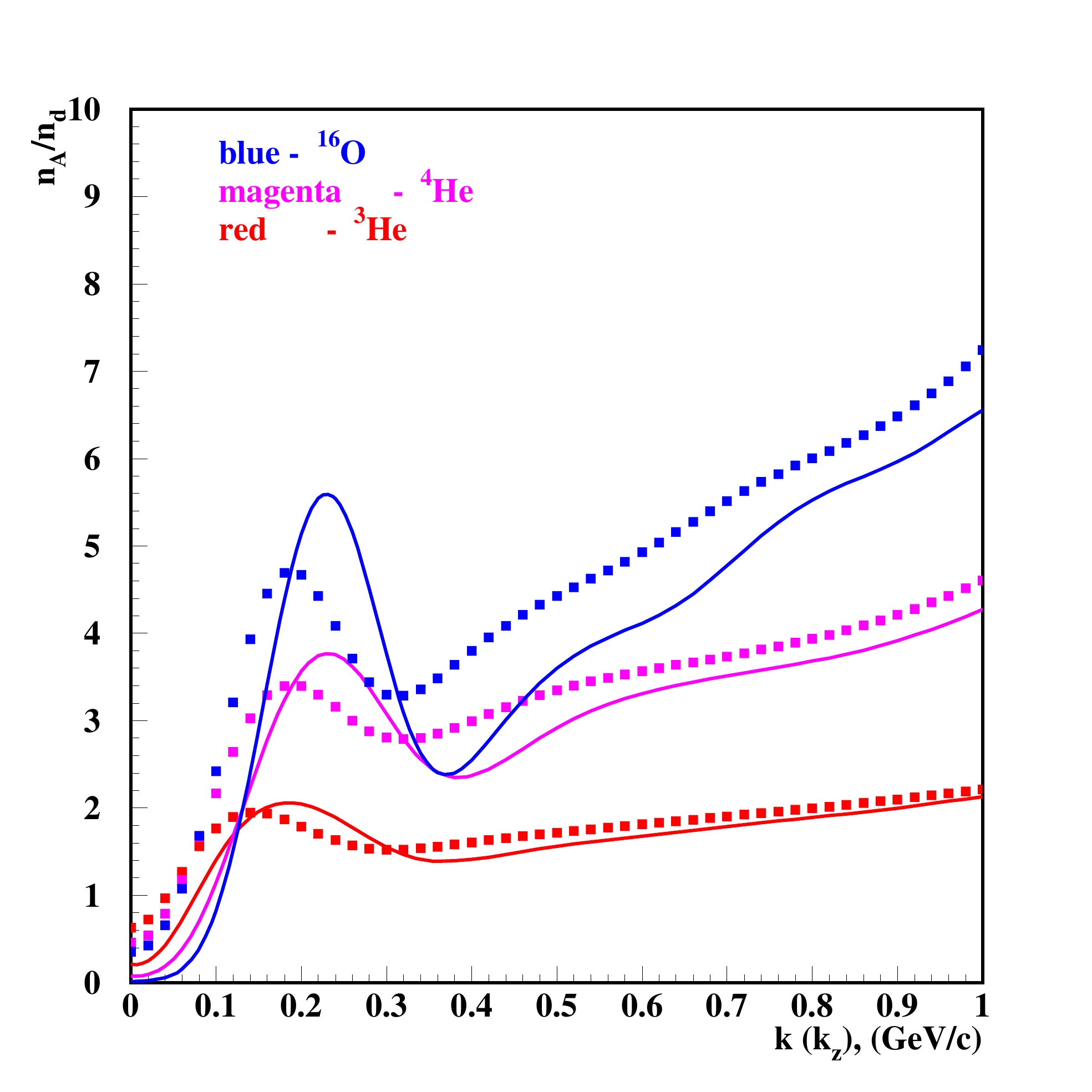}
\caption{Curves describe  momentum dependence of the ratio, ${n_A(k)\over n_d(k)}$.
Points represent  calculated  $k_z$ dependence of the ratio ${n_A(k_z)\over n_d(k_z)}$.}
\label{nkr1}
\end{figure}

\begin{figure}[ht]
\centering\includegraphics[height=8cm,width=12cm]{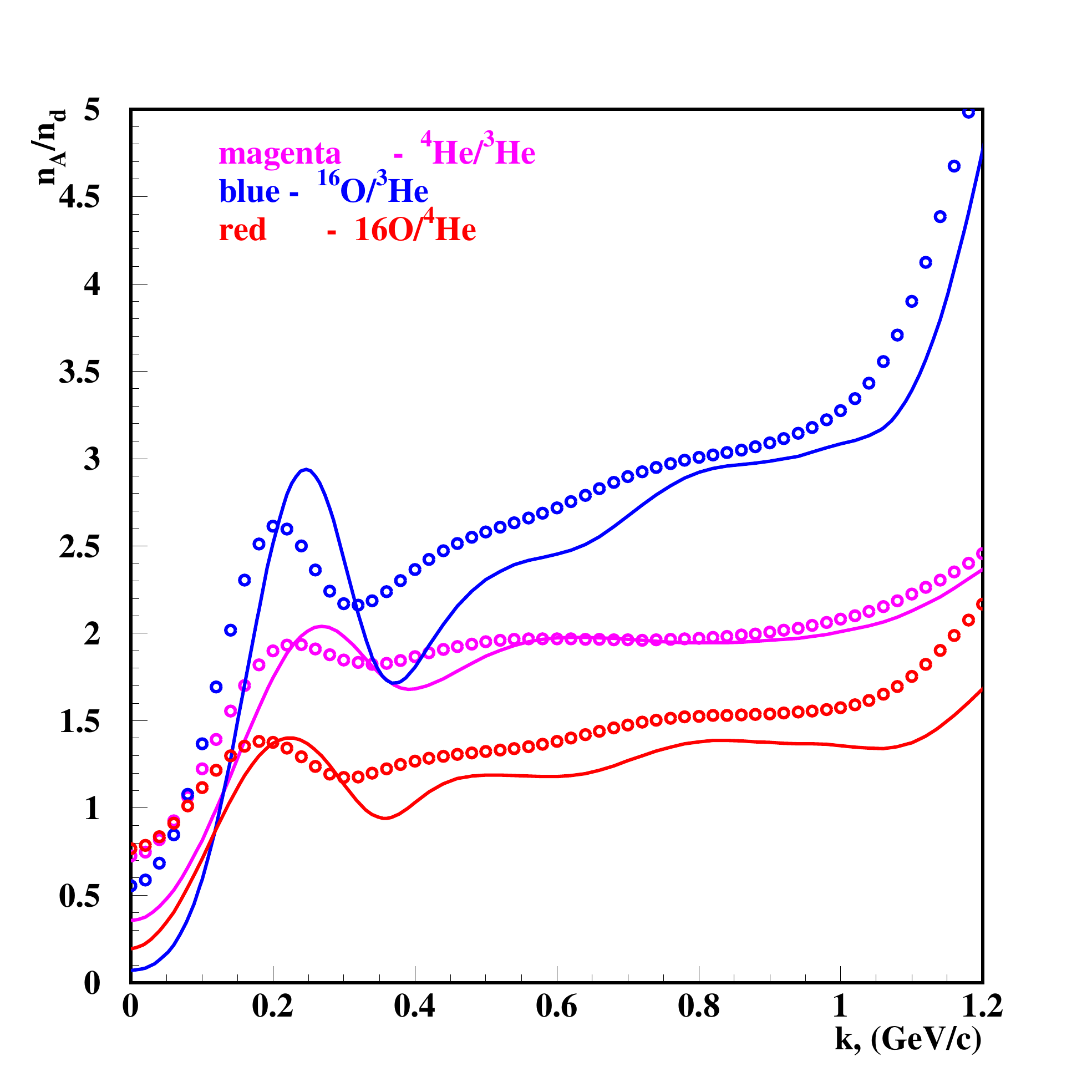}
\caption{Curves correspond to the momentum dependence of ratio, ${n_A(k)\over n_{^3He}(k)}$ 
and ${n_A(k)\over n_{^4He}(k)}$ .
Points are the calculated  $k_z$ dependence of the ratio
 ${n_A(k_z)\over n_{^3He}(k_z)}$. 
and ${n_A(k_z)\over n_{^4He}(k_z)}$.}
\label{nkr2}
\end{figure}

The ratios of nucleon momentum distributions are given in 
Fig.\ref{nkr1} and Fig.\ref{nkr2}.  It is spectacular that while the absolute magnitude of momentum distributions drop by  
three orders of magnitude in $0.3 < k < 1$~GeV/c range, the ratios 
$n_{A}\over n_d$ for a given nucleus does not change significantly.  
It appears, that the non-uniformity of $n_{A}\over n_d$ ratios in 
Fig.\ref{nkr2} is related to  the difference between $NN$ potentials for spin $0$ and $1$ states 
and to the center of mass motion of NN pair in nucleus $A$.
This  can be seen in Fig.\ref{nkr2}, where the ratios 
$n_{A}\over n_{^3He}$ and $n_{A}\over n_{^4He}$ show significantly 
weaker momentum dependence at $k> 400$~MeV/c. 

Other calculations of nuclear momentum distribution, $n(k)$,  within 
nonrelativistic nuclear theory with realistic NN potentials are   
also consistent with the dominance of two-nucleon correlations for momenta above 
300~MeV/c. Moreover numerical studies~\cite{Forest,Neff03,Alvioli05}  confirm 
that the dominant contribution in the momentum range 
$350 \le k \le 700 $MeV/c  is due to the tensor forces which dominate 
in NN channel with isospin $0$ and spin $1$.  Independent evidence for the dominance of $pn$ 
correlations comes from the study of two nucleon 
momentum distributions in nuclei~\cite{Schiavilla,Alvioli07}.

Note that quantity often used for the analysis of data is the  partially integrated momentum distribution:
\begin{equation}
n(k_z) = \int n(k) d^2k_\perp.
\label{kz}
\end{equation}
One can see from Figs.\ref{nkr1},\ref{nkr2} that
the  scaling is even more pronounced for these quantities.

It is worth mentioning that although $n_{A}(k)$ is the simplest function which can 
be constructed from the ground state nuclear wave function, it  
cannot be observed directly in any  processes.
The spectral function is the simplest quantity which is related to the cross sections 
of physical processes namely to $A(e,e'N)X$ and $A(e,e')X$ processes\footnote{ 
In principle, one could infer $n_A(k)$ from 
the sum rule $n_A(k)=\int dE_R S(k,E_R)$.}. At the same time as we will see below 
the $A(e,e')X$ cross section at large $Q^2$  is expressed through the light-cone nuclear density matrix 
which illustrates the nontrivial relation between light-cone and nonrelativistic quantities.

\subsection{Spectral Function}

The asymptotic behavior of 
$n_{A}(k)$  for  $k\to \infty$, according to Eq.(\ref{momlargek}), contributes only to the range of 
$\alpha< 2$ for the light-cone nucleon density $\rho_{A}(\alpha,k_t)$. This result follows 
from the account of the energy-momentum conservation for the whole process which is contained in the 
spectral function.

Reaching the region of $\alpha > 2$ is possible only if at least three 
nucleons are involved in the process. This does not necessarily require  specific 
three  body forces - an iteration of two nucleon interactions is in principle sufficient.
However in a wide range of the nucleon momenta, $k$ above the Fermi momentum  and for 
recoil nuclear energies close to ${k^2\over 2m_{N}}$ the two nucleon SRC approximation 
with the motion of NN pair in the mean field of nucleus taken into account  provides a 
good approximation for both nonrelativistic and light-cone description of nuclei.
This approximation is effective in the calculation of the nuclear spectral function:
\begin{equation} S_A(p_{i},E_R)=\left|\left<\phi_{A-1}\left|\delta(H_{A-1}-E_m)a(k)\right|
\psi_A\right>\right|^2,
\label{spectral}
\end{equation}
which represents a product of the probability of finding 
a nucleon in the nucleus with initial momentum $p_{i}$, and the probability that  
after instantaneous removal of this nucleon the   residual system will 
have recoil energy $E_R$. Note that in the traditional definition one separates 
the recoil energy $E_R$ into the sum of two terms  - $E_m$, the excitation energy of 
the $A-1$ system in its center of mass,  and the  
kinetic energy of the center of mass  itself -  ${p_{i}^2\over 2m_{A-1}}$. 
Although such definition is very convenient for  the case of nucleon removal from nuclear shells, 
it gives a less transparent pattern of  properties of the spectral function in the case of removal 
of nucleons from SRCs. Indeed,  removal of the nucleon from say a two nucleon SRC leads to 
nearly universal $E_R$ distribution for the spectral functions for different nuclei, while 
$E_m$-distributions strongly depend on $A$ especially for light nuclei.

Within the  plane wave impulse approximation the spectral function  defined above   is related 
to the differential cross section of $A(e,e'N)X$ reaction as follows:
\begin{equation}
{d\sigma\over d\Omega_{e^\prime}dE_{e^\prime} d^3p_f d E_{R}} = 
{j_{N}\over j_{A}}\sigma_{eN}\cdot S(p_{i},E_R).
\label{een}
\end{equation}
Here $j_N$ is the flux calculated for moving bound nucleon with momentum $p_{i}$, and 
$\sigma_{eN}$ represents the cross section of electron- ``bound-nucleon'' scattering.
In the limit of $Q^2\ge 1~GeV^2\gg p_{i}^2$ the above form of factorization
can be used also for more realistic case in which final state interactions~(FSI) are taken 
into account. However in this case the spectral function is modified to 
$S^{DWIA}(p_f,p_{i},E_R)$ which now contains the factors that account 
for FSI as well as modification of the flux factor in the 
rescattering part of the spectral function. This approximation is usually 
referred to as distorted wave impulse approximation\footnote{It is possible also to perform 
unfactorized calculation of FSI in which case  the interpretation of 
the scattering cross section through the spectral function is not possible. However 
at the kinematics in which FSI is a correction, effects due to unfactorization are 
insignificant.}

\begin{figure}[ht]
\centering\includegraphics[height=4cm,width=12cm]{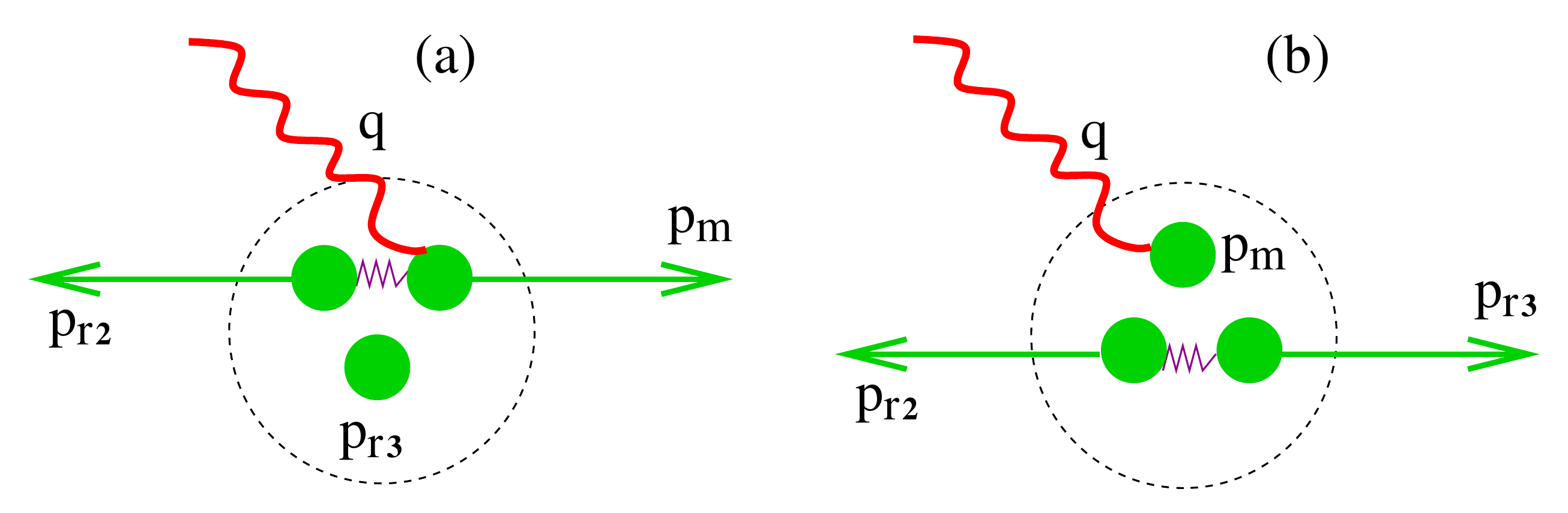}
\caption{Interaction of virtual photon with three nucleon system in configurations in which 
two of the nucleons are in SRC.}
\label{2Nsrc}
\end{figure}

\begin{figure}[ht]
\centering\includegraphics[height=8cm,width=12cm]{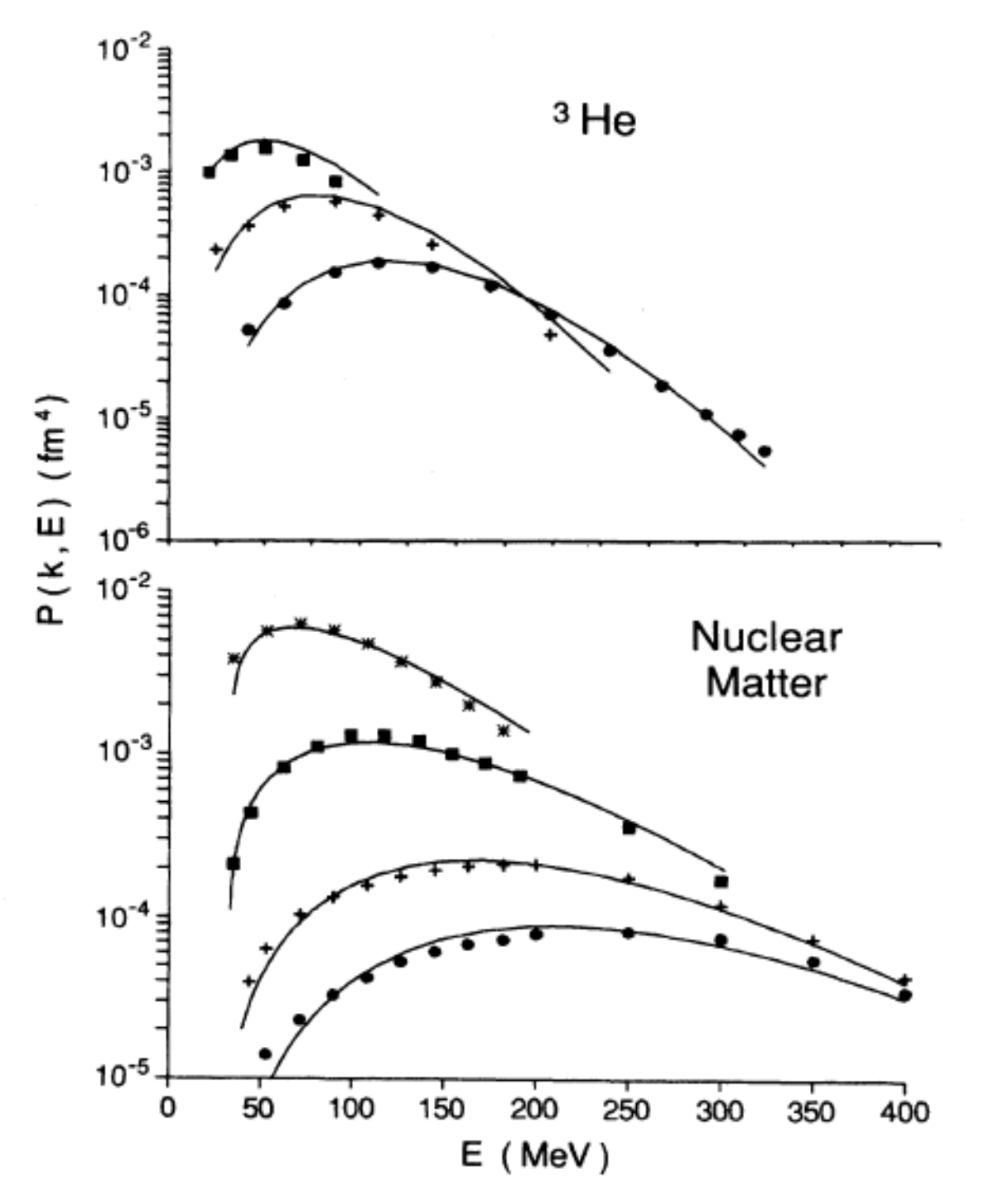}
\caption{Comparison of the two nucleon correlation model for the spectral function with direct 
calculations of A=3, and nuclear matter spectral functions.}
\label{twonciofi}
\end{figure}

In the case of spectral function
SRCs manifest themselves clearly in the properties of the recoil nuclear system  
when a fast nucleon is removed from the nucleus. 
In this case the second nucleon is effectively removed from the nucleus as the potential between 
this and the struck nucleon is destroyed  instantaneously.
This mechanism of breaking of correlations we will refer hereafter as  type 2N-I SRC mechanism  
of breaking SRC (Fig. \ref{2Nsrc}a).
Since momenta of the nucleons in NN correlation in average are equal and opposite, one finds for the average 
recoil  energy of the  residual system \cite{FS88}
\begin{equation}
<E_R> ={p_{i}^2\over 2m_N},
\label{averrec}
\end{equation}
as the $A-2$ system is essentially not perturbed during removal of the  nucleon from 2N SRC. 
Eq.(\ref{averrec}) agrees well with numerical studies of spectral functions for $A=3$ and 
infinite matter \cite{FSCS}. Moreover the distribution over $E_R$ calculated  
in Refs. \cite{Ciofi3Aa}--\cite{Fantoni} 
is well described by the model which takes into account 
the motion of the $NN$ pair in the mean field\cite{FSCS,CiofiSimula}(see Fig.\ref{twonciofi}).

In many cases spectral functions  are modeled based on 
type 2N-I SRCs  modified only by taking into account the mean 
field momentum distribution of the  center of mass motion of SRC in the 
nucleus~(see e.g. Refs. \cite{FS88,FSCS,CiofiSimula}) . 
Such approximations describe well the existing  data and  agree reasonably well with numerical  
calculations based on two-nucleon potentials only~(see e.g. Ref. \cite{Fantoni}).  
Note however, that  these calculations 
did not include contribution  of three particle three hole excitations so a good agreement 
of the models of Ref.\cite{FSCS} and \cite{Fantoni} may reflect deficiency of both models.

If the nucleon with small momentum is removed, the residual system is predominantly in one 
of the lowest $A-1$ nucleon states.  The  contribution of SRCs into large recoil
energy range is strongly suppressed in this case as compared to the expectation based on the 
total probabilities of SRCs.  In the simplest case of $A=3$ system to observe suppression of 
the large  recoil energies  for the case of removal of a nucleon with momentum $p_{n}\sim 0$ 
(Fig.\ref{2Nsrc} b) one needs to take into account the difference between "off-energy  shell"  and 
"on-energy-shell"  t-matrices of $NN$ scattering. For further discussion see Sec.2.3.

Note that mere observation of the correlation 
given by Eq.(\ref{averrec}) in  $A(e,e'N)X$ reactions will 
not allow conclude unambiguously that the spectral function, $S(p_{i},E_R)$ is 
sensitive to the SRC.
In general,  Eq.(\ref{averrec}) is satisfied for any reaction dominated by $two-nucleon$ 
processes 
with two nucleons not necessarily belonging to a SRC (for example contribution due to meson 
exchange currents).
However, if an additional kinematic conditions such as Eq.(\ref{minenergy})
are satisfied allowing to suppress  long range two-nucleon processes, one could use 
the relation Eq.(\ref{averrec})  to check the dominance of the SRCs.

\begin{figure}[ht]
\centering\includegraphics[height=8cm,width=12cm]{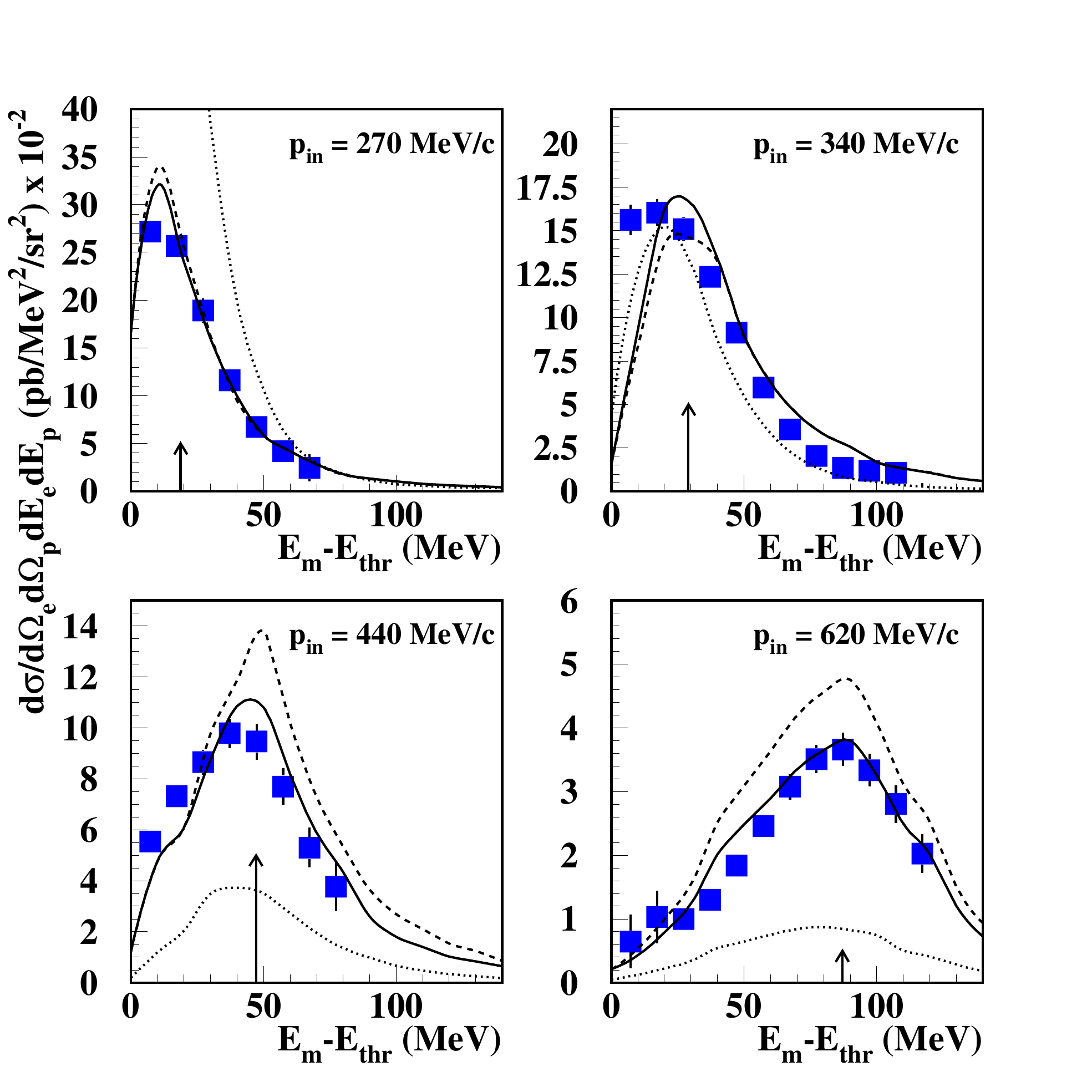}
\caption{The dependence of the differential cross section on the missing energy, for 
$^3He$ three-body break up reactions at different values of initial nucleon 
momenta. Dotted, dashed and solid curves corresponds to PWIA, PWIA + single rescattering 
and PWIA + single + double rescatterings. Data are from Ref.33. Arrows
define the correlation according to Eq.(\ref{averrec}).
Similar description of the data is achieved in Refs.34 and 35.}
\label{3bbu}
\end{figure}

In this respect it is worth  mentioning the recent measurement~\cite{Benmokhtar} of three-body 
break  up reaction of $^3He$ in kinematics satisfying condition of  Eq.(\ref{minenergy}). These 
experiment observed clear correlation consistent with Eq.(\ref{averrec})~(Fig.\ref{3bbu}). 

The comparison  of calculations based on DWIA\cite{eheppn1,eheppn2} 
with the data\cite{Benmokhtar} demonstrates
that  a substantial contribution from final 
state reinteraction not only preserves the pattern of the correlation of Eq.(\ref{averrec}) but 
also reinforces it~(Fig.\ref{3bbu}).
This indicates a rather 
new phenomenon, that 
in high energy kinematics sensitive to SRC, FSI is dominated by single rescattering 
of struck nucleon with a spectator nucleon in SRC. As a result this rescattering 
does not destroy the correlation property of the spectral function.

\begin{figure}[ht]
\centering\includegraphics[height=5cm,width=10cm]{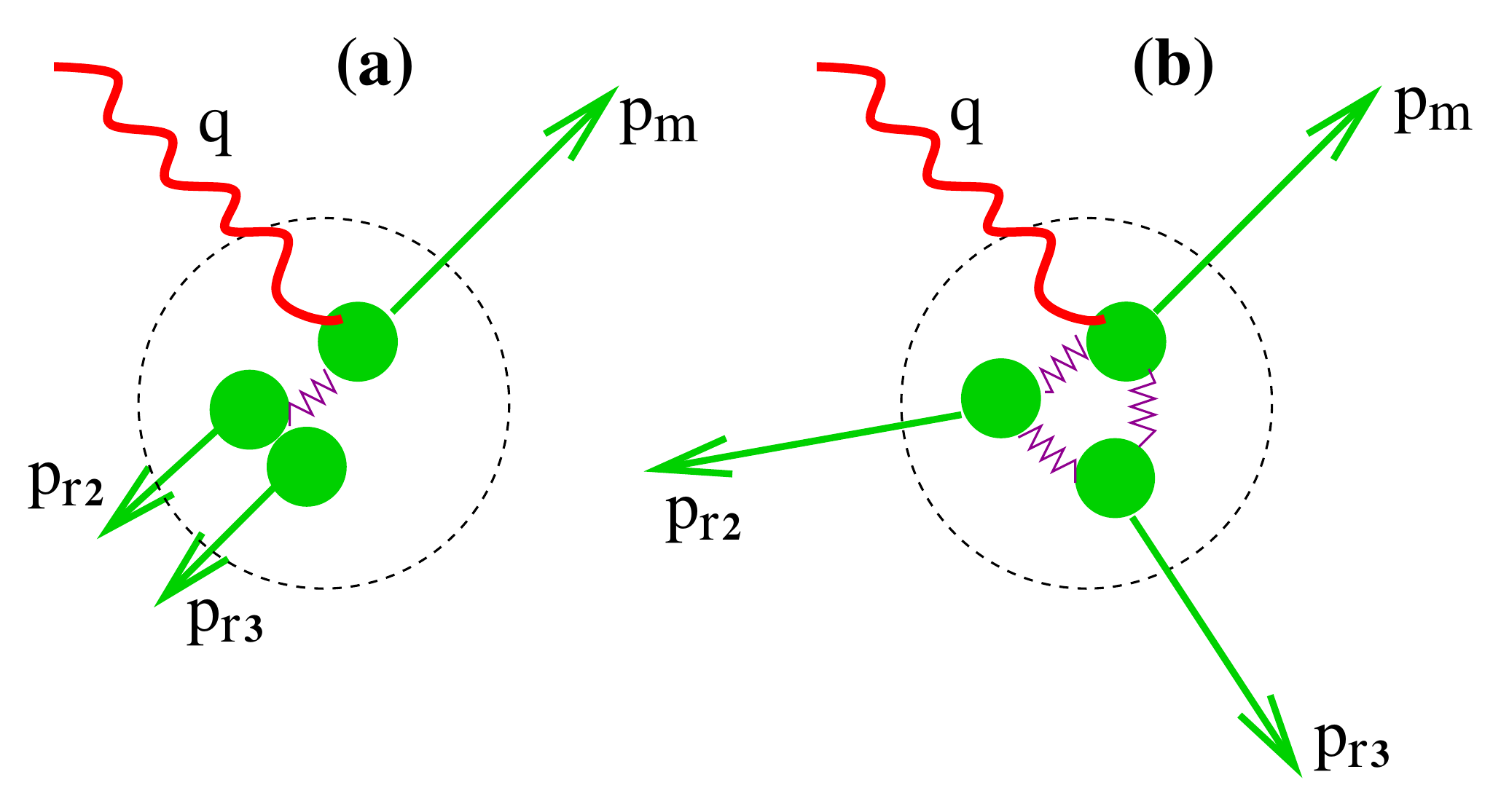}
\caption{Two different scenarios of probing  NNN correlations.}
\label{3Nsrc}
\end{figure}

Three nucleon~(3N) SRCs also contribute to the spectral function. 
In Fig.\ref{3Nsrc} we  consider two scenarios for 3N SRCs that can be evolved from 
2N correlations with an increase of the c.m. momentum of 2N-SRCs.
The one, which we refer to as type 3N-I SRC,
(Fig.\ref{3Nsrc}a) corresponds to the situation in which 
 initial momentum of struck nucleon (typically $p_{i}\ge 600$~MeV/c) is shared by 
two spectator nucleons with invariant mass close to $2m_N$\footnote{In reality
integral over the recoil momentum should give slightly larger 2N recoil mass, 
$m_{23} > 2m_N$
}. 
Such configurations  at $p_{i}\ge 600$~MeV/c 
should dominate for average recoil energy of
$E_R\sim p_{i}^2/4m_N$ in the spectral function \cite{FS88,eheppn2}, 
which for the case of $A=3$  
 corresponds to a rather constant value of missing energy 
$E_{m} \approx |\epsilon_A| + \epsilon_{2N}$. Here
$\epsilon_A$ and $\epsilon_{2N}$ 
represent  nuclear binding and 2N excitation energies.
Thus effects of type 3N-I SRCs should 
 be manifested in 
the  strength of the spectral function at $p_{i}\ge 600$~MeV/c 
and recoil energies  which (similar to the 2N SRCs) 
 have universal, $A$-independent 
values: 
\begin{equation}
 <E_R>\sim  {p_{i}^2\over 4m_N}.
\label{er3n-i}
\end{equation}
It follows from Eq.(\ref{er3n-i}) that the value of $<E_R>$ in this case is  approximately equal to
one half of the recoil
energy that characterizes type 2N-I correlations~(Eq.(\ref{averrec})).
These configurations give dominant contribution to $A(e,e')X$ cross section at  $2< x < 3$ and 
at large $Q^2$.
 
Another type of 3N SRCs (Fig.\ref{3Nsrc}b) (referred as type 3N-II SRCs) 
can originate from 2N SRCs  in situations in which
the center of mass momentum of NN correlation
becomes comparable with relative momentum of nucleons in 
the NN correlation with momenta of all three nucleons
 considerably exceeding Fermi momentum. 
This corresponds to an average recoil energy:
\begin{equation}
<E_R> \sim {p_{i}^2\over m_N},  
\label{er3n-ii}
\end{equation}
which are roughly twice as large as  recoil energies  characteristic to 
type 2N-I SRCs (Eq.(\ref{averrec})). Type 3N-II SRCs in general are more rare than 
type 3N-I SRCs, since they correspond to much larger recoil energy of residual nucleus, though 
it may be easier to observe them experimentally (See Sec.8.4).

As it follows from Eqs.(\ref{er3n-i},\ref{er3n-ii}) 3N SRCs generate correlations between $p_{i}$ 
and $E_{R}$, below and above the average recoil energy values characteristic to type 2N-I SRCs 
(Eq.(\ref{averrec}).
However since  the correlation observed for type 2N-I SRCs in Eq.(\ref{averrec}) has rather broad $E_R$
distribution it fully overlaps with 
$p_{i}\ - \ <E_{R}>$ correlations followed from 
Eqs.(\ref{er3n-i},\ref{er3n-ii})  up to rather large nucleon momenta.
This situation indicates the limited capability of the spectral function  to reveal 
details of three nucleon SRCs.
One needs to study decay products of the residual nucleus to observe such configurations.

Thus we  conclude, that even though 2N and 3N SRCs lead to a distinctive 
structure in the recoil energy  distribution of the spectral function, the dominance of type 2N-I 
correlations overshadows the $p_{i}-<E_R>$ correlation expected from 3N SRCs. The latter means that 
$p_{i}$ and $E_R$ variables are not enough to isolate 2N and 3N SRCs.
However the important observation is that all correlation relations of 
Eqs.(\ref{averrec},\ref{er3n-i},\ref{er3n-ii}) 
have a local character and at large $p_{i}$ they should be  manifested through the approximate 
universality of the spectral function distributions over $E_R$ for different nuclei.

It is worth noting that situation is different for the relativistic case for which 
use of the  light-cone momentum fraction $\alpha_i$ (Eq.(\ref{alphai})) allows 
to separate 2N and 3N SRCs. 
Indeed choosing $\alpha_i > 2$, (see Ref. \cite{FS88}) will significantly 
enhance the contribution of 3N SRCs in the spectral function.

\subsection{Decay Function}

Going one step further beyond the spectral function one can ask a question how the 
recoil energy $E_R$ is shared between the decay products of the residual nucleus. One can introduce 
the nonrelativistic decay function of the nucleus as follows\cite{FS88}
\begin{equation}
D_A(k_1,k_2,E_R)=\left|\left<\phi_{A-1}a^\dagger(k_2)
\left|\delta(H_{A-1}-(E_R-T_{A-1}))a(k_1)\right|\psi_A\right>\right|^2,
\label{decay}
\end{equation}
which is the  probability that after a nucleon with momentum $k_1$ is instantaneously removed from the 
nucleus the residual A-1 nucleon system will have excitation energy $E_m=E_R-T_{A-1}$ 
and contain a nucleon with momentum $k_2$. 
This function can be extracted from the differential cross section of double-coincidence 
experiments in which knocked-out fast nucleon (with momentum ${\vec p_f}$) is detected in 
coincidence with a slow nucleon~(with momentum ${\vec p_r}$)  which is produced in the recoil 
kinematics. In this case within PWIA differential cross section is expressed through the decay 
function as follows:
\begin{equation}
{d\sigma\over dE^\prime_e d\Omega^\prime_e, d^3p_f d^3 p_{r}} = {j_N\over j_A}
\sigma_{e,N}(p_f,p_{i},Q^2)\cdot D_A(p_{i},p_r,E_r).
\label{eenn}
\end{equation}
Comparing Eq.(\ref{een}) and (\ref{eenn}) one observes that:
\begin{equation}
S^{r}_A(p_{i},E_R) = \int D_A(p_{i},p_r,E_R)d^3p_r,
\label{sumrule}
\end{equation}
where $S^{r}_{A}$ represents the part of the complete spectral function of 
Eq.(\ref{spectral}) corresponding to the case of the break-up of $A-1$ residual nucleus 
with at least one nucleon in the continuum state. 

Eq.(\ref{decay}) represents the lowest order nuclear decay function in which 
only one recoiling nucleon  with large momentum is detected from the residual nucleus.
In principle one can consider decay function containing more than one fast nucleons 
at recoil kinematics. This will require an introduction of additional factor in the r.h.s. 
of Eq.(\ref{decay}) due to presence of several nucleons in the recoil system.
Note that in analogy with spectral function, the decay function defined above  can be 
generalized to $D_{A}^{DWIA}$ for the case  distorted wave impulse approximation 
in which final state interactions are taken into 
account and  factorization of nucleon electromagnetic current is  justified.

To investigate the basic features of the decay function related  to the SRC properties of nuclear 
ground state wave function, we analyze $D_A({\vec p_{i}}, {\vec p_{r}},E_R)$ function in 
the impulse approximation limit in which we neglect the final state  interaction of the 
struck  nucleon with spectator nucleons in the reaction. 
In this case the decay function can be represented as follows:
\begin{eqnarray}
D_A({\vec p_{i}}, {\vec p_{r}},E_m) &  & =  \nonumber \\
& & {1\over 2s_A + 1}\sum\limits_{s_A,s_{f},s_{r}}\sum\limits_{A-2}
\int \prod\limits_{i=3}^{A-1}d^3 p_{i}\cdot 
\delta(E_R - T_r - E_{A-2} - |\epsilon_A| - T_{A-2}) \nonumber \\
& & \times \left| \int d^3p_{r,A-2} 
\Psi_{(p_{r,A-2})}(p_{3},...,p_{A})\cdot \Psi_A(p_{i},p_{r},p_{3},...,p_{A})\right|^2, 
\label{dfif}
\end{eqnarray}
where $\Psi_{p_r,A-2}$ represents the wave function of recoil nucleon and spectator (A-2) system 
and $p_{r,A-1}$ is the relative momentum of the recoil nucleon with respect to the c.m. of the (A-2)
system. The sum $\sum\limits_{A-2}$ accounts for the different configurations
of the (A-2) system.
Within impulse approximation, the angular dependence of the decay function is defined by 
the  relative angle between ${\vec p_{i}}$ and ${\vec p_{r}}$.

It follows from Eq.(\ref{dfif}), that in order to a residual nucleus to decay into the state that 
contains a fast nucleon with momentum $p_r$, it should have  a sufficient recoil energy 
$E_R> {p_r^2\over 2m_N} + |\epsilon_A|$. 
This energy  should be  transfered to the residual nucleus during the process of removal of 
the struck nucleon to compensate the binding  energy of the struck nucleon with other nucleons
\footnote{Final state interaction of the struck nucleon which is not contained in the definition of 
the decay function in Eq.(\ref{dfif}) can transfer this energy only after the removal of the struck nucleon.}. 
SRCs provide an effective mechanism of such  energy transfer 
before the removal of struck nucleon. In this case an average energy transferred to residual 
(A-1) system is $\sim {p_{i}^2\over 2m_N} + |\epsilon_A|$. 

The above discussed dynamics creates an additional 
suppression factor for kinematic situation in which recoil fast nucleons 
are produced in the situation in which 
the initial momentum of struck nucleon $p_{i}\approx 0$ 
(see also  discussion in Sec.7.4 of Ref.~\cite{FS81}). 
Note that within  the  generalized eikonal model of $D_{A}^{DWIA}$, the recoil 
energy of the residual system is provided by 
the final state interaction of struck nucleon with nucleons of the residual 
system\cite{eheppn1,eheppn2}.

Similar suppression exists for $A\ge 4$ nuclei for the kinematics in which  a removal of a 
fast nucleon from one SRC is accompanied 
by an emission of a fast nucleon from another 
SRC separated by distances exceeding  average internucleon distances in the 
nucleus ($\ge 1.5 fm$). Such decay is additionally suppressed by
the short-range nature of $NN$ interaction.

In Refs.~\cite{Schiavilla} and \cite{Alvioli07} the nuclear double  momentum distribution 
was considered. In the kinematics in which momentum of a proton $k_1\gg k_F$ this quantity 
shows strong correlation with presence of a neutron  with momentum $k_2= - k_1$  reflecting 
presence of SRC and dominance of $pn$ correlations.  
Contribution of uncorrelated (A-2) nucleons  in this kinematics  is small since the factor (A-2) 
in the normalization of the double momentum distribution does not compensate (if A is not very 
large) the small probability of 
2N SRC per nucleon. However away from this kinematics uncorrelated contribution 
which is enhanced by a factor (A-2) becomes increasingly more important and difference from 
the decay function in which only correlated pairs contribute becomes large. As a result, 
it is difficult to  use the double momentum distribution to calculate effects of c.m. motion 
of 2N SRC in the mean field of nucleus and dependence of the pp/pn ratio on this motion.

\begin{figure}[ht]
\centering\includegraphics[height=9cm,width=12cm]{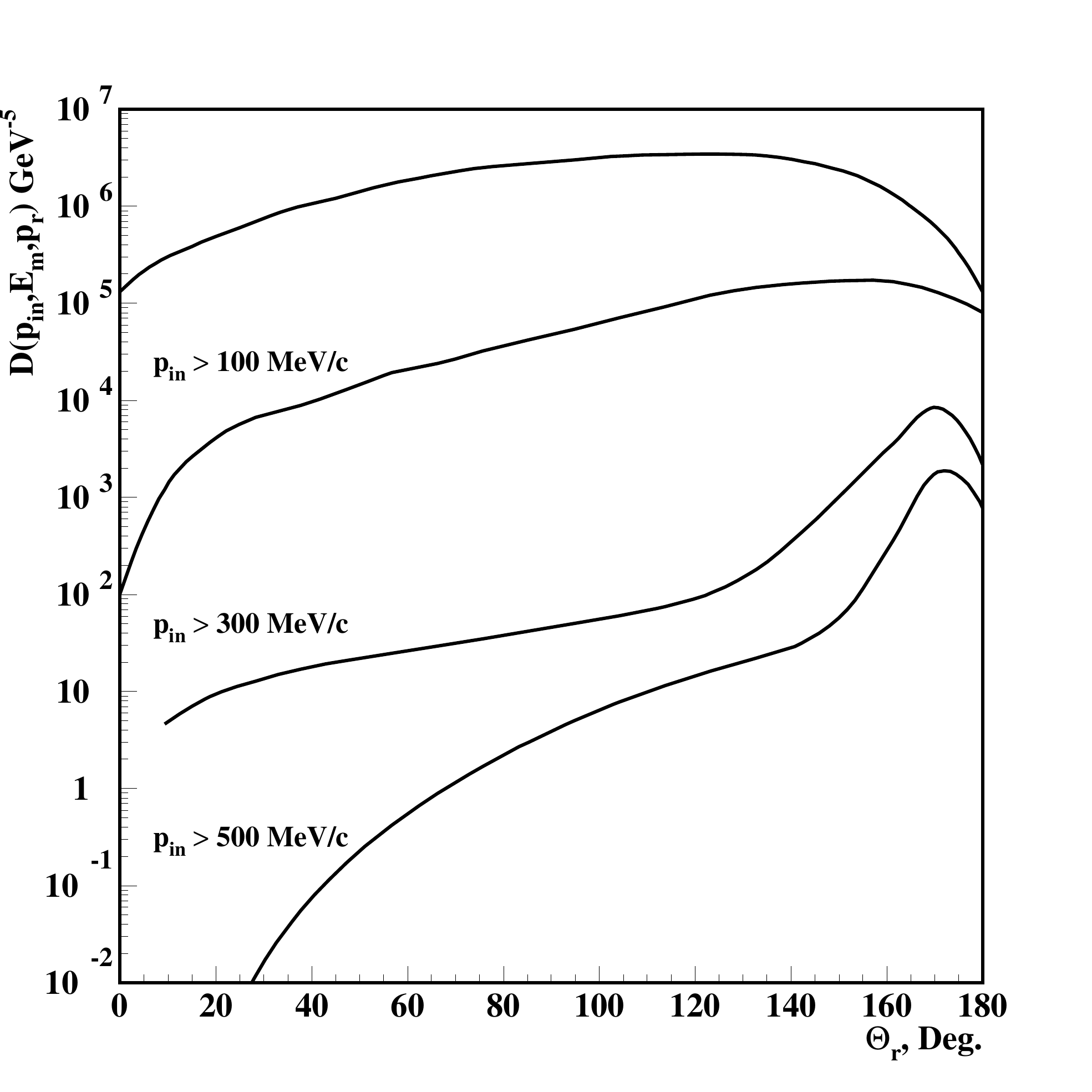}
\caption{Dependence of decay function on the  relative angle of initial and recoil 
nucleon momenta for different values of cuts imposed  
on initial nucleon momenta.}
\label{ang_corr}
\end{figure}

In following, we focus on kinematics, in which the removed nucleon momentum is large 
and therefore the above discussed suppression does not arise in the decay of the 
residual nucleus containing  at least one fast nucleon.

Overall, one expects decay function to  exhibit  much stronger sensitivity to the SRC 
structure of nucleus than the spectral function. 
For example, for situation of Fig.\ref{2Nsrc}a,
because of interaction within SRC is local  as compared to the 
average scale of internucleon distances it is natural to expect that after one nucleon of 
2N SRC  is removed the second one will be produced on the mass shell with momentum 
approximately equal and opposite to the one it had in the correlation: 
$\vec p_{r}\approx -\vec p_{i}$. Such correlations are clearly seen  in 
Fig.\ref{ang_corr} where the  dependence of  the decay function strength is given as a function 
of the relative angle of initial and recoil nucleons  for different 
values of cuts imposed on initial nucleon momentum\footnote{If SRCs are located at the center 
of the nucleus the escaping nucleon may rescatter in the nuclear medium leading to distortions 
of the spectrum which in kinematics of Eq.(\ref{minenergy}) can be taken into account in 
eikonal approximation (see e.g. Ref.~\cite{eheppn1}).}

Such pattern resulting from the breaking of SRC by instantaneous removal of one of the 
correlated nucleons by energetic projectile was suggested in Ref.~\cite{FS77} as a spectator 
mechanism for production of nucleons in the reaction of Eq.(\ref{fbn}).  
This pattern was experimentally confirmed in high momentum transfer triple coincidence 
$A(p,2pN)X$ experiment\cite{eip1,eip2} in which clear correlation between $p_{i}$ and $p_r$ 
was observed.

\begin{figure}[ht]
\centering\includegraphics[height=9cm,width=12cm]{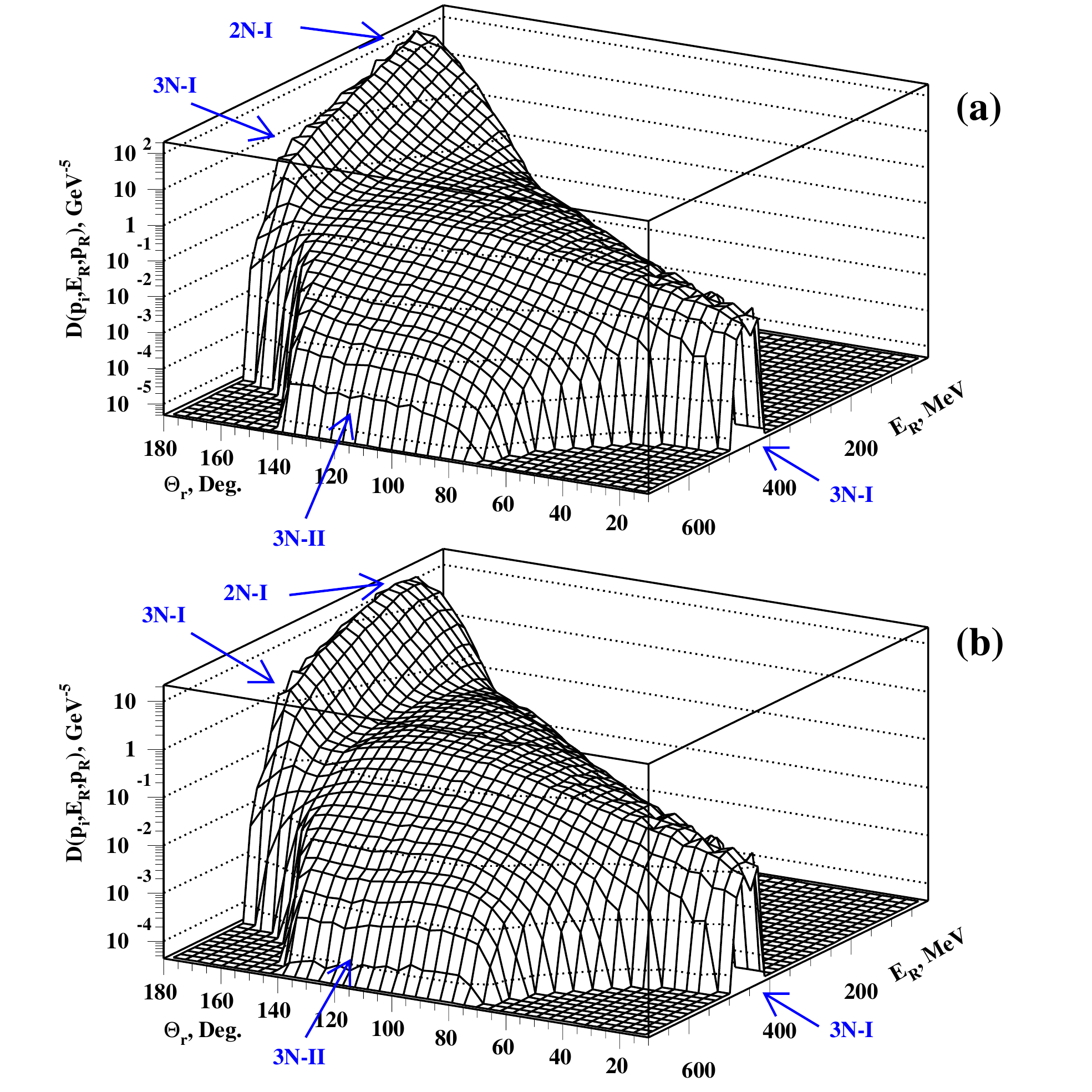}
\caption{Dependence of the decay function on residual  nuclei energy, $E_R$ and 
relative angle of struck proton and recoil nucleon, $\Theta_r$. 
Figure (a) neutron is recoiling against proton, 
 (b) proton  is recoiling against proton. Initial momentum of struck nucleon as well as 
recoil nucleon momenta is restricted to  $p_{i},p_r\ge 400$~MeV/c. }
\label{er_thr}
\end{figure}

Already this example demonstrates that moving from spectral to decay function 
provides 
an additional tool for  probing SRCs, such as correlation between initial and recoil nucleon 
momenta.

Another advantage of the decay function is the possibility to isolate three-nucleon SRCs 
and to probe their dynamics. 
Fig.\ref{er_thr} shows the dependence of the decay function on the 
angle between  initial, ${\vec p_{i}}$, and  
recoil, ${\vec p_{r}}$ nucleon momenta, 
and the recoil energy, $E_R$ for $p_{i},p_r\ge 400$~MeV/c. 
This figure shows 
kinematical domains where it is possible to 
separate
2N and 3N correlations by
varying 
energy of the recoil system.
In the calculation presented above the minimal recoil energy for type 2N-I SRCs (Fig.\ref{2Nsrc}a) 
will be $\sim {p_{n,min}^2\over 2m_N}\approx 80$~MeV, while for type 3N-I SRCs (Fig.\ref{3Nsrc}) 
the minimum  of  recoil energies  is twice as large.  
The upper left side of the figure demonstrates how  type 2N-I SRCs evolve to a type 3N-I SRC 
with the third nucleon also recoiling against the removed nucleon.
One can also see from the figure that with an increase 
of recoil energy type 3N-II correlations start to dominate.  The important signature 
in this case is the relative angle of recoil nucleon emission being close to $120^0$  
which is characteristic for  
type 3N-II SRCs. The lower right part of the figure shows also 
different realization of 3N-I SRCs in which both struck and recoiled nucleons  share/balance  
the momentum of the  third nucleon which has roughly twice the momentum of $p_{i}$ or $p_{r}$.  

\begin{figure}[ht]
\centering\includegraphics[height=8cm,width=10cm]{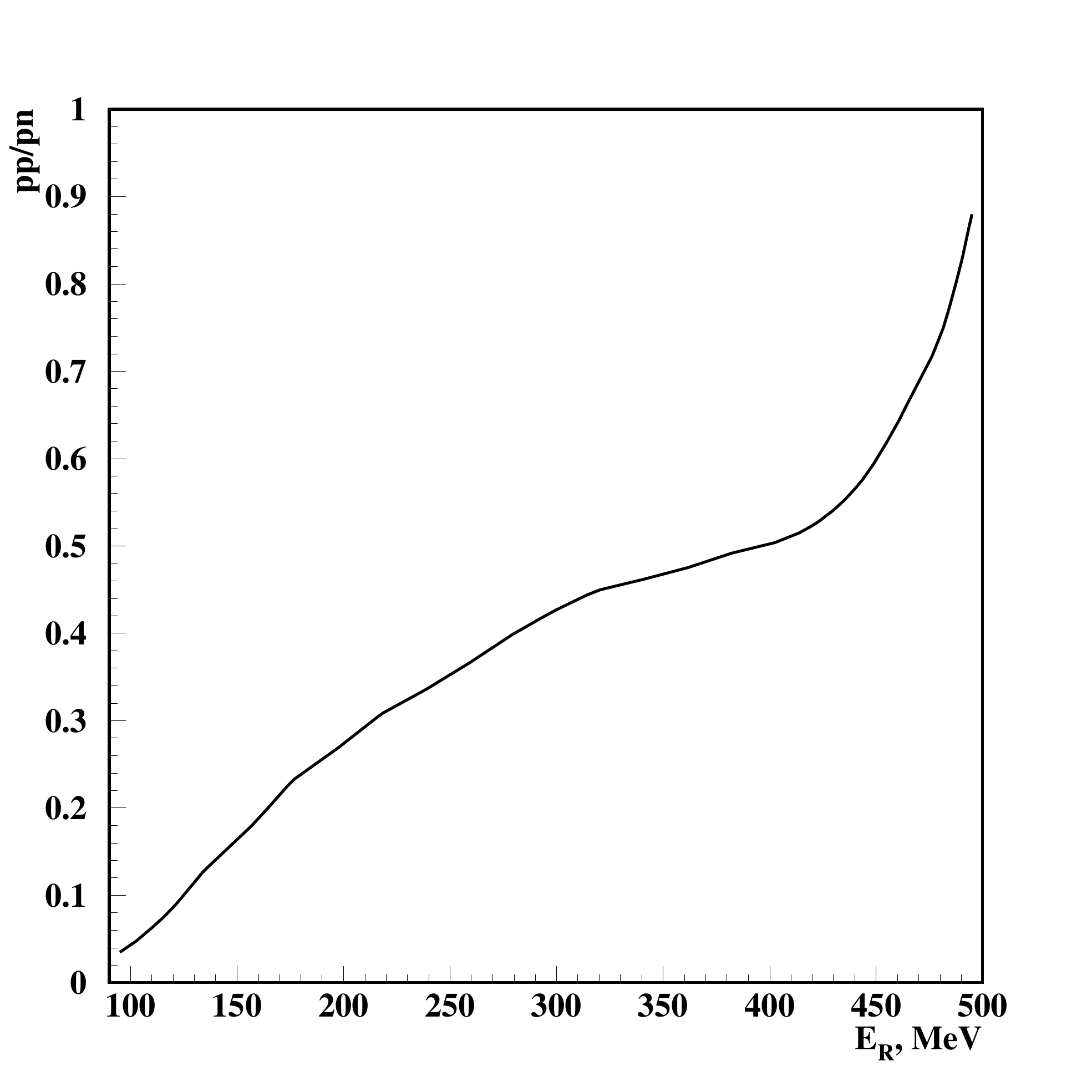}
\caption{Recoil energy dependence of the ratio of the decay functions when a proton in $^3He$ is
struck and the proton on one case and the neutron in other  is produced in the decay.
Both initial momenta of struck and recoil nucleons
are  set to be larger than $400$~MeV/c.
Also, the relative angle between initial and recoil nucleon momenta is restricted to  
$180\ge \theta_r \ge 170^0$.}
\label{pp_pn}
\end{figure}

Figs.\ref{er_thr}a and \ref{er_thr}b present the decay functions for proton 
removal with production of either proton or neutron  in the decay.
Comparison of these two cases shows 
(see upper left part of the graph) that in type 2N-I SRCs 
the strength of $pn$ correlation is larger than the strength of  $pp$ correlation  
by factor of ten.  This feature reflects the dominance of tensor 
interaction in $S=1$, $T=0$ channel of NN interaction at short distances and 
was confirmed experimentally, both for hadron- and electron- induced triple coincidence
reactions on carbon\cite{eip3,eip4}.  
Interesting consequence of the onset of 3N SRCs is that these two rates become practically 
equal once recoil energy increases. 
More detailed view of relative strengths of $pp$ and $pn$ decay functions is given 
in Fig.\ref{pp_pn}. 
The increase of the ratio of $pp$ to $pn$ strengths with 
an increase of  the recoil energy represents  
an unambiguous indication of the dominance of type 3N-I SRC effects.

As we mentioned before, the notion 
of the decay function can be extended to 
the situations in which more than two nucleons are detected in the products of 
the decay of the residual nucleus.  One such extension is the 
study of two recoil nucleons without detecting the struck nucleon. In this situation nucleons 
with approximately equal momenta will  be emitted predominantly  at large relative angles to minimize 
the momentum of the struck nucleon.

Concluding this chapter we would like to emphasize that experimental possibility of 
measuring nuclear decay function in high momentum transfer triple coincidence 
$A(e,e'N_f,N_r)X$ or  $A(h,h'N_f,N_r)X$ reactions opens up completely new perspectives in 
studying the dynamics of 2N and 3N SRCs.

\section{Scaling of the ratios of cross sections of  $A(e,e')X$ reactions at $x> 1$.} 

\subsection{Introduction}
Here we consider $A(e,e')X$ reactions at  kinematics: 
\begin{equation}
x=A{Q^2\over 2m_Aq_0}> 1, \, \ \ \mbox{and} \ \ 4\ge Q^2 \ge 1.5~ \mbox{GeV}^2,
\label{kin}
\end{equation}
which  have been   measured recently at JLab\cite{Kim1,Kim2}. 
They complement and improve  the previous measurements which were performed at SLAC in 
the 80's~(see Ref.~\cite{FSDS} and references therein).

\begin{figure}[ht]
\centering\includegraphics[height=8cm,width=10cm]{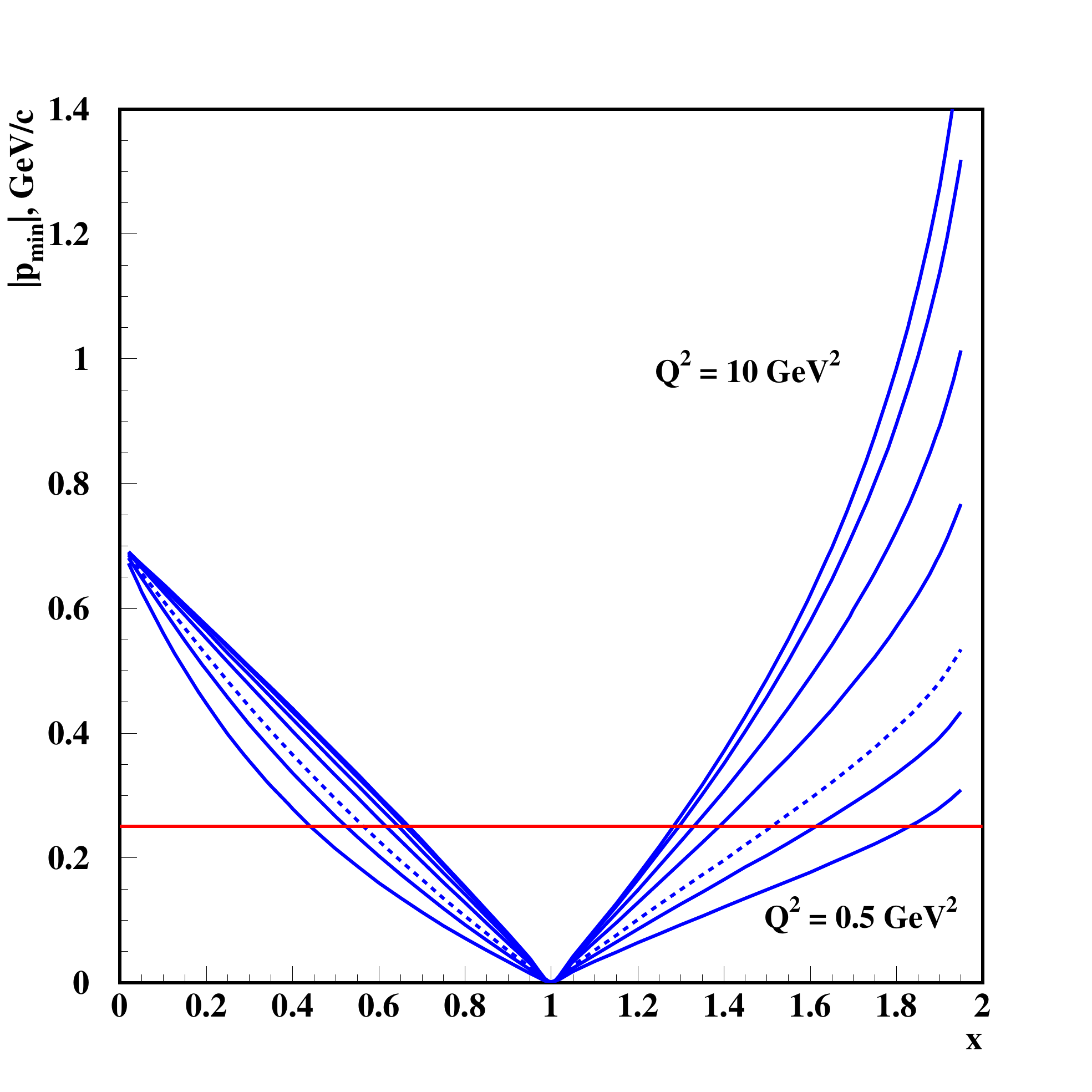}
\caption{The $x$ Dependence of $\left|p_{min}\right|$ for 
different values of $Q^2$, with recoil energy given by the 
two nucleon approximation.
}
\label{x_pmin}
\end{figure}

Before presenting  
a more formal discussion we review an intuitive picture of the reaction.
 It follows essentially from the definition of $x$ that  
its magnitude  cannot be larger than the  number of nucleons in a given nuclei.
This can be seen from the definition of the produced mass in the reaction
 \begin{equation}
 W^2= Q^2 (-1+ {m_T\over x m_N}) +m_T^2 \ge m_T^2,
 \end{equation}  
which leads to $x\le m_T/m_N$.
In the impulse approximation  the process is described as  
an absorption of the virtual photon 
 by a nucleon which had a momentum 
opposite to the direction of virtual photon momentum.

 The kinematics of the process resembles that of the deep inelastic scattering off massive partons, 
and therefore  in the limit of large $Q^2$ we expect that $x=\alpha_i$, where $\alpha_i$  is the 
light-cone momentum fraction  of the nucleus carried by the 
initial nucleon (defined in Eq.(\ref{alphai})) which is struck by virtual photon in 
quasielastic scattering. Therefore the  relation $x=\alpha_i$ indicates that for 
$x\ge j$ at least $j$-nucleons should be involved in the process. Note here that inequality $x\le j$ for the scattering off nucleus consisting of $j$ nucleons 
is valid for  all $Q^2$. 
To see how the relation $x=\alpha_i$ 
emerges in the  discussed process and to estimate 
the deviation from 
this relation   for finite $Q^2$, it is convenient to introduce 
four-momentum of the struck nucleon $p^\mu_{i}=p^\mu_A-p^\mu_{R}$ where 
$p^\mu_A$ and $p^\mu_{R}$ are 
four-momenta of target and  residual nuclei, and $m_{i}^2=
p_{i\mu}p_{i}^{\mu}$. In the impulse approximation the requirement that the produced nucleon 
is on-mass-shell leads to the  relation 
\begin{equation}
(q+p_A-p_{R})^2=m_N^2.
\label{scat}
\end{equation}
Using Eq.(\ref{scat}) and definition of $x$ one finds
\begin{equation}
x = {\alpha - {m_N^2-m_i^2\over 2 m_Nq_0}\over 1 + {2p_i^z\over q_0+q_3}},
\label{xal}
\end{equation}
where $q_3=|{\bf q}|$.
It follows from Eq.(\ref{xal}) that  in  the limit ${\bf q}\gg p_i$,  with $x$ being kept constant,
$x = \alpha_i  + O(1/Q^2)$.

One can see from Fig.\ref{x_pmin} 
that for discussed $Q^2$ the minimum momentum of the struck nucleon calculated in impulse 
approximation, (with recoil energy estimated based on type 2N-I SRC picture of high momentum component of 
nuclear wave function) increases with $x>1$ and becomes 
significantly larger than $k_F\sim 250$~MeV/c for x =1.5 at $Q^2\ge 1.5$~GeV$^2$. 
Therefore with the gradual increase of $x$ virtual photon should  first  probe the most
abundant high momentum configuration which is 2N SRC and then
with an increase of $x$ above two, only high momentum nucleons whose
removal is associated with a recoil energy smaller than  
the characteristic recoil energy for the interaction with 
two nucleon correlation (Eq.(\ref{averrec})).
This can be achieved if struck nucleon momentum is  balanced by  momenta of 
two nucleons  that exceed $k_{F}$.
This picture corresponds to the type 3N-I SRCs~(see Fig.\ref{3Nsrc} and discussion in Sec.2.2)  
with average recoil energy defined according to Eq.(\ref{er3n-i})\footnote{Note that for 
intermediate values of $Q^2\sim 2-3$~GeV$^2$ and $x\le 2.5$ 
the average  value of  $\alpha_i$ for the struck nucleon could still be less than two.}. 
Hence we expect that the natural mechanism of  quasielastic processes 
at large $Q^2\ge 1$~GeV$^2$ and $x>1$ is the scattering 
off 2, 3,.. nucleon SRCs which have universal (A-independent ) properties. 

Since the nucleus is a dilute system, in kinematics in which for example the  scattering 
off $j\times N$ SRC is  possible, the scattering off $(j+1)\times N$ SRC should be a 
small correction.  Therefore for two and three nucleon correlation kinematics,
for $Q^2$ range where quasielastic scattering gives dominant contribution to the cross 
section\footnote{For $Q^2 \ge 6$~GeV$^2$ contribution from inelastic scattering becomes 
significant leading to an increase of the typical  momenta of nucleons which dominate at 
given $x$. This leads to an increase of the ratios with an increase of $Q^2$\cite{FSDS}.} 
we expect~\cite{FS81,FS88}  inclusive cross section ratios to scale as follows:
\begin{equation}
\left. {2\over A}{\sigma(eA\to e' X) \over \sigma(e \ ^2H\to e' X)}\right|_{ 2> x\ge 1.5}= a_2(A),
 \ \mbox{and} \  \ 
\left. {3\over A}{\sigma(eA\to e' X) \over \sigma(e \ ``A=3'' \to e' X)}\right|_{ 3> x\ge 2}= a_3(A),
\label{ratio}
\end{equation}
where it is assumed that the ratios are corrected for difference of 
the electron-proton and electron-neutron cross sections.

\begin{figure}[ht]
\centering\includegraphics[height=8cm,width=10cm]{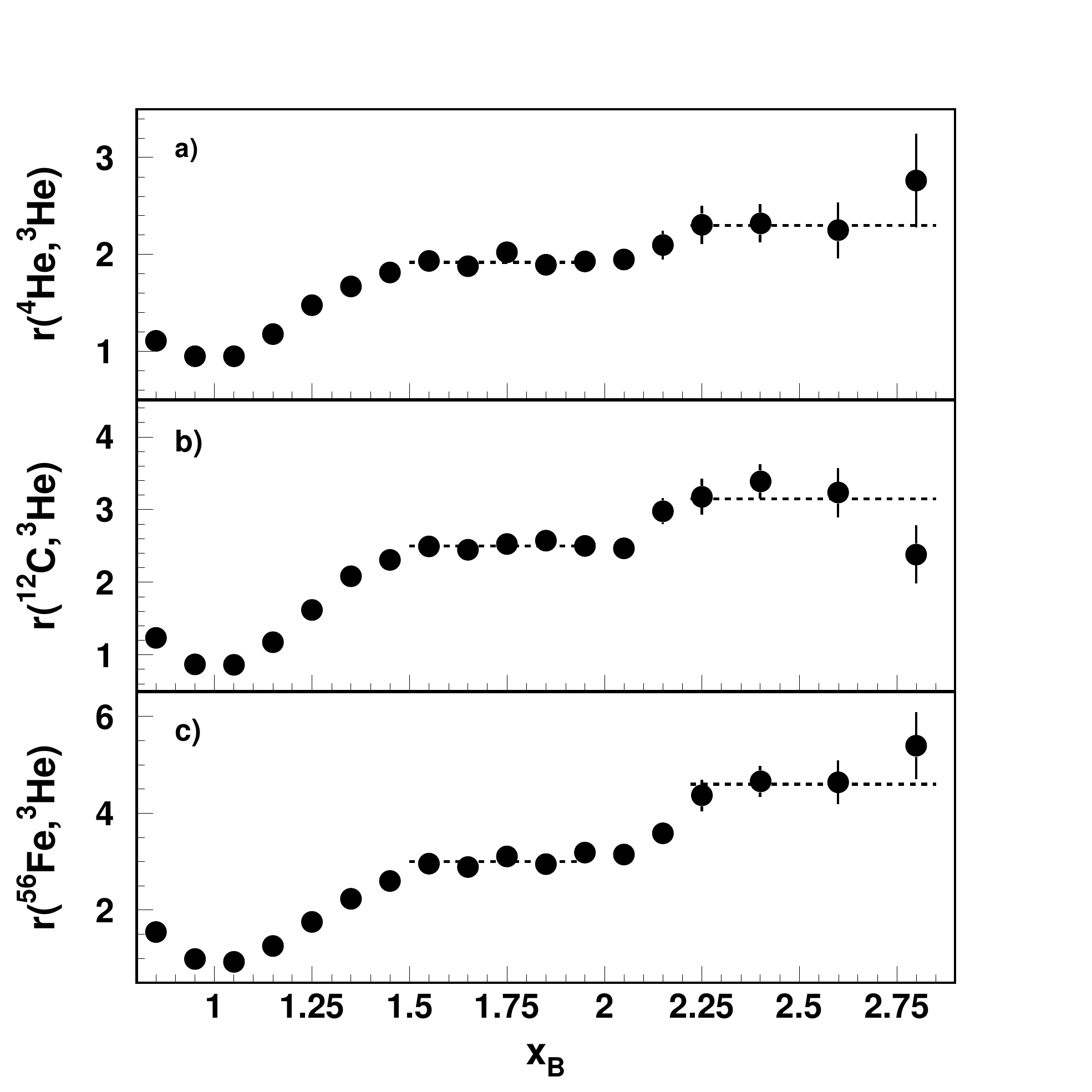}
\caption{The $x$ dependence of the ratios of inclusive cross sections. Dashed curves indicate 
the scaling expectations from  2N- and 3N- SRCs.}
\label{ratiodata}
\end{figure}
The most recent data from Jefferson Lab\cite{Kim2} which confirm the prediction of Eq.(\ref{ratio}) are shown 
in Fig.\ref{ratiodata}. 

\begin{figure}[ht]
\centering\includegraphics[height=8cm,width=12cm]{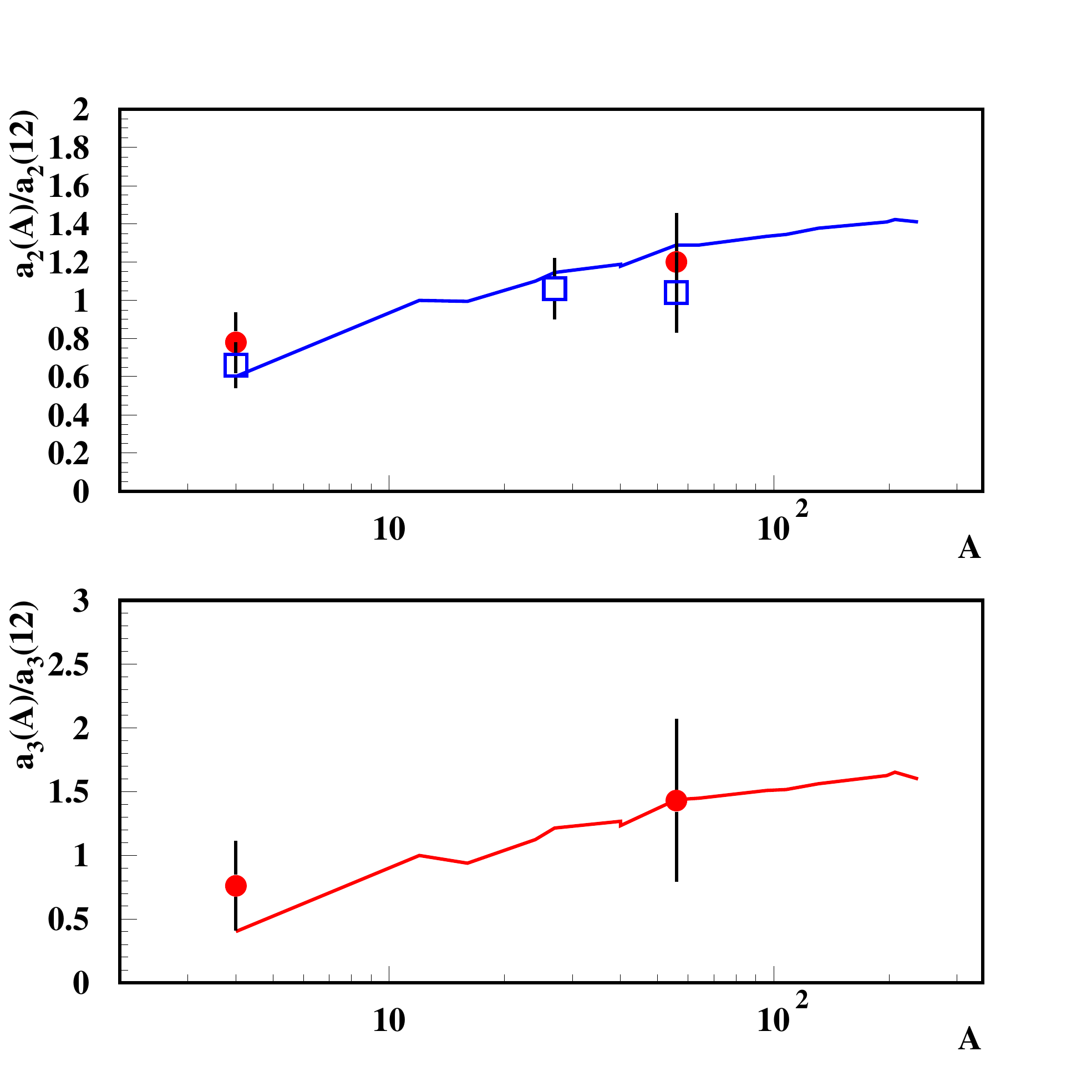}
\caption{The $A$ dependence of $a_2$ and $a_3$ calculated based on Eq.(\ref{a_j}) and compared 
with the data from Refs.11,13,14. Both calculations and data 
are  normalized to the corresponding values of $a_2(C)$ and $a_3(c)$.}
\label{an_adep}
\end{figure}

The quantities, $a_2(A)$ and $a_3(A)$ represent the excess of 
per nucleon probabilities of finding  2N and 3N SRCs in nucleus, as 
compared to the deuteron and A=3 nucleus respectively. 
The fact that SRCs represent  high density fluctuation of 
the nuclear matter and  constitute only a small part of nuclear wave function 
allows us to calculate the $A$ dependence of $a_2$ and $a_3$ through the 
nuclear matter density function evaluated within mean-field approximation.
Indeed, the fluctuation character of SRCs allows us to justify the estimate\cite{FS81}:
\begin{equation}
a_j\propto \int \rho_{A}(r)^{j}d^3r \approx \int\rho^j_{A,mf}
\left(1 + j{\rho_{A,SRC}\over \rho_{A,mf}}\right)d^3r. 
\label{a_j}
\end{equation}
Above we 
expressed the nuclear matter density function~($\rho_{A}(r)$) through 
the sum of mean field ($\rho_{A,mf}$) and SRC ($\rho_{A,SRC}$) density 
functions ($\rho_{A} = \rho_{A,mf} + \rho_{A,SRC}$).
Using the fact that for nuclei with $A\ge 12$ 
the contribution to the normalization due 
to SRCs is much smaller then unity~($\int \rho_{A,SRC}(r)d^3r \ll 1$), for 
not very large $j\ll A$ the second term of the integrand in right hand part of 
Eq.(\ref{a_j}) can be neglected.  As a result one can estimate the $A$ dependence
of $a_j$ using mean-field nuclear density function, $\rho^{A,mf}(r)$.
Fig.\ref{an_adep} compares the prediction of Eq.(\ref{a_j}),  using  
the Skyrme-Hartree-Fock model\cite{Reinhard} for $\rho_{A,mf}(r)$,   with the experimental data 
of $a_2(A)$\cite{Kim2,FSDS}\footnote{Note that in Ref.~\cite{FSDS} the values of 
$a_2(A)$ were extracted from the data 
assuming similar momentum dependences for $pn$, $nn$ and $pp$ momentum distributions
in SRC region. Since recent studies\cite{eip3,eip4} demonstrated that $pp$ and $nn$ SRCs
are significantly suppressed as compared to that of $pn$, the estimates of $a_2$ in 
 Ref.~\cite{FSDS} should be reevaluated for nuclei with large excess of neutrons, 
such as $^{197}Au$. Thus we do not include in  Fig.\ref{an_adep}
the values of $a_2(197)$ estimated in Ref.\cite{FSDS}.}
and $a_3(A)$\cite{Kim1}. It is interesting that above estimates of $a_2$ work even for 
lightest nucleus such as $^3He$. 
However in the  $^4He$ case the estimate clearly fails for $a_3$ since one cannot use the
 mean field approximation for estimating correlation of three out of four nucleons.

It is worth mentioning that $A$ dependence of $a_2(A)$ obtained above 
is significantly  slower than the $A$-dependence one  would infer from the 
experimentally measured number of quasideuteron pairs in the nucleus,  ${LNZ\over A}$,
as it is determined from nuclear photoabsorption reaction at $E_{\gamma} \sim 100 MeV$.
Here $L$ is the   Levinger factor. 
Also, the absolute value of $a_2(C)$ which follows 
from the value of Levinger factor for Carbon is by factor of two smaller than the one 
which follows from high energy data.

The extracted values of $a_2(A)$ allow us to estimate the absolute magnitude of 
high momentum component of nuclear wave function using information about the 
high momentum component of the deuteron wave function. 
For realistic deuteron wave functions  the  probability per nucleon to have 
momentum above $300$~MeV/c is about 4$\div$ 5\%. 
Hence the above estimates of $a_{2}(A)$  corresponds to   the probability to find a 
nucleon with momentum $k\ge 300 \,MeV/c$,  say  in the iron, of $\sim 20\div 25\%$.

\subsection{Space-time structure of high $Q^2$ and $x>1$ quasielastic scattering}
\label{space-time}

For a more formal analysis of the process it is convenient to start with consideration 
of the expression for the cross section as a Fourier transform of the 
commutator of electromagnetic currents  $J_{\mu}$
between wave functions of the nucleus 
in its rest frame:
\begin{equation}
2m_{A}q_3 \sigma^{(r)}= \int e^{iqy}\langle A\mid [J_{\mu} (y), J_{\lambda} (0)]
\mid A\rangle
\epsilon_{\mu}^{(r)}
\epsilon_{\lambda}^{(r)} d^4y.
\label{eq:emc}
\end{equation}
where $q_3 = \sqrt{Q^2+Q^4/4m^2x^2}$, and $\epsilon^{(r)}$ is the polarization vector of the 
virtual photon. Strong oscillations in the exponential lead to 
the condition that in the discussed kinematic range
\begin{equation}
y_+ \sim  \frac{1}{q_-} < 1 \, {\rm fm} ,\;
y_- \sim \frac{1}{q_+} < 0.2 \,{\rm fm} , 
\label{lcscale}
\end{equation}
where we introduced the LC variables $y_{\pm}=y_0 \pm y_z$. Also
it follows from causality (i.e. from condition that commutator of 
electromagnetic currents is 0 for space-like intervals) that  $y_t^2\le 1/Q^2$.  To reach this 
we use an approximation in which nucleons are point-like  
and  nucleon  structure is accounted for in  terms of form factors.  Account of meson 
currents leads  to a nonlocality of electromagnetic  current at $Q^2=0$ and  restricts
the size of the probed region  to the  radius of a nucleon.  Challenging question is how this 
nonlocality depends on $Q^2$.
The analysis of quark models of a nucleon shows that a selection of 
$Q^2 \approx few$~GeV$^2$ squeezes 
an  effective size of the    nucleon \cite{FMS}.
It is quite difficult to observe this phenomenon directly because of an expansion of 
small size configurations\cite{FFS} after it was produced in a hard subprocess.

\subsection{SRCs and Final State Interaction}
{}From the general considerations of Sec.3.2 one observes that the interaction between the knock 
out nucleon and residual system which may contribute to the total cross section in the 
kinematics of Eq.(\ref{kin}) 
is dominated by distances  less than about $1$~fm which corresponds to the interaction 
within the SRC and therefore it is canceled in the ratios like Eq.(\ref{ratio}).

\begin{figure}[ht]
\centering\includegraphics[height=4.6cm,width=10cm]{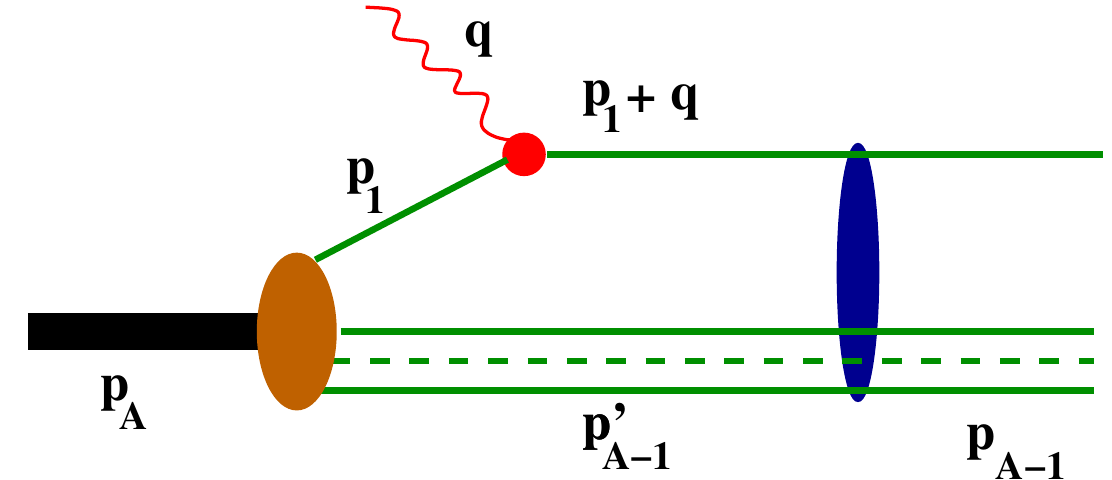}
\caption{General diagram representing final state interaction.}
\label{FSI_gen}
\end{figure}

This conclusion can be reinforced by considering the rescattering diagram  
(Fig.\ref{FSI_gen}) and treating it as a Feynman diagram.
In this case one can calculate the virtuality of  
struck nucleon at the intermediate state (before the FSI blob). We find \cite{FSDS}
that if the momentum of struck nucleon ${\bf p_1}$ is 
significantly different from the momentum, ${\vec p_i} = {\vec p_f} - {\vec q}$, 
corresponding to the initial momentum in  impulse 
approximation, then the virtuality of struck nucleon becomes large. Indeed on can 
estimate the virtuality of struck nucleon as follows: 
\begin{equation}
\Delta M^2 \equiv m^2 - (p_1 + q)^2 \approx m^2 + Q^2(1-{1\over x}) - \tilde m^2,
\label{virt}
\end{equation}
where $\tilde m^2 = p_1^2= (p_A-p^\prime_{A-1})^2$ and $p^\prime_{A-1}$ is the 
four momentum of the recoil nucleus at the intermediate state (see Fig.\ref{FSI_gen}).
For  small momenta of the  initial nucleon, 
$p_1\approx 0$~($m^2-\tilde m^2\sim 0$) 
the virtuality grows linearly with $Q^2$ at fixed $x\ne 1$. 
It grows also with $x$ moving away from $x=1$. 
Thus for kinematics of Eq.(\ref{kin}) the rescattering amplitudes 
of Fig.\ref{FSI_gen} for a struck nucleon with small initial momenta are 
suppressed due to large virtuality of the struck nucleon in 
the intermediate state. 
For example, for $x=1.5$ and  $Q^2= 2$~GeV$^2$ virtuality is $\sim 1$~GeV$^2$.

\begin{figure}[ht]
\centering\includegraphics[height=8cm,width=10cm]{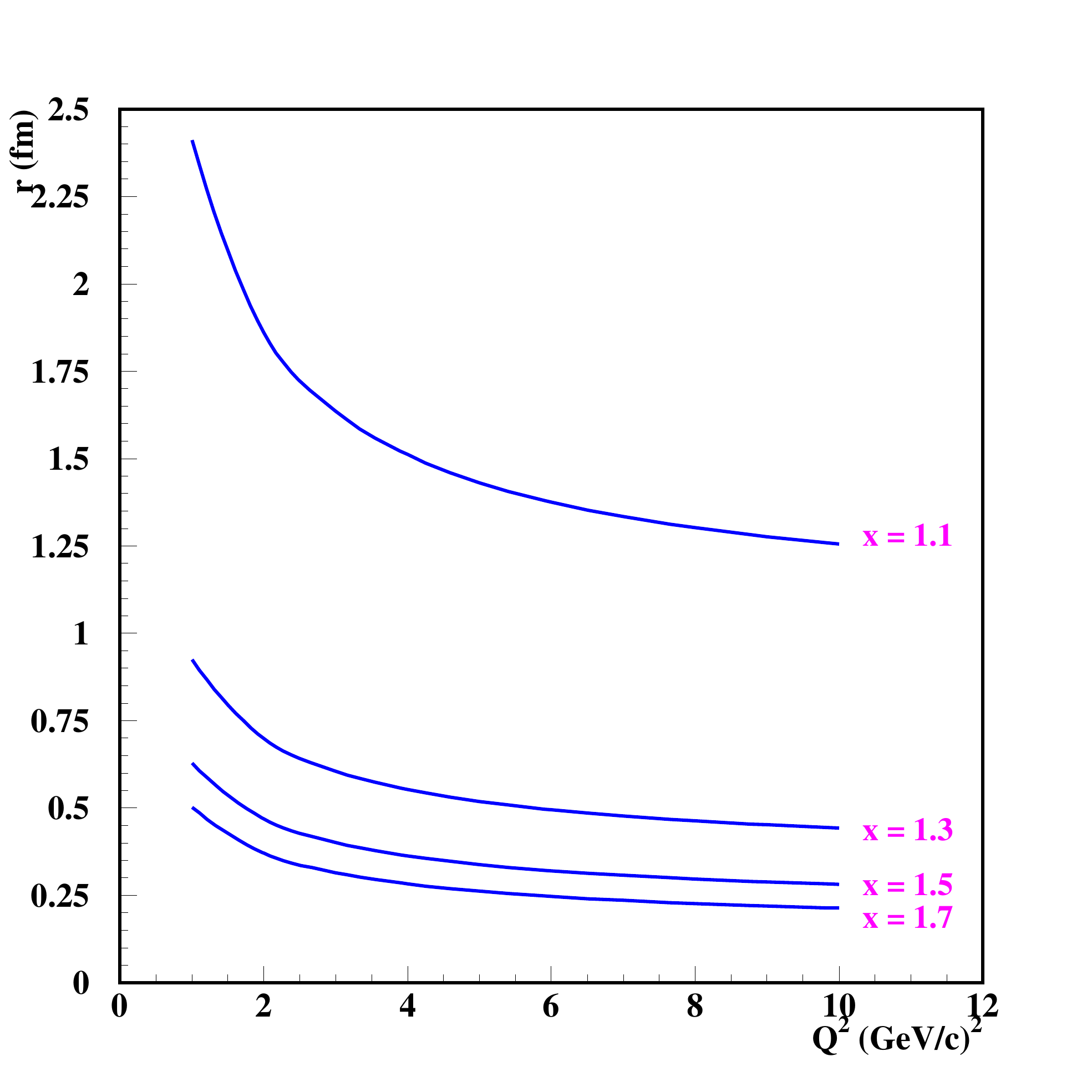}
\caption{Characteristic distance from $\gamma^*N$ interaction point which  
struck nucleon (having $p_1\le p_{Fermi}$) can propagate before reinteraction 
that contributes to the total cross section of $\gamma^*A$ scattering.}
\label{rint}
\end{figure}

To understand what physical phenomena cause  this suppression, it is convenient 
to represent the FSI amplitude of Fig.\ref{FSI_gen} within noncovariant theory 
in which time-ordering is explicitly present 
which allows to consider space-time evolution of the process\footnote{Note that relativistic 
effects in this case can be included within light-cone non-covariant theory.}. In this 
case the FSI amplitude can be represented as follows:
\begin{equation}
A^{FSI,\mu}(e A\to e X)\sim\int d^3 p_1 \psi_A(p_1)J^\mu_{em}(p_1,q){1\over 
\Delta E + i \epsilon }t_N(p_1+q,p_f),
\label{ncfsi}
\end{equation}
where $J^\mu_{em}$ is the electromagnetic current, $t_N$ represents the rescattering amplitude 
of struck nucleon and $\Delta E$ is the energy difference between intermediate and 
initial states:
\begin{equation}
\Delta E  = - q_0 - M_A + \sqrt{m^2 + (q+p_1)^2} + \sqrt{\tilde M_{A-1}^{2} + p_1^2}.
\label{deltaE}
\end{equation}
Within this representation we can estimate the characteristic distances that struck nucleon 
propagates as:
\begin{equation}
r\approx {v\over \Delta E},
\label{rd}
\end{equation}
where $v$ is the velocity of the struck nucleon in the intermediate state.
Averaging momentum  $p_1$ over the  $p_1\le p_F$ region and using realistic 
nucleon momentum distribution  one obtains the characteristic length, for $x>1$ 
kinematics in which struck nucleon with small initial momentum propagates before 
it rescatters with  nucleons from the residual nucleus. 
For $^{27}Al$ target  the $Q^2$ dependence of $r$  
for  different values of $x$ are 
presented in Fig.\ref{rint}.

This estimate demonstrates that FSI in kinematics of Eq.(\ref{kin}) overwhelmingly 
takes place at distances which are within SRC. Additionally, since 
the rescattering amplitude, $t_N$ for $p_1\sim 0$ is highly off-energy shell 
(this is equivalent to large virtuality of interacting nucleon in the covariant 
formalism) it is strongly suppressed as compared to the on-shell amplitude.

\subsection{FSI in SRC and Generalized Eikonal Approximation}

The distances relevant for FSI can be investigated in more detail 
by considering FSI in  the eikonal approximation
\cite{FGMSS,FSS97,MS01,Sabine1,Ciofi_gea,Ciofi_Kaptari,Sabine2,Ciofi_Kaptari_2}.

We now introduce an additional restriction to the kinematics of 
Eq.(\ref{kin}) to ensure 
that the momentum of knocked-out nucleon, as it emerges from SRC, is large 
enough ($p_f \ge 500$~MeV/c) 
relative to all nucleons, including those in the SRC,
so that the 
eikonal approximation can be applied 
for evaluation of 
final state reinteractions\footnote{For simplicity we discuss here 
$x< 2$ kinematics where scattering off two nucleon correlation dominates, though our consideration 
can be easily extended to the case of scattering off  $N>2$ 
SRC's.}. Hence 
 we require that 
\begin{equation}
W_{NN} - 2m_N  = \sqrt{4m_N^2 + {Q^2(2-x)\over x}} - 2m_N \ge 60 \ MeV.
\label{kinadd}
\end{equation}

\begin{figure}[ht]
\centering\includegraphics[height=4.0cm,width=12cm]{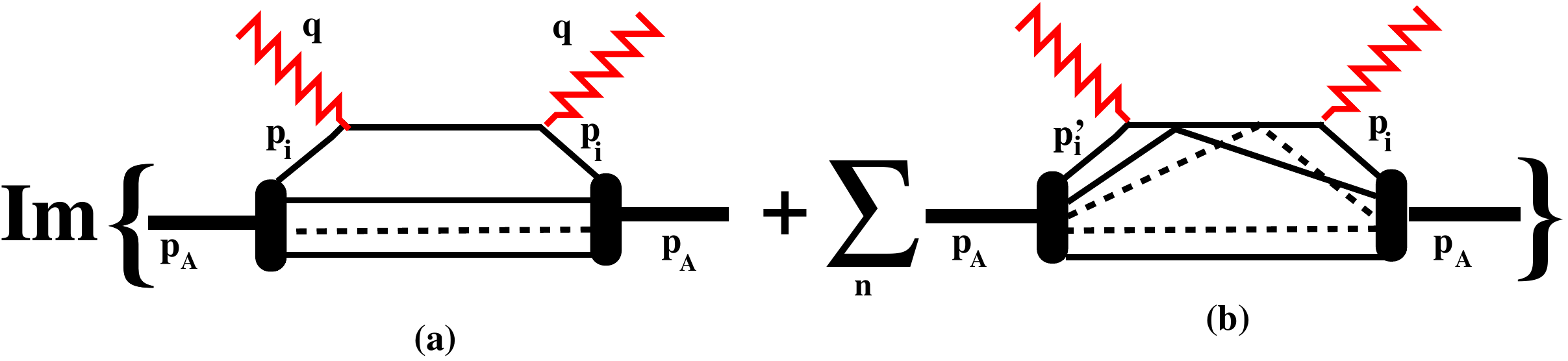}
\caption{Imaginary part of $\gamma^*A$ forward Compton  scattering amplitude defines 
the nuclear structure function of inclusive $A(e,e')X$ scattering.}
\label{im_compton_fwd}
\end{figure}

It is worth noting that to satisfy unitarity one should include  both elastic and inelastic
rescattering of knocked-out nucleon with spectator nucleons in the nucleus. This is technically 
very difficult to realize within eikonal approximation at finite energies 
(see e.g. Ref.~\cite{BT})\footnote{If one would try to apply 
the Glauber theory for description of the cross section of 
inclusive processes by including 
only elastic reinteractions, 
the unitarity 
would be  
violated resulting in a strong overestimate of the role of the FSI.}.
However 
the unitarity condition will be automatically fulfilled 
if we calculate the inclusive $A(e,e')X$ cross section through the imaginary part of
the amplitude of virtual Compton scattering off the nuclei at forward direction as in 
Fig.\ref{im_compton_fwd}.

In this case the nuclear matrix of inclusive scattering can be expressed through 
the forward  Compton scattering amplitude as follows:

\begin{equation}
W_{A}^{\mu\nu}(Q^2,x) = {1\over 2\pi M_A} {\cal I}m  A^{\gamma^*A\rightarrow \gamma^*A}
(P'_A = P_A).
\end{equation}

 Let us now estimate the imaginary part of the forward Compton scattering 
amplitude within virtual nucleon approximation, in which the electromagnetic 
interaction takes place off the virtual nucleon and  the motion of which  
is described by nonrelativistic nuclear wave function.

\medskip
\medskip

\noindent{\bf Impulse approximation:} We first evaluate the impulse approximation part of the 
forward compton scattering amplitude at  high $Q^2$ and  $x>1$.
One can calculate 
contribution of 
the diagram of Fig.\ref{im_compton_fwd}a by  applying effective 
Feynman diagrammatic rules for high energy electro-nuclear reactions 
(see e.g. Ref.~\cite{MS01}) and then performing  nonrelativistic reduction procedure, to  
relate $(nucleus, A) \rightarrow  (A\times nucleons)$ transition 
to the  nonrelativistic nuclear wave function.
We obtain:
\begin{eqnarray}
& &{\cal I}m A^{\gamma^*A\rightarrow \gamma^*A}_0(P^{\prime}_A = P_A) = 
 -A \sum\limits_{Em}{\cal I}m 
\sum\limits_{s_1,s_1^{\prime},s_f}\int 
\Psi^\dagger_{A}(p_1,s_1;p_2;...;p_A)\times \nonumber \\
& & J_{eN}^{\dagger, \mu}(p_f,s_f,p_1,s_1){1\over p_f^2 - m_N^2+ i\epsilon}
J_{en}^{\nu}(p_f,s_f,p_1,s_1^{\prime})\Psi_{A}(p_1,s_1^{\prime};p_2;...;p_A)\times \nonumber \\
& & d^3p_i \prod\limits_{j=2}^{A} d^3p_j\delta^3\left(\sum\limits_{i=1}^A p_i\right),
\label{compt_ia}
\end{eqnarray}
where $\sum\limits_{E_m}$ accounts for the sum and integration over the excitation energy of 
recoil nuclear system.

To see how the process at high $Q^2$ and $x>1$  evolves in 
the space and time  we transform Eq.(\ref{compt_ia}) to
the coordinate representation. To simplify 
following
discussions we consider 
the kinematic limit:
\begin{equation}
q_0\gg E_m, \ \ x \gg {p_i^2\over 2m_N^2},
\label{cord_lim}
\end{equation}
in which case
\begin{equation}
p_{f}^2 - m_N^2  + i\epsilon = (p_1+ q)^2 - m_N^2 + i\epsilon = 
2q(p_{iz} - p_{1z} + i\epsilon),
\label{piz}
\end{equation}
where 
\begin{equation}
p_{iz} = {2q_0(m_N-E_R) - Q^2\over 2 q},
\label{piz0}
\end{equation}
and $E_R$ represents the total kinetic energy of the recoil system as it was 
defined in Sec.2.
 
Using 
\begin{equation}
\Psi_A(p_1,p_2,....,p_A) = {1\over ((2\pi)^{3\over 2})^A} 
\int \Psi_{A}(r_1,r_2,....,r_A)\prod\limits_{i=1}^A
e^{-i{\bf p_ir_i}}d^3r_i \delta^3(\sum r_i-r_A),
\label{wfr}
\end{equation}
and
\begin{equation}
{1\over p_{iz}-p_{1z} + i\epsilon} = -i\int \Theta(z_0)e^{i(p_{iz}-p_{1z})} dz_0,
\label{theta}
\end{equation}
we can rewrite Eq.(\ref{compt_ia}) in the coordinate space representation as:
\begin{eqnarray}
& & {\cal I}m A^{\gamma^*A\rightarrow \gamma^*A}(P^\prime_A-P^A) = 
{A\over  2 q} \sum\limits_{E_m}\sum\limits_{spins} J_{eN}^\mu(Q^2) J_{eN}^{\nu}(Q^2)
\times\nonumber \\
& & \ \ 
\sum\limits_{E_m}\int \Psi^\dagger_A(b_1,z_1;r_2; ...;r_A)\Theta(z_1-z_1^\prime)e^{ip_{iz}(z_1-z^\prime_1)}
\Psi_A(b_1,z^\prime_1;r_2;...;r_A)\times \nonumber \\
& & \ \ \ \ \ \ \ \ dz_1 dz^\prime_1 d^2b_1 \prod\limits_{j=2}^A
d^3r_j.
\label{compt_ia_r}
\end{eqnarray}
The above expression allows a rather transparent interpretation of the space-time evolution of 
the inclusive process. First, the $\Theta$ function enforces the condition that 
in longitudinal direction absorption of virtual photon happens before its emission. 
Secondly, one can see that at large $p_{iz}$,  the longitudinal distance which knocked-out 
nucleon propagates is 
proportional to $\sim {1\over p_{iz0}}$. Thus one observes that starting at 
$p_{iz}\ge 300$~MeV/c
the struck nucleon propagates the distances of $\le 1$~Fm. This estimate is in accordance 
with that of Sec.\ref{space-time}.

\medskip
\medskip

\noindent{\bf Final State Interactions:} We now discuss the diagrams  of Fig.\ref{im_compton_fwd}b
representing final state interactions.
In the $W_{NN}\gg 2 m_N$ limit
the imaginary part of the sum of all rescattering diagrams in Fig.\ref{im_compton_fwd}b
cancels  out due the the closure condition for the intermediate states.
However it follows from Eq.(\ref{kinadd}) that  for 
$x>1$ the  mass of the produced two nucleon system, $W_{NN}$,  is close to  $2m_N$ and condition 
for  application of closure is not satisfied.  
Thus in this situation the explicit evaluation of the rescattering part of 
the Compton scattering amplitude is required.

We evaluate rescattering  diagrams  within the Generalized Eikonal approximation~(GEA)
\cite{FSS97,MS01,Ciofi_gea} which uses the set of effective Feynman diagram rules
to calculate a given $n^{th}$-order rescattering of the struck-nucleon off the spectator nucleons 
in nuclei.

Within GEA nuclear Compton scattering amplitude corresponding to the 
$n$-fold rescattering of energetic knocked-out nucleon off the 
spectator nucleons can be expressed as follows:
\begin{eqnarray}
& &  {\cal I}m A^{\gamma^*A\rightarrow \gamma^*A}_n(P^{\prime}_A = P_A) =  
A \cdot {\cal I}m \sum\limits_{E_m}\sum\limits_{spins}\sum\limits_{perm}\int 
\Psi^\dagger_{A}(p_1;...;p_l;..;p_{l+n-1};..p_A)  \nonumber \\
& & {J_{eN}^{\dagger, \mu}(p_f,s_f,p_1,s_1)\over 2q(p_{iz}-p_{1z}+\Delta + i\epsilon)}
\prod\limits_{j=1}^n 
{\xi(s_j) f_{NN}(p^\perp_{1,j}-p^\perp_{1,j-1})\over 
2(p_{iz}-p^z_{1,j}+\Delta_j + i\epsilon)}{d^3p_{1j}\over (2\pi)^3}\nonumber \\ 
& &  J_{en}^{\nu}(p_{f,n},s_f,p_{1,n},s_{1,n})
\Psi_{A}(p_{1,n};...;p^\prime_l;...;p^\prime_{l+n-1};..p_A)
{d^3p_1}  \prod\limits_{k=1}^{A} d^3p_k\delta^3\left(\sum\limits_{i=1}^A p_i\right),
\label{compt_n}
\end{eqnarray} 
where $\Delta\sim \Delta_j \approx {q_0\over q}E_m$,  
$p^\prime_{l+j} = p_{l+j} - (p_{1,j}-p_{1,j-1})$ and 
$\xi =  {\sqrt{s_j(s_j-4m_N^2)}\over 2q m_N}$, where $s_j$ is the c.m. invariant energy 
square of two nucleons for the $j^{th}$  rescattering.
 
We transform the above equation to
coordinate representation estimating all 
propagators at their pole values.
Again considering the kinematical limit of Eq.(\ref{cord_lim})
and using coordinate representations of Eqs.(\ref{wfr},\ref{theta}), after several 
straightforward steps 
we arrive at:
\begin{eqnarray}
& & {\cal I}m A^{\gamma^*A\rightarrow \gamma^*A}(P^\prime_A-P^A)_{(n)} = 
{A\over  2 q}\sum\limits_{E_m} \sum\limits_{spins} J_{eN}^\mu(Q^2) J_{eN}^{\nu}(Q^2)
 \sum\limits_{l_1\ne l_2...\ne l_n=2}^A\times\nonumber \\
& &\int \Psi^\dagger_A(b_1,z_1;r_2;..;r_{l_1};..;r_{l_n};..;r_A)
\prod\limits_{i=1}^n\Theta(z_1-z_{l_i}){\cal I}m \Gamma_{NN}(b_i-b_{l_i})\Theta(z_{l_i}-z^\prime_1) \nonumber \\
& & e^{ip_{iz}(z_1-z^\prime_1)}
\Psi_A(b_1,z^\prime_1;r_2,..;r_{l_1};..;r_{l_n};..;r_A) dz_1 dz^\prime_1 d^2b_1 \prod\limits_{j=2}^A
d^3r_j. 
\label{compt_n_r}
\end{eqnarray}
where
\begin{equation}
{\cal I}m \Gamma_{NN}(b) = -{1\over 2} \int {\cal I}m f_{NN}(k_\perp)e^{-i{\bf k_\perp b}}
{d^2k_\perp\over (2\pi)^2}.
\end{equation}

Eq.(\ref{compt_n_r}) shows clearly how the scattering develops in the coordinate space.
First of all the reinteractions take place  over
the longitudinal distances 
confined between the points of absorption~($z_1$) and emission~($z^\prime_1$) of the virtual 
photon, this is enforced by the product of two $\Theta$ functions: 
$\Theta(z_1-z_{l,i})\Theta(z_{l,i}-z^\prime_1)$.
Secondly, due to the exponential factor $e^{ip_{iz}(z_1-z^\prime_1)}$,
the overall longitudinal distance is limited by the longitudinal momentum probed in the 
reaction - in the same way as it was for the case of  impulse approximation.
That is $z_1-z^\prime_1\sim {1\over p_{iz}}$.

According to this result starting at  $|p_{iz}|\ge 300$~MeV/c the rescatterings 
will be confined at longitudinal distances of $1$~Fm. In such case 
the first rescattering will take place predominantly with a
nucleon in SRC. Contributions from 
double and higher order rescatterings become numerically negligible since it is rather
improbable to find two and more nucleons at longitudinal distances $\le 1$~Fm 
from the struck nucleon.
The latter observation reinforces the expectation that the higher order rescatterings play 
negligible role in Eq.(\ref{ratio}) as soon as eikonal regime is established for final 
state reinteraction processes at $x>1$ inclusive kinematics.

Large value of FSI  and  the violation of Eq.(\ref{ratio}) were suggested in 
Ref.~\cite{BN} in which dominant FSI effects have been generated at  $x\ge 1$ due to  
electron scattering off almost stationary nucleon within a nucleus. 
However within eikonal approximation such rescattering  corresponds to the non-pole 
contribution in the rescattering amplitude, such as Eq.(\ref{compt_n}), which 
has only  real part once the unitarity of 
eikonal amplitude is restored (see above) and therefore does not contribute to the 
inclusive cross section.

\subsection{FSI in near threshold kinematics}

The above discussed picture of FSI  for 2N SRC kinematics gradually changes with an increase 
of $x$ if the kinematic threshold is approached. For example if we consider the  
$x\rightarrow 2$ limit for the scattering off the deuteron for fixed $Q^2$,
the   final state  mass of NN system is close to that of the deuteron. In this case, 
similar to  the case of 
deuteron form factor one gets comparable contributions from small  relative momenta 
in  the initial and final states. In the nonrelativistic calculations \cite{Arenhoevel}  
for $Q^2~\sim~few$~GeV$^2$ this contribution was found 
to be important for $W-m_D\le 50 MeV$. 
For heavier nuclei this effect for $x\sim 2 $ is washed out by the motion of the 
nucleon pair in the mean field and by the contributions of three nucleon correlations.

\subsection{Measuring light cone momentum distribution of nucleons in nuclei}
The above discussions allow us to conclude that for kinematics of Eq.(\ref{kin}) when additional 
condition for FSI being in the eikonal regime, (Eq.(\ref{kinadd}) is 
satisfied, the FSI is predominantly confined 
within  SRC and as a result it   should mostly
cancel out in the ratios of Eq.(\ref{ratio}). 

\subsubsection{The $\alpha$ distribution and FSI}

Next question which we would like to address is whether in addition to observing the onset of SRC at 
$x>1$ kinematics through the ratio of Eq.(\ref{ratio}) one can extract information about SRC which 
is less affected by the FSI.

For this we discuss several important advantages we gain by using 
light-cone momentum fraction of  interacting nucleon, $\alpha_i$, as it is defined 
in Eq.(\ref{alphai}).

Using $\alpha_i$ we can rewrite the denominator of knocked-out nucleon's
propagator,  which enters  in Eq.(\ref{compt_ia}), in the following form:
\begin{equation}
p_f^2 - m_N^2 +i\epsilon = mq_+\left(\alpha_i - {Q^2\over m q_+} + {q_-\over m_Nq_+}
(M_A - p_{R+}) + {m_i^2 - m_N^2\over q_+m_N} + i\epsilon\right),
\label{foralpha}
\end{equation}
where we use energy and momentum conservation: 
$p_f^\mu = q^\mu + P^\mu_A - P^\mu_R$ and define  $q_\pm = q_0\pm {\bf q}$ and $m_i^2 = (P_A - P_R)^2$.
In Eq.(\ref{foralpha}) $p_{R+}$ represents the $''+''$ component of the recoil nucleus 
four-momentum and it is a light-cone analog\cite{FS88} of the
recoil energy $E_R$ (discussed 
in Sec.2) which characterizes the total kinetic energy of recoil system in the nuclear 
lab frame\footnote{In light cone representation the sum over $E_R$ in the closure relation 
for $n_A(k)$ is replaced by the sum 
over $p_{R+}$ in the closure relation for $\rho_A^N(\alpha,p_t)$.}.

It is instructive to compare Eqs.(\ref{piz},\ref{piz0}) with Eq.(\ref{foralpha}).
It follows from  Eqs.(\ref{piz},\ref{piz0}) that the fast nucleon propagator depends on $E_R$ 
in large momentum transfer limit $q_0\sim q$. As  a result it cannot be factored 
out of the sum over the recoil system's excitations.

However situation is quite different for representation based on the light-cone variables. 
It follows from Eq.(\ref{foralpha}) that in 
the limit, $Q^2\to \infty, x=const$,
$$q_+\gg q_-\, \ \ \ \  
\mbox{and}  \ \ \ \ q_+m_N\gg (m_i^2-m_N^2).$$
As a result, the  denominator is practically independent of the  excitation energy of residual  
nucleus, $p_{R+}$. Consequently, $\alpha_i$  is factored out 
from the sum - $\sum\limits_{p_{R+}}$ 
(which replaces $\sum\limits E_R$
in Eq.(\ref{compt_ia}) in  the light cone representation) with 
$\alpha_i\approx {Q^2\over m_Nq_+}$, which  depends only on $x$ and $Q^2$.

The light cone
factorization considerably simplifies the expression for 
the inclusive cross section.
Using the correspondence relation in nonrelativistic limit between nonrelativistic and light-cone 
wave functions of the nucleus (see Eq.(\ref{nr-lc})),   and 
above discussed factorization of the 
knocked-out nucleon's propagator out of
the integral of Eq.(\ref{compt_ia}) for 
forward Compton scattering amplitude in impulse approximation one obtains:
\begin{equation}
{ \cal I}m A_{(0)}^{\gamma^* A\rightarrow \gamma^* A}(P^\prime_A = P_A) = 
{\pi A \over q_+} W_N^{\mu,\nu}(Q^2,x,\alpha_i)\cdot \rho_{A}(\alpha_i),
\label{IA_rho}
\end{equation}
where $W_N^{\mu,\nu}\sim J_N^{\mu,\dagger}J_N^\nu$ represents the electromagnetic tensor of 
nucleonic currents and $\rho_A(\alpha)$ represents light-cone density matrix of nucleus 
integrated over the transverse momentum of the nucleon
 which  
is defined as:
\begin{eqnarray}
\rho_{A}(\alpha) & =  &  \sum\limits_{p_{R+}}\sum\limits_{spins}\int 
|\Psi_{A}(\alpha_1,p_{1t};\alpha_2,p_{2t}; ....\alpha_A,p_{At})|^2\times \nonumber \\
& & \delta(\alpha -\alpha_1) \prod\limits_{j=1}^A{d\alpha_j\over \alpha_j}d^2p_{jt} 
\delta(\sum\limits_k\alpha_k-A)\delta(\sum\limits_k p_{kt}).
\label{denfun}
\end{eqnarray}
According to this result, within the impulse approximation,
in high $Q^2$ and $x\ge 1$ kinematics, the  inclusive 
$eA$ scattering probes the light-cone density matrix, $\rho_A(\alpha)$ of the nucleus.

\medskip
\medskip

We are in position now  to analyze how the final state interaction alters this
factorization. The key point here 
is that in 
eikonal approximation the propagator of fast rescattering nucleon 
is approximately independent of the excitation energy of the recoil system.

This can be seen from Eq.(\ref{compt_n})
 in which the denominator of the fast nucleon 
propagator can be rewritten in the following form
\begin{equation}
p_{iz0}-p_{iz} + \Delta + i\epsilon = m [\alpha_1 - \alpha_i + 
{q_0-q_3\over q_3 m}E_R + i\epsilon] \approx 
m[\alpha_1 - \alpha_i - {Q^2\over 2 q_3^2}{E_R\over m} + i\epsilon],
\label{delta_alphs}
\end{equation}
where $\alpha_1 = {p_{1-}\over P_{A-}}$ and $E_R = m + E_{A-1}-M_A$.
It follows from Eqs.(\ref{delta_alphs},\ref{compt_n}) that 
in the limit 
${Q^2\over 2 q_3^2}{E_m\over m}\ll 1$ the rescattering part of the amplitude is 
independent of 
the excitation energy of recoil nucleus.
Hence 
using relation between 
nonrelativistic and light-cone nuclear wave functions (Eq.(\ref{nr-lc}) we obtain:
\begin{eqnarray}
& &  {\cal I}m A^{\gamma^*A\rightarrow \gamma^*A}_{(1)}(P^{\prime}_A = P_A) =  
A \cdot {\cal I}m \sum\limits_{p_{R+}}\sum\limits_{spins}\sum\limits_{l=2}^A\int \nonumber \\
& & \Psi^\dagger_{A}(\alpha_1,p_{1t};\alpha_2,p_{2t};..\alpha_l,p_{lt};..\alpha_A,p_{At})  
J_{eN}^{\dagger, \mu}(\alpha_f,p_{ft},s_f,\alpha_1,p_{1t},s_1) \nonumber \\
& & {\sqrt{s_{NN}(s_{NN}-4m_N^2)}\over 8m_Nq_v^2} 
{f_{NN}(p^\prime_{1,\perp}-p_{1,\perp})\over (\alpha_1-\alpha_i + i\epsilon)
(\alpha^\prime_1 - \alpha_i + i\epsilon)} 
J_{en}^{\nu}(\alpha^\prime_f,p^\prime_{ft},s_f,\alpha^\prime,p^\prime_{1t},s_1^{\prime})
\nonumber \\
& & \Psi_{A}(\alpha^\prime_1,p^\prime_{1t};\alpha_2,p_{2t};..\alpha^\prime,p^\prime_{lt},.\alpha_A,p_{At})
{d\alpha_1,d^2p_{it}} {d\alpha^\prime_1,d^2p^\prime_1\over (2\pi)^3}  \nonumber \\
& & \prod\limits_{j=2}^{A} {d\alpha_j\over \alpha_j}d^2p_{jt}\delta(\sum\limits_{k=1}^A\alpha_k - A)
\delta^2(\sum\limits_k p_{kt}).
\label{compt_1lc}
\end{eqnarray} 
The imaginary part of the above expression is estimated through the 
imaginary part of $NN$ scattering amplitude, $f_{NN}$ and the pole values of 
the propagators. The latter yields  the conservation of light-cone momentum 
fraction of the struck nucleon, i.e. $\alpha_i = \alpha_1 = \alpha^\prime_1$. 
If factorization of electromagnetic current is assumed then
the overall effect of FSI can be represented through the correction part to the 
light cone density matrix\footnote{This approximation is generally referred as  
Distorted wave impulse approximation} in such a way that expression (\ref{IA_rho}) will 
still describe the inclusive cross section replaced with modified density matrix:
\begin{equation}
{ \cal I}m A^{\gamma^* A\rightarrow \gamma^* A}(P^\prime_A = P_A) = 
{\pi A \over m q_+} W_N^{\mu,\nu}(Q^2,x,\alpha)\cdot \rho^{DWIA}_{A}(\alpha_i)
\label{IA_rhom}
\end{equation}
where 
\begin{equation}
\rho^{DWIA}_A(\alpha) = \rho_A(\alpha) + \Delta \rho_A(\alpha)
\end{equation}
with 
\begin{eqnarray}
\Delta\rho(\alpha) & = & -{1\over 4}\sum\limits_{p_{R+},spins}\sum\limits_{l=2}^A\int 
\Psi^\dagger_{A}(\alpha_i,p_{1t};\alpha_2,p_{2t};..\alpha_l,p_{lt};..\alpha_A,p_{At})  
\times \nonumber \\
& & {\cal I}mf_{NN}(p^\prime_{1,\perp}-p_{1,\perp}) 
\Psi_{A}(\alpha_i,p^\prime_{1t};\alpha_2,p_{2t};..\alpha^\prime,p^\prime_{lt},.\alpha_A,p_{At})
\times \nonumber \\
& & {d^2p_{it}} {d^2p^\prime_1\over (2\pi)^2}  
\prod\limits_{j=2}^{A} {d\alpha_j\over \alpha_j}d^2p_{jt}\delta(\sum\limits_{k=1}^A\alpha_k - A)
\delta^2(\sum\limits_k p_{kt}).
\label{del_rho}
\end{eqnarray} 
In this case the inclusive cross section  can be represented as follows:
\begin{equation}
{d\sigma^{eA}\over dE_{e^\prime} d\Omega_{e\prime}} = K \bar\sigma_{eN}\rho^{DWIA}_A(\alpha_i),
\label{rhoalpha}
\end{equation}
where $K$ contains kinematic factors.
Eq.(\ref{rhoalpha}) leads again to the scaling prediction in the form
\begin{equation}
{{d\sigma^{eA}\over dE_{e^\prime} d\Omega_{e\prime}}\over K \bar\sigma_{eN}} = 
\rho^{DWIA}_A(\alpha_i).
\label{rhoalpha_norm}
\end{equation}
Thus in high energy limit the dependence of the cross section on the nucleus structure 
is reduced to the dependence on modified LC nuclear density matrix which is a function of 
one variable, $\alpha_i$. This result leads to a prediction of scaling law 
which is an analogue of the Bjorken scaling in deep inelastic scattering for the case of elastic 
scattering off nucleonic constituents in nucleus.

The structure of the rescattering term is similar to that in the deuteron break up 
at same $x$ and $Q^2$. 
This allows us to estimate $\Delta \rho(\alpha)/\rho(\alpha) \le 2\%$ in the region of sufficiently 
large $W_{NN}$ for which eikonal approximation is applicable.
A detailed numerical study of this effect will be presented elsewhere.

\subsection{Light-cone scaling of the ratios}
\begin{figure}[ht]
\centering\includegraphics[height=7cm,width=12cm]{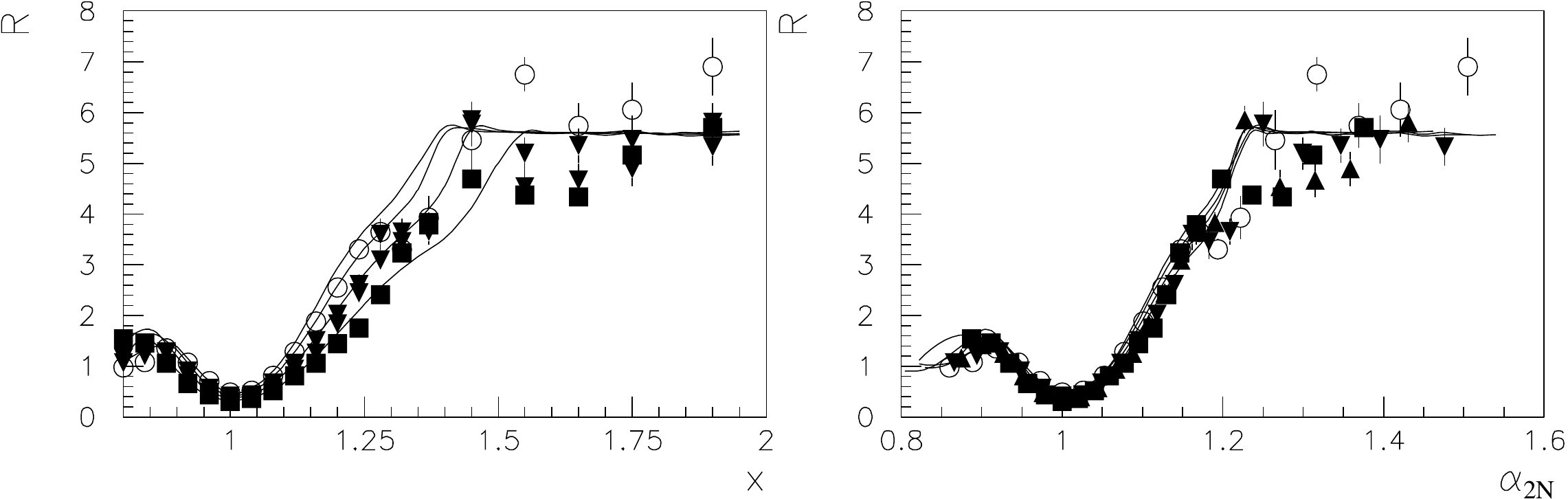}
\caption{The $x$ and $\alpha_{2n}$ dependence of the ratio 
$R={2\sigma^{56}\over 56 \sigma^d}$ for different values of $Q^2 = 1.2\div 2.9$~GeV$^2$.}
\label{alpha_ratio}
\end{figure}

We demonstrated
above that in kinematics of Eq.(\ref{kin})  FSI is predominantly 
acting within the SRC and therefore cancels out in the ratios of the cross sections at fixed 
$x$, and  $Q^2$. 
However it follows from the discussion in section 3.1 that the same $x$ corresponds to different 
struck nucleon momenta for different $Q^2$. Since in the large $Q^2$ limit $x\to \alpha_i$ and cross 
section does not depend on the other three components of nucleon momentum (or equivalently 
$p_{R}$) it is natural to consider ratios for the same $\alpha_i$~(see Eq.(\ref{rhoalpha_norm})).
In the region of $x< 2$ where scattering off two nucleon correlation is allowed we observe that the 
dominant contribution comes from configurations with recoil energy similar to that in the 
deuteron~(see Sec.~2.2). 
The motion of the pair leads to smearing of the recoil energy distribution.
However the maximum of the distribution over $\tilde m^2$ is close to the value of  $\tilde m^2$ for 
the deuteron for the same $\alpha_i$.
 Moreover we checked in Ref.~\cite{FSDS} that variation of 
$\tilde m^2$ around the deuteron value leads to a small variation of 
$\alpha_i$. Accordingly, one can integrate  over the $p_{R, +}$, and 
$p_{i, t}$  leading to conclusion that  for $x < 2$,
  \begin{equation}
  {2\sigma_{eA}(x,Q^2) \over A\sigma_{e^2H}(x,Q^2)}=
  {\rho_A(\alpha_{2N})\over \rho_{^2H}(\alpha_{2N})},
\label{alpha_scale}
  \end{equation}
    where 
    \begin{equation}
 \alpha_{2N} = 2 - {q_-+2m\over 2m}\left(1 + \sqrt{W^2-4m^2\over W}\right)
\label{alphatn}
\end{equation}
is the  minimal value of $\alpha_i$ for the scattering off the deuteron for 
kinematics of  Eq.(\ref{kin}).  Here $W^2=4m^2+Q^2(2/x-1)$ 
is the  invariant mass squared for the   electron scattering off the 
deuteron in the kinematics of  Eq.(\ref{kin}).
Obviously, the relation of Eq.(\ref{alpha_scale}) will be violated at large $Q^2$ due to  the 
contribution of inelastic processes (see Fig.7 in Ref~\cite{FSDS}).
This relation predicts that even though the ratios of cross sections plotted as a function of $x$ 
somewhat change with $Q^2$, they should yield a much better scaling if plotted as a function 
of $\alpha_{2N}$. Indeed the SLAC data are consistent with this prediction, 
(see  Fig.\ref{alpha_ratio}). One can see from the figure that the data cover  the region 
$\alpha\le 1.55$ corresponding to the internal nucleon momenta $\le 0.6$~GeV/c. 
For larger $Q^2$  and $x$ sufficiently close to $2$
corresponding to the minimal values of
$\alpha_{2N}$ being close to two, 
a new pattern may be expected.
This is related to the observation of   Ref.~\cite{FS81}  
that 3N SRCs appear 
to become  important for $\alpha\ge 1.6$. 
Therefore  when this kinematics 
is
reached one expects that the increase of the 
ratio and onset of the new plateau  should start below $x=2$.  

One can also define typical $\alpha_{3N}$ for scattering off a three nucleon correlation. It is rather 
insensitive in a wide $Q^2$ and $x$ range to the value of the recoil mass of the two nucleon 
system. 
Hence it will be interesting to check at which values of $x$ the scaling of the ratios as a function of 
$\alpha_{3N}$ will set in  for  $x> 2$.
At the same time it is worth emphasizing that for the region of $\alpha_{3N}  < 2$ and  $x>2$ 
the  cross section is not related directly to the LC density matrix but to the integral  
of the spectral function for approximately constant $\alpha_{3N} $  over a restricted range of the 
recoil masses.
These masses will  exclude values  typical for  two nucleon correlations and 
therefore will not allow a closure approximation. 
A comparison of the $x> 2$ and  $1< x < 2$ data for $\alpha_{2N}=\alpha_{3N}$ would provide a 
unique information about relative importance of different contributions to the spectral function 
for $\alpha_i \sim 1.5 \div 1.7$. Such experiments are planned in JLab though the current plans for 
6 GeV do not cover sufficiently high values of $Q^2$.

\section{Breaking SRCs in Hard Semi-Exclusive Reactions}

We explained in Sec.2 that a removal of a nucleon from type 2N-I SRC leads to an emission of a 
nucleon in the direction opposite to the  direction of initial momentum of struck nucleon.
Therefore  if  projectile removes say a proton from 2N SRC the decay function mechanism 
leads to an emission of a neutron with momentum distribution proportional to the square of 
the deuteron wave function at high momenta.
 This mechanism suggested in Ref.~\cite{FS77} allowed to explain the shape of the 
spectrum of fast nucleons emitted backward in reactions of Eq(\ref{fbn}) and led to the value of 
$a_2(C) \sim 4\div 5$ which is very close to the one obtained from $A(e,e')X$ reactions at 
$Q^2\ge 1.5$~GeV$^2$ and $x> 1$ described in Sec.3.

A limitation of the reactions 
listed
in Eq.(\ref{fbn}) is that due to 
their  inclusive nature it is impossible to  fix the 
momentum of the nucleon (nucleons)  which was removed in the process of producing 
fast backward nucleon.  It can be partially alleviated by studying 
fast backward  production in deep inelastic scattering in which case predominance of 
the scattering from  forward moving nucleons leads to a reduction of average Bjorken $x$
proportional to the light cone fraction of this nucleon\cite{FS77}. 
This effect is observed in several neutrino bubble chamber experiments, 
(see e.g. Refs.~\cite{nuA1} and \cite{nuA2}), although the production of fast backward nucleons due to  
hadronic reinteractions leads to a reduction of this effect.

Much more stringent tests of the structure of SRC can be performed in coincidence experiments  
in which a nucleon is removed from the nucleus  with known momentum, and the second nucleon from 
the decay of the SRC is detected \cite{FS88,FLFS89}.
An example of such process is the reaction, 
$p+ A\to p+p+(A-1)^*$, in which both protons are detected at large c.m. angles
\footnote{Initially, the interest in experimental study of this reaction
was prompted by suggestion of Refs.~\cite{Mueller} and \cite{Brodsky} that 
these  processes can be used for studies  of color transparency phenomenon.}.  

For the purposes of studies of SRC it is sufficient to reach the range of large 
energy momentum transfers   corresponding to $E_p \ge 5$~GeV in which case color transparency 
effects appear to be  small. 
Detection of two forward protons allows to determine with  high precision 
the light-cone fraction of the interacting nucleon,  $\alpha_{i}$,  and therefore to measure 
the light-cone density matrix of the nucleus. 
The analysis of Ref.~\cite{isr}  
of EVA data\cite{eip1} on $\alpha_i$ distribution found that the data agree 
well with calculation which includes SRC with the strength determined from $A(e,e')X$ data
at  $Q^2\ge 1.5$~GeV$^2$ and $x>1$.

Since the cross section of elementary $pp\to pp$ reaction decreases very rapidly with an 
increase of the energy:
$${d\sigma^{pp\to pp}(s_{pp}, \theta_{c.m.})\over d \theta_{c.m.}}\propto s_{pp}^{-10},$$ 
for $\theta_{c.m.}\sim 90^o$, the scattering 
preferentially occurs off the forward moving proton which have $\alpha_{i} <1$ since in 
this case $s_{pp}^{i}\approx \alpha_{i} s_{pp}< s_{pp}$.

If a forward moving proton ($\alpha_i$) belonging to a type 2N-I SRC is removed from the  nucleus, 
correlated nucleon from the decay of SRC should be emitted backward. 
Indeed, for the  case of 2N SRC the
light-cone  momentum fraction for the  correlated spectator is:
$\alpha_{s} \approx 2- \alpha_{i}  > 1$. Expressing $\alpha_s$ through the 
lab frame energy $E_s$ and $z$-component of momentum $p_s$: $\alpha_{s} = {E_s-p_{s,z}\over m_N}$
one observes that condition $\alpha_s > 1$ indeed 
corresponds to a production of nucleon in the  backward hemisphere~($p_{sz}<0$). 
Based on this observation it was predicted\cite{FS88,FLFS89}  that  in the process of  
$p+A\to p+p+(A-1)^*$ there should be a strong correlation  between knock-out of the fast forward 
proton and emission of a fast backward nucleons, mostly neutrons.

\begin{figure}[ht]
\centering\includegraphics[height=9cm,width=12cm]{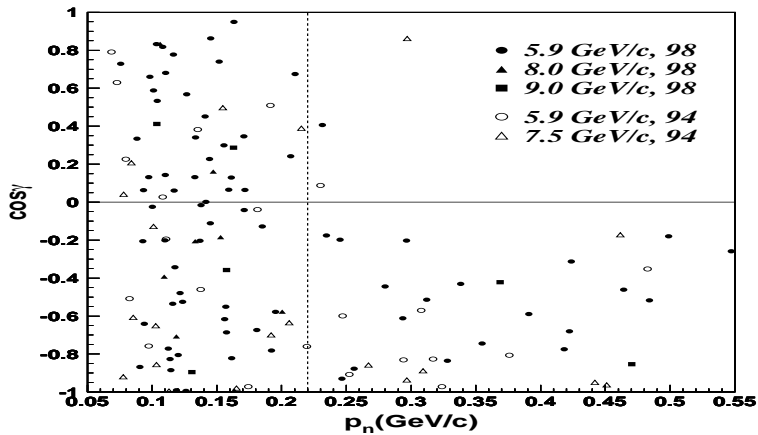}
\caption{The correlation between~$p_n$ and its direction $\gamma$ relative 
to $\vec p_i$. Data labeled by $94$ 
and $98$ are from Refs.15 and 16 respectively. The momenta
on the labels are the beam momenta.  The dotted vertical line corresponds to 
$k_F=220$~MeV/c.}
\label{evaexp}
\end{figure}

The BNL experiment E850 (EVA)\cite{eip2} was supplemented with  neutron detectors which  covered 
a large fraction of the backward angles. The experiment discovered  
\cite{eip1,eip2}  that in the process of $p+^{12}C \to p+p+n +X$, a removal of a proton with 
momentum $|p_i|>k_F=220$~MeV/c (the Fermi momentum  for carbon) produces 
strong back-to-back directional correlation between $p_{i}$ and momentum of the neutron, $p_n$ (Fig.\ref{evaexp}). 

The experiment\cite{eip2} extracted  the following quantity:
\begin{equation}
F = \frac{\mbox{Number of (p,ppn) events ($p_i,p_n>k_F$)}}
{\mbox{Number of (p,pp) events ($p_i> k_F$)}}, 
\label{F}
\end{equation} 
which represents the measure of correlation of backward neutrons with 
initial momentum of struck proton. 
For initial momentum range of  $250-550$~MeV/c experiment extracted 
$F = 49\pm 13\%$, which means about half of the events with  $|p_i|>k_F$ had 
directionally correlated neutrons with $ |p_n|>k_F$.

Further analysis of these data was performed in Ref.~\cite{eip3}, which observed that in 
high momentum transfer kinematics, above 
mentioned ratio can be related to the quantity $P_{pn/pX}$ which represents the relative 
probability of finding $pn$ correlation in the $''pX''$ configuration that contains a proton 
with $p_i> k_{F}$. Such relation reads:
\begin{equation}
P_{pn/pX} = {F\over T_n R},
\label{PpnSRC}
\end{equation}
where $T_n$ accounts for the attenuation of the neutron in the nucleus and 
\begin{equation}
R \equiv {\int\limits_{\alpha_i^{min}}^{\alpha_i^{max}}
\int\limits_{p^{min}_{ti}}^{p^{max}_{ti}}
\int\limits_{\alpha^{min}_n}^{\alpha^{max}_n}
\int\limits_{p^{min}_{tn}}^{p^{max}_{tn}}
D^{pn}(\alpha_i,p_{ti},\alpha_n,p_{nt},P_{R+}) {d\alpha\over \alpha}d^2p_t {d\alpha_n\over \alpha_n}
d^2p_{tn}dP_{R+}\over 
\int\limits_{\alpha_i^{min}}^{\alpha_i^{max}}
\int\limits_{p^{min}_{ti}}^{p^{max}_{ti}}
S^{pn}((\alpha_i,p_{ti},P_{R+}) {d\alpha\over \alpha}d^2p_t dP_{R+}}.
\label{R}
\end{equation}
Here $D^{pn}$  is light-cone generalization of the decay function discussed in Sec.2, corresponding to 
the situation in which removal of a proton with light-cone momentum $\alpha_i, p_{ti}$ from the nucleus 
was followed by an emission of neutron with momentum ($\alpha_n, p_{nt}$). The recoil energy 
is described by $P_{R+}$ (see discussion in Sec.3). The spectral function  $S^{pn}$  
in Eq.(\ref{R}) represents the part of the spectral function $S^{p}$ related to $pn$-correlation only and 
it is related to the decay function by the  relation analogous to Eq.(\ref{sumrule}).

In Ref.~\cite{eip3} decay function was modeled based on 2N SRC model which includes 
the motion of the center of mass of the correlation in the nucleus mean field.  Within 
this approximation the decay function is represented through the convolution of two density 
matrices representing relative ($\rho_{SRC}$) and center of mass ($\rho_{cm}$) momentum 
distributions as follows:
\begin{eqnarray}
D^{pn}  & =  & \rho^{pn}_{SRC}(\alpha_{rel},\vec p_{t,rel})\cdot
\rho^{pn}_{c.m.}(\alpha_{c.m.},\vec p_{t,c.m.}){\alpha_n\over
\alpha_{c.m.}}\times
\nonumber \\
& & \delta\left(P_{R+}- {m^2+p_{t,n}^2\over m\alpha_n} - {M^2_{A-2}+
p^2_{t,c.m.}\over m(A-\alpha_{c.m.})}\right),
\label{D_c.m.}
\end{eqnarray}
where $\alpha_{rel} = {\alpha_i-\alpha_n\over \alpha_{c.m.}}$, 
$p_{t,rel} = p_{ti} - {\alpha\over \alpha_{c.m.}}p_{tn}$, 
$\alpha_{c.m.} = \alpha_i+\alpha_n$, and $p_{t,c.m.} = p_t + p_{tn}$. 
Within  2N-SRC model~\cite{FS81}, $\rho^{pn}_{SRC}$ is related 
to the LC density matrix of the deuteron as:
\begin{eqnarray}
 \rho_{SRC}(\alpha,p_t) = a_{pn}(A){\Psi^2_D(k)\over 2-\alpha}\sqrt{m^2+k^2},
\label{rc.m.}
\end{eqnarray}
where  $\Psi_D(k)$  is the deuteron wave function, and for  $0< \alpha < 2$
\begin{equation}
k = \sqrt{{m^2+p^2_t\over \alpha(2-\alpha)} - m^2}.
\label{klc}
\end{equation} 
The parameter $a_{pn}(A)$ is the probability (relative to the deuteron) of
having a $pn$ SRC pair in nucleus $A$, which is analogous to $a_2$ quantity 
discussed in Sec.3 but only for proton-neutron correlation.

The c.m. motion of the SRC relative to the $(A-2)$ spectator system is 
described  by a Gaussian ansatz similar to Ref.~\cite{CiofiSimula} with $\sigma$ 
being a parameter.
This distribution can be expressed through 
the LC momentum of the c.m. of the SRC as follows:
\begin{equation}
\rho_{c.m.}(\alpha,p_{t}) = 2m\left({1\over 2\pi \sigma^2}\right)^{3\over 2}
e^{-{m^2(2-\alpha)^2+p_{t}^2\over 2\sigma^2}}.
\end{equation}
It is normalized as $\int \rho_{c.m.}(\alpha,p_{t}) {d\alpha\over
\alpha}d^2p_t = 1$.
The parameter, $\sigma$ describing the width of the momentum 
distribution of the c.m. of 2N SRC was determined  experimentally: $\sigma = 143\pm 17 $MeV/c. 
It was found to be in excellent agreement with the 
value calculated in Ref.~\cite{CiofiSimula} for the carbon spectral function within the 2N-SRC 
model -  $\sigma=139$~MeV/c.

It  was demonstrated that above  approximation for the decay function describes major 
characteristics of the 
$A(p,2pn)X$ data \cite{eip3,isr}. 
Based on this model of decay function and constraining the integration region of Eq.(\ref{R}) by 
kinematical cuts of experiment:
\begin{eqnarray}
\mbox{struck proton: }&&  0.6 < \alpha_i < 1.1;  \  p_i > p^{min}=0.275~\mbox{MeV/c}    
\nonumber \\ 
\mbox{recoil neutron: }&& 0.9 < \alpha_n < 1.4;  \ p^{min}< p_n < 0.55
\nonumber \\
&&  72^0 < \theta_n < 132^0,
\label{cuts}
\end{eqnarray}
for which the recalculated value of correlation strength was $F=  0.43_{-0.07}^{+0.11}$ it was possible 
to estimate:
\begin{equation} 
P_{pn/pX} = 0.92^{+0.08}_{-0.18}.  
\label{P_pn_exp}
\end{equation}
The physical interpretation of this result is that the removal of a proton from the nucleus with initial momentum 
$275-550$~MeV/c is $92^{+8}_{-18}\%$ of the time is accompanied by the emission of a correlated 
neutron that carries momentum roughly equal and opposite to the initial proton momentum. 
Using this result, and assuming
dominance of 2N SRCs in the high momentum component of the nuclear wave function at $250 < p_i < 550$~MeV/c, 
it is possible to estimate an upper limit for the ratio of absolute probabilities of $pp$
and $pn$ SRCs as\cite{eip3}:
\begin{equation}
{P_{pp}\over P_{pn}} 
\le {1\over 2} (1-P_{pn/pX})= 0.04_{-0.04}^{+0.09}.
\label{pp_pn_BNL}
\end{equation}
This result indicates that probabilities of $pp$ or $nn$ SRCs in carbon are at least by  factor of six smaller than 
that of $pn$ SRCs. This provided the first estimate of the isospin structure of 2N SRCs in nuclei  
and may have important implication for modeling the equation of state of asymmetric nuclear matter.

Further studies were performed very recently at  JLab~\cite{eip4,eip5} for electro-nuclear reactions in which case  
$^{12}C(e,e'p)X$,  $^{12}C(e,e'pp)X$, 
and $^{12}C(e,e'pn)X$ scatterings at $Q^2 = 2$~GeV$^2$ and $x_B = 1.2$ were studied  
in the range $300\le p_{i} \le 600$~GeV/c. Since for the chosen kinematics the $\Delta$-production 
is suppressed and $p_{i}> k_{Fermi}$, the type 2N-I SRC has been predicted to give the 
dominant contribution to the cross section.
Experiment also observed a  recoiling partner, proton/neutron,  from SRC 
produced
back-to-back 
to the momentum $p_i$. When interpreted within the same model as used 
for the $A(p,ppn)X$ reaction it was found that only in 
$(9.5 \pm  2)\%$ of the $^{12}C(e,e'p)X$ events, a recoiling nucleons were protons.
The ratio of the recoil neutron and recoil proton production cross section was found to be 
$8.2\pm 2.2$.
 
If one considers the ratio of  probabilities of emission from the $pp$ pair 
with respect to  $pn$ pair the identity of the protons leads 
to a factor of two larger emission probability - in the pp case one of the protons is always 
has a momentum in the forward direction while in the case of the $pn$ correlation this 
probability is 50\%.  Therefore for the ratio of the probabilities of $pn$ and $pp$  pairs one 
finds, $16.4\pm 4.4$. This has to be corrected for the feeding of $pp$ events from the $pn$ events
due to the charge exchange which yield the following final result for ratios of $pp$ to $pn$ SRC probabilities: 
 \begin{equation}
 {P_{pn}\over P_{pp}} = 18.0\pm 5.0.
\label{pp_pn_JLab2}
\end{equation}
corresponding to
 \begin{equation}
 {P_{pp}\over P_{pn}} = 0.056\pm 0.018,
\label{pp_pn_JLab}
\end{equation}
The pp/pn 
ratio was found to be practically constant in the whole studied momentum range: 
$300 \le p_N \le 600 MeV/c$. Also the sum of the absolute $pp$ and $pn$ probabilities was found to be 
close to one, indicating that practically all nucleons in 
this momentum range (with accuracy of the order 10\%) belong  to two nucleon SRC.

The ratio found in Eq.(\ref{pp_pn_JLab}) is in a good agreement with the one in Eq.(\ref{pp_pn_BNL}).
This is certainly not trivial since the mechanism of the reactions in two cases are very different 
(for example a virtual photon could couple to the exchange currents), 
and also in the proton projectile case  a forward moving proton was struck, while in the electron 
experiment virtual photon was absorbed by a backward moving proton. 
In addition the invariant transferred momentum for  $(p,2p)$ reaction was 
$-t\sim 5 $~GeV$^2$ 
exceeding by far $Q^2\sim 2$~GeV$^2$ for $(e,e'p)$ process. 
The observed ratio appears to be somewhat smaller than a naive expectation of a factor of ${1\over 9}$  
based on the pion exchange.

Note also that in the kinematics of JLab experiment invariant mass of the two nucleon system is 
rather small, while transverse momenta are significant. As a result the final state rescatterings 
between two outgoing nucleons of SRC are rather large. 
However our studies indicate that this does not affect 
the ratios and overall probabilities discussed above.

To summarize, the study of hard semi-exclusive correlation processes  demonstrate high discovery  
power in probing different aspects of SRCs.
Already first experiments\cite{eip3}---\cite{eip5} confirm the   SRC origin of 
high momentum nucleons in nuclei. For carbon case SRCs appear to dominate  starting at momenta 
which are close to the Fermi momentum. Most of the SRCs are due to $pn$ correlations. Qualitatively 
these expectations are consistent with an expectations of potential models of the nucleus.  
For more quantitative comparisons it would be necessary to perform calculations of the decay 
functions of nuclei.  
One of the most important aspects of these studies is that it opens up a new venue in probing the isospin 
structure of SRCs.  Such possibilities are especially important for studies of structure of asymmetric 
nuclear matter at high densities such as neutron stars.

\section {Short-range nucleon correlations and neutrino emission by   neutron stars}

In this section we will discuss an example which demonstrates how our understanding of SRCs can be 
used in studies of the properties of cold dense 
nuclear matter. We will concentrate on  the discussion of equilibrium issues of neutron star and related neutrino luminosity.


\subsection{Introduction}
A normal neutron star is bound by gravitational interactions.
Global characteristics of neutron stars follow from the equations 
for the hydrostatic equilibrium in the general relativity, see Ref.\cite{Tolman1,Tolman2}.
A neutron star can be divided into several layers: the crust, the outer 
and the inner cores. The outer core extends up to the densities 
$\rho\sim (2-3)\rho_0$,  where $\rho_0\approx 0.16 nucleon/fm^3$ 
is the nuclear matter density.  The inner core extends to the center 
of the neutron star where densities can be significantly larger 
$\sim 5-10 \rho_0$ and may contain muons, hyperons, and exotic matter.  
Due to inverse $\beta$ decay, the nuclear matter dissolves into a uniform liquid composed of neutrons  at the density 
$\sim 1/2\rho_0 $, with 
\begin{equation}
x=N_{p}/N_{n} \sim 5\div 10\% , 
\end{equation}
admixture of protons and equal admixture of electrons and tiny admixture of muons, see 
Refs.~ \cite{Baym,PanHeis,Sedr}.
In the inner core the value of proton fraction is probably larger:  
$ \sim 10 \div 13\%$ \cite{Yakovlev}.  The most efficient 
neutrino cooling  reactions are due to direct  URCA processes involving neutron $\beta$ decay:
\begin{equation}
n\to p+e+\bar \nu_e,
\label{beta}
\end{equation}
and $\beta$ capture  in
\begin{equation}
e+p\to  n+\nu_e.
\label{inverse} 
\end{equation}
Thus it is worth to analyze how internucleon interactions 
influence thermally excited direct URCA processes within 
 cold neutron stars. Standard cooling scenario assumes that direct URCA  processes can occur in 
the inner core only \cite{Lattimer2}.

In the ideal gas approximation the zero temperature neutron star   is described as the system of 
degenerate   neutron, proton
and electron gases with the ratio of proton and neutron densities, 
$x\ll 0.1$.  For any positive neutron density  the Pauli blocking in 
the electron and   proton  sectors guarantees stability of a neutron 
star to the neutron $\beta$-decay  cf. Refs.~\cite{Weinberg} and \cite{S.T}.  
The number densities of protons and 
electrons are equal to ensure electrical neutrality of the star, so $k_F(e)=k_F(p)$.  
The neutron Fermi momentum is significantly larger than the proton Fermi momentum because of the 
larger number of the neutrons:
\begin{equation}
x^{1/3} k_F(n)=  k_F(p). 
\label {gas}
 \end{equation}

The internucleon interaction produces nucleons with momenta 
above Fermi surfaces,   cf. Eqs.(\ref{neutron},\ref{proton}). 
To guarantee conservation of the electric and the baryon charges
nucleon occupation numbers  below the corresponding Fermi 
surfaces - $f_i(k,T=0)$ should be smaller than unity especially
for protons.  The nonrelativistic Schr\"{o}dinger equation with 
realistic nucleon-nucleon  interactions gives occupation numbers for protons 
with zero momenta $\approx 70\%$ for the nuclear matter density.  
Even a larger depletion of occupation numbers  is found for protons with momenta near the  
Fermi surface \cite{pholes}.

The Landau Fermi liquid approach \cite{L.L}  in which momentum  distribution of quasiparticles 
coincides with the Fermi distribution for the ideal gas of fermions  is effective starting 
approximation for  describing strongly interacting 
liquid.  It has been explained by A.B.Migdal that nucleon distribution at zero temperature 
should exhibit the  Migdal jump at $k=k_F$  which justifies applicability of the Fermi step 
distribution at zero temperature. The value of the  Migdal jump is
equal to the renormalization factor $Z<1$  of the 
single-particle Green's  function in the nuclear matter.  The condition $Z< 1$ follows from 
the probability conservation \cite{Migdal1,Migdal2} and implies that occupation numbers for nucleons with momenta 
$k < k_F$ are below one. In the limit  of small proton concentration 
Fermi surface nearly disappears since proton neighborhood is   
predominantly strongly interacting with neutron medium. So the 
height
of 
the Migdal jump for the proton distribution should decrease 
$\propto x$ for $x\to 0$. (Decrease of the Migdal jump due to a large probability of SRC 
has been discussed a long time ago 
for  the liquid $^3He$ in Ref.\cite{Dygaev}. ). 
Thus   for a highly asymmetric mixture of  protons and neutrons the interaction tends to 
extend proton  momenta  well beyond $k_F(p)$.

We show that for   the temperatures 
$T\ll 1$~MeV the presence of the high momentum proton tail  leads 
to a  different value and temperature dependence of  
URCA processes for $x\ge 1/8$,  cf. Eq. \ref{enhancement} as compared to that in  
Refs.~\cite{Bahcall,Friman,Lattimer} where the Fermi momentum distribution for 
quasiparticles was used. As the consequence of the presence of the high momentum proton tail 
the neutrino luminosity due to direct URCA processes differs from zero even for 
$x < 1/8$  i.e. in  the  region forbidden in  the ideal gas approximation for quasiparticles 
by the Pauli blocking and the  momentum conservation.

The electron gas within neutron star is ultrarelativistic. So the Coulomb parameter $e^2/v\ll 1$. 
Here $e$ is the electric charge of electron and $v=p/E\approx c$ is its velocity. Thus approximation of 
the free electron gas is justified. The Coulomb interaction  between protons with momenta $k\ge k_F(p)$  
and electrons produces  electrons with momenta above the electron Fermi surface,  although with a 
very small
probability cf. Eq.(\ref{electron}).  So  the occupation probability for electrons:  
$f_e(k_e \le k_F(e),T=0)$ is slightly less than one. 

Thus the interaction produces holes in  all Fermi  seas removing the absolute Pauli blocking for the 
direct neutron, muon, hyperon  
$\beta$-decays.  We show however that the account of the Pauli 
blocking in the electron sector ensures stability of a neutron to the direct $\beta$ decay in the 
outer core of a  neutron star. 
Condition of stability may be  violated in the inner core where however use of nucleon degrees of freedom is questionable.

If hyperon  stars exist (for the review of this subject and references see Ref.~\cite{Bethe07}), 
neutrino luminosity due to direct $\beta$- decay may appear significantly larger than for a neutron star.

\subsection{The role  of the interaction}
 
High momentum nucleon component of the wave function of a 
neutron star follows directly from  the Schr\"{o}dinger equation in the limit  $k\gg k_F$ where 
$k_F$ is Fermi momentum. The derivation 
of the formulae is similar to that in \cite{FS81}.

At the leading order in $(k_F^2/k^2)$ the occupation numbers for protons and neutrons with momenta 
above Fermi surface are:  
\begin{equation}
f_{n}(k,T=0) \approx \left({\rho_{n}}\right)^2(\left(\frac{V_{nn}(k)}{k^2/m_N}\right)^2+ 2x\left(\frac{V_{pn}(k)}{k^2/m_N}\right)^2),
\label{neutron}
\end{equation}
and
\begin{equation}
f_{p}(k,T=0) \approx  \left({\rho_{n}}\right)^2(x^2 \left(\frac{V_{pp}(k)}{k^2/m_N}\right)^2 +
2x \left(\frac{V_{pn}(k)}{k^2/m_N}\right)^2).
\label{proton}
\end{equation}
Here $\rho_{i}$ is the density of constituent $i$ .   The factor  $V_{NN}(k)$  describes the high 
momentum tail of the potential of 
the $NN$ interaction.  The factor 2 in the above formulae accounts 
for the number of spin states. In the first term, this factor is canceled due to the identity of 
nucleons within the pair. In the derivation of the formulae for the probability of SRCs we used
the approximation of nucleon density uniform in coordinate space
to  describe the uncorrelated part of the wave function. Thus, the 
value of the high momentum tail depends strongly on the nucleon density in the core of a neutron star.  
Since $k_F(p) $ is significantly  smaller than $k_F(n)$,  the probability to find a proton with 
$k\ge k_F(p)$ for a  neutron density close to the nuclear density should be  significantly larger than 
in  nuclei where $x \approx 1$. 
Note also that the analysis of the recent data on SRCs in the symmetric nuclear matter found a 
significant  $\sim 20\%$ probability of nucleons above  the Fermi surface in nuclei which is 
predominantly due to $I=0$ SRCs\cite{eip3,eip5}.

The Coulomb interaction between protons from SRCs and electrons produces electrons with momenta above the 
electron Fermi surface.   Such electrons are ultrarelativistic  so Feynman diagrams  approach should be 
used to evaluate the high momentum electron component rather than the  nonrelativistic 
Schr\"{o}dinger equation. We find for  the  high momentum electron component approximate expression:
\begin{eqnarray}
\label{electron}
f_{e}(k_e\ge k_F(e),T=0) \approx (1/2)\int (d^3k_p/(2\pi))^3 f_p(k_p) 
\theta(k_p-k_F(p))\rho_{e} \cdot   & &  \\ \nonumber \cdot 
(1-f_p(k_p,T=0))\left(\frac{k_e+{3\over 4}  k_F(e) }{ \sqrt{k_e}\cdot \sqrt{{3\over 4}  k_F(e)}}\right)
\left(\frac{V_{Coulomb}(k)}{k_e-k_e^2/2m_N-{3\over 4}  k_F(e)}\right)^2. 
\end{eqnarray}
The factor $1-f_p(k_p,T=0$ is the number of proton holes which prevent Pauli blocking for the proton after 
interaction with the electron. Effectively, Eq.(\ref{electron})  gives  the probability for triple (e-p-n) 
short range correlations.  This equation can be simplified for  applications by using average quantities: 
\begin{eqnarray}
\label{electron2}
f_e(k_e\ge k_F(e),T=0)\approx 
(1/2) P_{pn} \left<H \right> \cdot
\nonumber \\
\cdot {\rho_{e}}
\left(\frac{k_e+{3\over 4}  k_F(e) }{ \sqrt{k_e}\cdot \sqrt{{3\over 4}   
k_F(e)}}\right)
\left(\frac{V_{Coulomb}(k)}{k_e-k_e^2/2m_N-{3\over 4}  k_F(e)}\right)^2. 
\end{eqnarray}
Here $P_{pn}$  is  the probability of pair nucleon correlation and $\left<H \right>  \approx P_{pn}$.

The factor $$1/2\left(\frac{\sqrt{k^2+m_e^2}+\sqrt{<k_e^2+m_e^2>}}{(k^2+m_e^2)^{1/4}<k_e^2+m_e^2>^{1/4}}\right)$$   
follows  from the Lorentz transformation of the electron e.m. current, conveniently calculable from the Feynman diagrams. 
Here  $<k_e^2>$ is the average value of the square of electron momentum within the 
electron Fermi sea.

\subsection{Impact of SRC on the direct and modified URCA processes 
at small  temperatures}

In the Landau Fermi liquid  approach at finite  temperature, $T$    
the direct  URCA process Eqs.\ref{beta} and \ref{inverse}
is  allowed   by  the energy-momentum conservation law if the proton 
concentration exceeds $x=1/8$ \cite{Lattimer}.  The restriction 
on the proton  concentration follows from the necessity to guarantee the  momentum triangle:
\begin{equation}  
k_{F}(p)+k_{F}(e)\ge k_{F}(n),
\label{Prakash}
\end{equation}
in the absorption of electrons  by the protons.

If proton concentration is below threshold or direct URCA process 
is suppressed   due to nucleon superfluidity neutrino cooling 
proceeds through the less rapid modified URCA processes:
\begin{equation}
{n+(n,p)\to p+(n,p)+e +\bar \nu_e}, 
\label{np1}
\end{equation}
and 
\begin{equation}
{e+ p+(n,p)\to  n +(n,p)  +\nu_e},
\label{np2}
\end{equation}
in which additional nucleon enables momentum conservation.

The neutrino luminosity resulting from the  direct and modified URCA 
processes, $\epsilon_{URCA}$, was  evaluated in Ref.\cite{Lattimer}
for $x\ge 1/8$
where the Fermi distribution:
\begin{equation}
f_{i, bare}(k,T)= \frac{1}{1+\exp \frac{E_i-\mu_i}{kT}},
\end{equation}
describes
 the  Pauli blocking factors $1-f_e(k,T)$ and $1-f_p(k,T)$ in 
the final state. After integration over the phase volume of the decay products it was found:
\begin{equation}
\epsilon_{URCA}=c(kT)^6 \theta(k_F(e)+k_F(p)-k_F(n)).
\end{equation}
Here $c(x\ge 0.1)$ has been calculated in terms of the square of the electroweak coupling constant relevant for  low energy  weak interactions and the phase volume factors.

In the case of realistic NN interactions  significant fraction of protons has momenta above the  proton Fermi momentum. So 
Eq.\ref{Prakash} is satisfied  for the proton large momentum tail even 
for x smaller than 0.1.  For the sake of illustrative 
estimate we substitute in the probability of 
neutron $\beta$-decay the Pauli blocking factor  $(1-f_{p,bare}(k,T))$,  by the actual distribution of 
protons within the core of a neutron star.  We account for the probability of additional neutron from (p,n) 
correlation  by the additional factor $P_{pn}$.

To simplify the discussion we will ignore here tiny probability for  electron holes at zero temperature and  parameterize neutrino luminosity as
\begin{equation}
\epsilon_{URCA}=c (kT)^6  R,
\end{equation}
where $R$ accounts for the  role of SRC in  neutrino luminosity at small temperatures.  We find  
 \begin{eqnarray}
R\approx \kappa_{pn}^2 \left[\int [(1-f_p(k_p,T))\theta(k_F(p)-k(p))+  \right. \nonumber \\ \left.+f(k_p,T)\theta(k_p-k_F(p)] \theta(k_F(e)+k(p)-k_F(n)) d^3 k_p/(2\pi)^3 \right] \cdot \nonumber  \\
\cdot \left[\int (1-f_{p,bare}(k_p,T)) \theta(k_F(e)+k_F(p)-k_F(n))d^3 k_p/(2\pi)^3 \right]^{-1}.
\end{eqnarray}
  Here $f_{p}(k_p,T)$ is the occupation number of protons  
accounting the interaction and    $f_{p,bare}(T,k)$  is the 
Fermi distribution function  over proton momenta at nonzero temperature. The factor   $\kappa_{pn}$ is the overlapping
integral between a component of the wave function of the neutron star
containing pair nucleon correlation and the mean field wave function  of the star.
      For the numerical  estimate we use approximation:
$\kappa_{pn}^2=P_{pn}$. 
For a rough estimate we neglect the first term
in the numerator of the above formulae and  put $T=0$ in the second term. Using for the estimate $V_{NN}(k)\propto 1/k^2$ for $k\gg k_F$  
and  Eq.(\ref{proton}) to evaluate large $k$ behavior of 
$f_p \propto (1/k)^8 $  we obtain:
 \begin{equation}
R\approx \frac{(P_{pn}^2/5) \rho_n } {(m_NkT)^{3/2}},
\label{enhancement}
\end{equation}
where $P_{pn}$ is the the probability for a proton to have momentum $k \ge k_F(p)$. For  the illustration, 
we numerically evaluate the enhancement factor $R$  for neutron density close to $\rho_0$, 
 $x=\rho_p/\rho_n=0.1$,
and $P_{pn}=0.1$.  
So,   
\begin{equation}
R\approx 0.16 P_{pn}(MeV/kT)^{3/2}.
\end{equation}
The enhancement is significant for $kT\ll 1 \textrm{MeV}$. Remember that after one year a neutron 
star cools to the temperatures $T\le 0.01 \ \textrm{ MeV}$. 

Neutrino luminosity due to direct URCA processes decreases with decrease of $x$
 but  differs from zero even 
for  the popular
option: $x\le 0.1$. So investigation of the neutrino luminosity of 
the  neutron stars may help to narrow down the range of the allowed values 
of the $x$ ratio.

\subsection{$\beta$ stability of neutron within the outer core of zero temperature neutron star}
\label{t=0}
Normal neutron star  is bound  by gravity.  Gravity does not forbid  decays of constituents of the star  
if energy and momentum are conserved in the decay (the equivalence principle).

Constraints due to the energy-momentum conservation law  and the  
Pauli blocking in the electron sector work in  the  opposite directions.  Indeed, the  maximal momentum of an 
electron from  $\beta$-decay of a neutron with momentum $k_n$ is 
$\approx 1.19 MeV /(1-k_n/m_p)$. Hence, an electron produced in the 
neutron $\beta$ decay may fill the electron hole with momentum 
$k\approx 1 MeV/c$ only.  The dominant process  which may lead to the formation of electron holes  is the elastic 
interaction of an energetic proton with electrons within the free electron gas.  Energy-momentum conservation is  
fulfilled   in the case of   nonrelativistic nucleons if electron in the Fermi sea kicked out  by proton 
has minimal energy in the range:
\begin{equation} 
E_{hole}(k)=((p-k_f+k)^2-p^2)/2m_N+k_f. 
\end{equation}
Here $p$ is the proton momentum and $k_{f}$ is the electron momentum in the final state.  Scattered electron has energy 
$k_f \ge E_F(e)$,  so it is legitimate to neglect by
the  electron mass.  Hence, the minimal energy of the hole (when electron and proton momenta are antiparallel in the initial state)  is
\begin{equation}
E_{hole} \sim (1/2\div 1/3)E_F(e),
\end{equation}
for the proton momenta around $p=0.4\div 0.5 GeV/c$ typical for SRC and decreases with increase of $p$.  Evident mismatch between energies of produced electron holes and electrons in the neutron decay guarantees that an electron from  $\beta$-decay of  a neutron can not fill an electron hole.  

In the case of  ultrarelativistic nucleon gas (inner core of a star?) 
energy-momentum conservation does not restrict energies of 
electron holes  produced in (e-p) interaction: 
\begin{equation}
E_{hole}(k)=\sqrt {(m_N^2+(p+k-k_f)^2}-\sqrt {(m_N^2+p^2)}+
\sqrt {(m_e^2+k_f^2)}
\end {equation}
In the limit $p/m_n\to \infty$ we obtain  expression for minimal 
energy of hole:
\begin{equation}
E_{hole}(k)=\sqrt {(m_e^2+k_f^2)}-k_f +k\approx k+m_e^2/k_f.
\end{equation}
However in this regime use of nucleon degrees of freedom would be questionable. We will not discuss further in this paper interesting question on the possible $\beta$  instability of neutron within the inner core of star.

Direct $\beta$ decay of muon produces electrons with momenta
up to $m_{\mu}/2$  which are not far from the  electron Fermi momentum.   So evaluation of Pauli blocking for muon, hyperon  
$\beta$-decays requires model building.

 It follows from above discussion that 
the reduction of the difference between neutron and proton 
momentum distributions  influences collective modes. The   most significant effect would be the tendency to suppress the superfluidity 
of protons (superconductivity)  due to  the deformation of the proton Fermi surface because of  an increase of the fraction of protons having
momenta above the Fermi surface.  Existence of SRC  will not strongly influence the  possible superfluidity of neutrons.  Note that superfluidity of neutrons will  further suppress neutron $\beta$ decay due to formation of neutron Cooper pairs near the Fermi surface.

Electrons and neutrinos in the $\beta$ decays  of hyperons, muons,     are vastly more energetic than in  neutron decay. Hence, if  hyperon 
or muon stars exist,  they  should decay  significantly more rapidly  than the neutron stars and produce larger neutrino flux.

\section{QCD and nuclear physics}

\subsection{Introduction}

Spontaneously broken chiral symmetry plays a critical role in the nuclear structure. 
Small mass of the pion (due to small mass of u and d quarks) ensures presence of a large 
distance scale in nuclei $-$ small density of nuclear matter as compared to the density within 
the nucleon. 
At the same time since pion is a pseudogoldstone meson of QCD, 
low energy theorems for interaction of pions with nucleons  ensure 
that probability of low momentum pionic degrees of freedom in nuclei is small.  
The spontaneously broken chiral symmetry is responsible 
practically for all of the  nucleon mass, leading to  a barrier  of $\approx 1$~GeV 
between nonperturbative and perturbative vacua\footnote{This property is absent in many  
bag models  which are often used to evaluate properties 
of nuclear matter and to derive equation of state.}.

However the chiral symmetry breaking alone
is not sufficient to answer questions concerning microscopic structure of  SRCs such as 
the  role of the mesonic and baryonic components in these configurations, 
as well as account of the relativistic motion of the nucleons in SRCs.
In fact studies of nuclear structure within chiral 
perturbation theory introduce a perturbation series over the pion interaction with 
subtraction terms arising from the  integration over the  high momenta characteristic 
for SRC.  As a result these approaches deal with effective wave functions of nuclei in which 
high momenta due to SRCs are absent (hidden in the subtractive terms).

To address the properties of nuclei at high resolution one needs to take into account 
several other fundamental properties of QCD  namely  the decrease of 
the coupling constant at small space-time intervals (asymptotic freedom)  and  
proportionality of strong interaction strength to the region occupied by 
color - the color screening phenomenon. These features of QCD lead to a number 
of   new phenomena in nuclei unexpected in pre QCD approaches. 
This includes transparency of nuclear matter for propagation of 
fast spatially small 
quark-gluon wave packets, increase of the radius of the region occupied by 
color  within a bound nucleon - a  precursor of color conductivity. Evidence for 
such phenomena  in moderate energy processes is discussed below.  
Search for color opacity phenomenon at ultrahigh energies is one of the goals of LHC.

We discussed in Secs.3 and 4  that hard processes indicate  the dominance of nucleon degrees 
of freedom in SRCs, even for higher densities present in nuclei 
for which nonnucleonic effects are expected to be enhanced as compared to  
average nuclear configurations.
Within meson  nucleon theories interaction grows with a decrease of the distance due 
to the presence of the Landau pole,
\begin{equation}
g^2(t)=\frac{g_o^2}{1-bg_o^2\ln(t/t_o)},
\label{invariantcharge}
\end{equation}
leading to very strong interactions within the SRC.  Here $g(t)$ is running 
coupling constant, $t$ is a virtuality and $b >0$.  Therefore the  dominance of nucleonic 
degrees of freedom  does not seem natural within these models. No such phenomenon is 
expected in QCD due to the phenomenon of asymptotic freedom.

Overall QCD dynamics is strongly different from the expectations of  preQCD field 
theory models: approximate Bjorken scaling for the cross sections 
of hard processes follows from asymptotic freedom ($b <0$ in Eq.(\ref{invariantcharge})) 
and color screening phenomenon. Such scaling is absent
in preQCD models due to increase of the invariant charge with virtuality i.e.  
due to the Landau ghost pole in the running coupling constant. 
Moreover, color fields are screened within spatially small wave 
packets of quarks and gluons  leading to a decrease of 
interaction of these quark-gluon packets with ordinary hadrons.  
All these features of QCD have been  observed in 
the numerous hard QCD and electroweak phenomena.  Experimentally approximate Bjorken scaling 
for  lepton-nucleon(nucleus) scattering sets in at  $Q^2\ge 1$~GeV$^2$. 

To summarize: nonrelativistic nuclear models are phenomenological 
approaches in which short-distance effects enter as a boundary condition 
and absorbed in the  parameters of the models. Hence it is not surprising 
that expectations of these models for hard nuclear phenomena differ strongly 
from the QCD expectations. The difference is especially large for  phenomena 
in which one deals with the coupling to the meson exchange currents at large virtualities.

\subsection{Implications  of  hard phenomena for the structure of 
 nonnucleonic configurations in nuclei }

Nucleons are composite particles, and therefore  internucleon interaction  
should lead to a deformation of the  bound nucleon wave function.  
A certain modification of the bound nucleon is manifested in the difference between 
quark-gluon distributions within a nucleon and a nucleus (usually referred as EMC 
effect  \cite{EMC,EMCB1,EMCB2}).  This conclusion follows from 
the  combined application  of the exact baryon 
charge and momentum sum rules\cite{FS88}.  
Observation of bound nucleon structure function modifications in the nuclear medium
rises question 
of how these modifications can be represented in terms of non-nucleonic degrees of 
freedom in nuclei and what is the probability of such components 
in the nuclear wave function.

PreQCD models predict enhancement of meson currents in nuclei and 
significant non-static meson fields in nuclei, onset of the meson  condensate 
regime at densities comparable to the average nuclear density, $\rho_0$.  Such a hypothesis leads 
also to the expectation of enhancement of the antiquark ratio 
$R_{\bar q}= \bar q_A(x,Q^2)/\bar q_N(x,Q^2)$ for $x\le 0.1$ of $\ge 10\%$ for $A\ge 40$ 
while the data find $R_{\bar q}( 0.05\le x \le 0.1) < 1$ with the typical error bars of 1\%.   
The enhancement of   $R_{\bar q}$ 
originates in these models
from pion virtualities, $p_{\pi}^2\sim 1$~GeV$^2$, which 
are far from 
the region where concept of meson exchanges can be 
justified:~($\left|p_{\pi}^2\right| \le \, few \,\, m_{\pi}^2$).

Presence of $\Delta$-isobars in the nuclear wave function  on the level of 
$few$~\%   is not excluded experimentally.  In fact presence of $\Delta$-isobars 
in $A=3$ nuclei is necessary to satisfy the exact QCD Bjorken sum rule~\cite{FGS}.

QCD suggests  possibility of direct role of color field in nuclear structure.  
Color screening phenomenon means that smaller is the  size of the quark-gluon wave 
packet smaller is its interaction with a hadron. As a result quark-gluon 
configurations of smaller than  average  size are suppressed within the bound 
nucleon \cite{FS85}. This  can be demonstrated by applying variational principle 
which requires  suppression of configurations with minimal attraction to increase 
binding energy. The strength of deformation, $\gamma$, is expected to increase 
with an increase of the momentum of nucleon, $k$   approximately as 
$\gamma \propto k^2$ 
(for simplicity we omit here a term related to the energy of the residual system due 
to which $\gamma$ is linear in virtuality of the interacting nucleon).
 As a result the   deformations should increase with an 
increase of  nuclear density. 
An analogous  effect of induced polarization of two interacting atoms is well known 
in the atomic physics where such deformation is found to be different 
for directions along and perpendicular to the axis between the atoms. Similarly the 
deformation of parton distributions in the bound nucleon may depend both on the 
absolute value of the nucleon momentum, $\vec{k}$, and its 
direction with respect to the momentum transfer $\vec{q}$.

To summarize: the EMC effect unambiguously demonstrates presence of nonnucleonic 
degrees of freedom in nuclei. Description of the effect as   deformation of bound 
nucleon  is dual to the presence of baryonic non-nucleonic degrees of freedom
within hadronic basis description of the modification  effects. 
In the 
color screening mechanism, described above, these baryonic components 
 are highly coherent with the nucleon components and their absolute 
probability is rather small. This is consistent with the experimental constrains 
on the admixture of such components coming from high energy data\cite{FS88}.
In Sec.8 we discuss some of these restrictions and how  further  
measurements of cross sections of the reactions: 
$e+A\to e' + N_{backward}+\pi +X$ could improve the limits or lead to 
discovery of these components.

\subsection{QCD and meson currents}

One of the  intriguing observations of low energy nuclear physics is that 
nucleon - nucleon  potentials decrease rather weakly with momentum transfer.  
Within one boson exchange models (OBEP) such behavior requires very hard 
(nearly point-like) meson - nucleon form factors.  Such a weak dependence of meson - 
nucleon form factors on $t$ seems to be difficult to reconcile
with the observations of 
quark structure of mesons and nucleons at high energies 
as well as with the t-dependence of the exclusive pion electroproduction by the 
longitudinally polarized photon which is dominated by the pion pole contribution.

If one  applies  the  OBEP model  to the calculation of antiquark content of the 
nucleon at $Q^2 \sim $ few GeV$^2$ one finds an inconsistency:  hard form factors generate 
too many antiquarks at $x\ge 0.1$. For example, if one uses an exponential parameterization 
of the $\pi NN$ form factor, $F_{\pi NN}(t)=\exp (\lambda t)$, one finds  
$\lambda \ge 1$~GeV$^{-2}$.  Note that   approaches like OBEP ignore the loop diagrams which 
are unavoidable in quantum field theories and which lead to 
the Landau pole in the running coupling constant at  momentum transfer in the vicinity of 
$-t\approx 1$~GeV$^2$. Still even in this form mesonic models do  not  match even 
qualitatively 
to the QCD pattern of the decrease of meson  fields as a function of momentum transfer 
as a consequence  
of the asymptotic freedom and color screening phenomenon. Another 
shortcoming of the models is the increase of the antiquark density with nuclear density which 
contradicts the data, see discussion in section 6.2.

\begin{figure}[ht]
\centering\includegraphics[height=6cm,width=12cm]{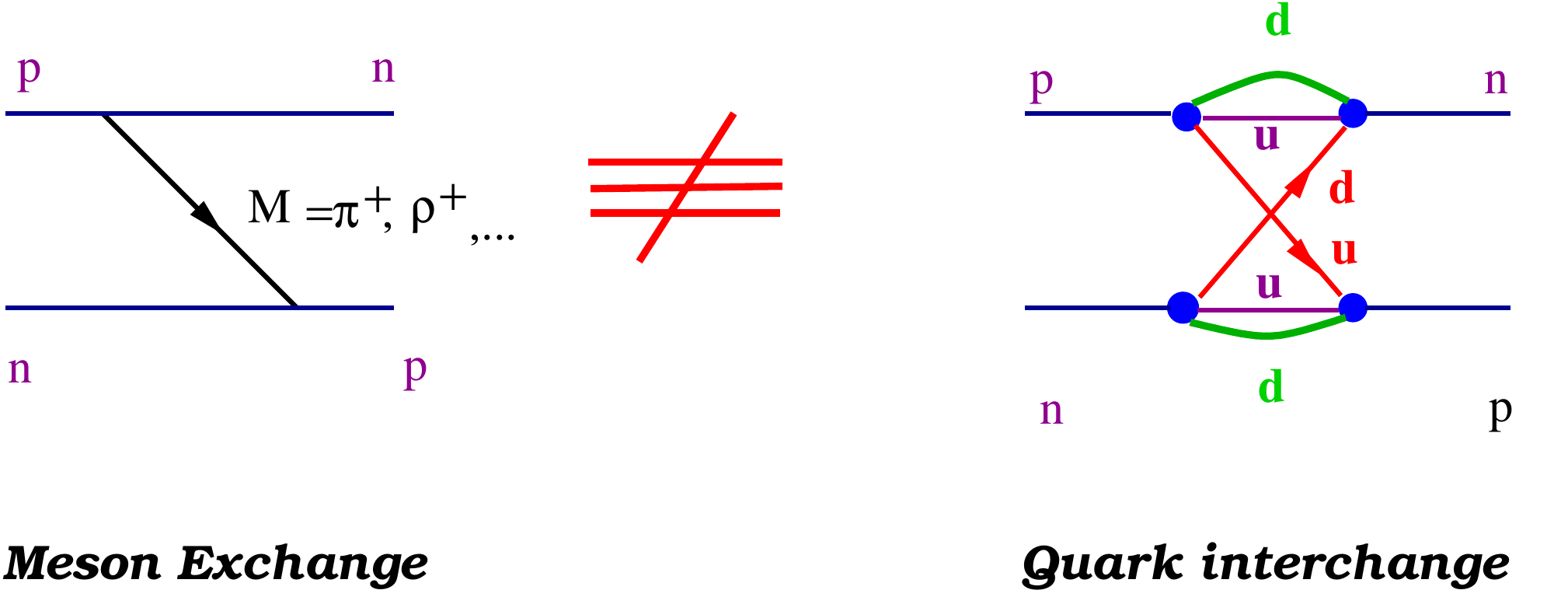}
\caption{Diagrams corresponding to meson and quark exchange mediated NN interactions.}
\label{meson}
\end{figure}

The listed paradoxes originate from the contribution of pion fields with momenta 
$\ge 0.5$~GeV/c in which case internucleon distances (distance 
between  pion and a nucleon) become substantially smaller than the nucleon size of 
$\sim 0.6 \, fm$\footnote{We use here nucleon radius as given by the axial form factor 
since this radius does not contain contribution of the soft pion fields. Classically one 
can fit a pion between two nucleons only if $r_{NN} \ge 2 r_N+2r_{\pi} \ge 2\, fm$.}  where 
geometrically one can hardly think of the emission of pion.  This is related to the 
observation 
that the pion exchange can be separated from the other   contributions only in the 
vicinity of  $t= m_{\pi}^2$.  At the same time an exchange by the meson quantum numbers 
in $t$-channel at small distances 
does not require physical presence of the exchanged mesons.   For example,  an 
exchange of quarks between two nucleons (see Fig. \ref{meson}) can also provide an 
exchange of same  meson quantum numbers in t-channel (see e.g. Refs.\cite{gdpn,gheppn}). 
A quark exchange does not lead to the change 
of the number of antiquarks in the intermediate state. This will  remove contradiction with the 
measurements of the A-dependence of  antiquark distributions,  and may have  relatively 
weak $t$ dependence in the discussed $t$-range. Hence a possible solution of the paradox maybe 
the matching of the interaction potential of nearly static pion exchange for small $t$ with the  
quark interchange at larger  $t$.

There is another constraint on the model of NN interaction coming from 
high energy behavior of the amplitudes that should match the Regge 
pole behavior. The  Reggezation of the meson exchange  leads 
to a strong modification of the energy dependence of corresponding amplitudes~($A$),  
for example from $A\propto s$ for a $\rho$-meson  exchange 
to $A\propto s^{1/2}$ for the $\rho$-meson Regge trajectory  at $t=0$. For 
the pion case the difference is very small for t=0, but becomes large with increase of  
$-t $.  It is interesting that the discussed change of energy dependence arises in the dual 
Veneziano type models as a result of strong cancellations between  s-channel resonances. An early  onset of the Regge type behavior at low $t$ limit for  two body processes  
is known as Dolen- Horn duality.

Another set of phenomena characteristic for QCD which is absent in  meson theories 
is the fluctuation of interaction strength of the hadrons.  
In particular, in meson models it is hardly possible to generate 
the characteristic QCD phenomenon of color transparency 
- suppression of interaction strength with nuclear media of those  
configurations in hadrons that have constituents occupying a volume much smaller 
than  the volume of average hadronic configuration\footnote{The dual  description of the color 
screening phenomenon is to represent spatially small 
quark-gluon configuration as a coherent superposition of hadrons \cite{FMSG}.}.
 The color transparency phenomenon  is observed in variety of experiments at collision 
energies ranging  from hundred GeV \cite{Ashery} to  few GeV \cite{JLabct}. 
In particular,   in the coherent process of dijet production by pions off the nuclei: 
$\pi ^+A \to \, two \, jets\, +A$  the observed ratio of the yields from platinum and 
carbon targets\cite{Ashery} is at least by factor of eight larger than the ratio
expected from preQCD estimates.

The same property of QCD is responsible for validity of QCD factorization 
theorem\cite{CFS} for exclusive processes which is confirmed by data on 
diffractive electroproduction of vector mesons at collider  energies\cite{HERA}.  

To summarize, the concept of the pion exchange currents which is popular in  
low energy nuclear physics for processes with low momentum transfer 
is qualitatively consistent with QCD. At the same time preQCD models predict that 
contribution of meson currents should 
increase with an increase of virtuality which contradicts to QCD prediction 
in which   meson currents should decrease with an increase of the 
virtuality. The theoretical approaches in description of SRCs should  
be consistent with this basic property of QCD.

\section{Light-cone description of high energy processes involving nucleons 
and nuclei.}

We discussed in introduction and in Sec.3 that large $Q^2$ inclusive $A(e,e')X$ 
reactions probe the light-cone~(LC) density matrix and ultimately  LC wave function of 
the nucleus.

Account of relativistic effects should be done in accordance with  
basic properties of QCD. One of the theoretical challenges for the 
relativistic quantum field theory  is to separate bound state wave function
from the background of vacuum fluctuations which are always present within 
a field theory. 

This can be easily done for LC  wave functions of bound state in 
kinematical  domain close to one in which nucleon motion is nonrelativistic. 
Thus a question arises about the possibility of an approximation 
in which motion of nucleons is treated relativistically while no additional 
degrees of freedom is included - the LC mechanics of nuclei. 
To estimate the relative role of different degrees of freedom 
in the nuclear wave function we  use the experimental data on NN 
interaction and the idea (implemented in QCD string models) that inelasticities 
in hadron-hadron collisions are related to the production of resonances. 
Within this picture, in the $I=0$ channel nucleon degrees of freedom should dominate 
up to  $k^2/m_{N}\sim (m_{N^*}-m_{N})\approx 600$~MeV. In the channel 
with isospin $T=1$ inelasticities may appear important at a lower energy scale:
$k^2/m_{N}\sim m_{\Delta}-m_{N}\approx 300$~MeV.

These estimates  indicate that up to the very large momenta on nuclear scale,
$k \le 800$~MeV/c for the deuteron and $k \le 550$~MeV/c for heavier nuclei  
an approximation in which only nucleonic degrees of freedom are accounted 
for the nuclear wave function is a legitimate approximation.
  
\subsection{Light cone quantum mechanics of nuclei}

\begin{figure}[ht]
\centering\includegraphics[height=3.6cm,width=12.4cm]{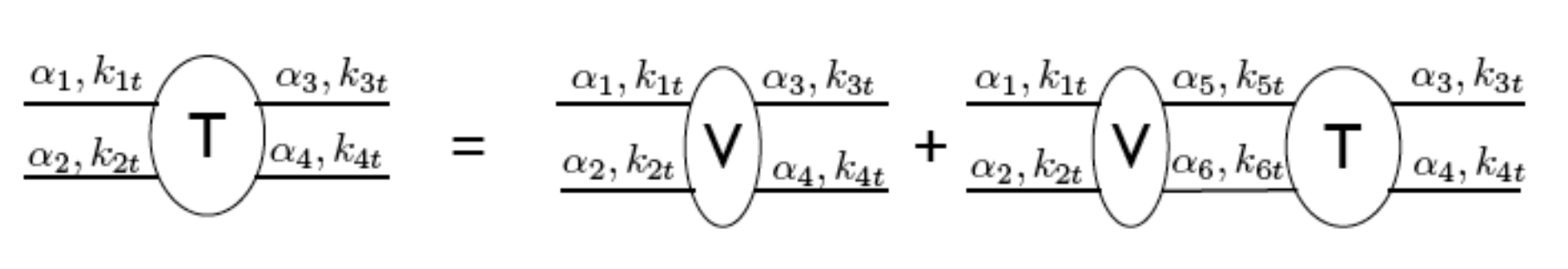}
\caption{LC equation for $NN$ scattering in  two nucleon approximation.}
\label{FSeq}
\end{figure}

As it was shown above for rather wide range of internal momenta in the nucleus the 
inelasticities in NN interaction is very small which can be considered as a small 
parameter in the problem. 
Due to the presence of such small parameter it makes sense to consider two nucleon 
approximation for LC wave function of the deuteron\cite{FS76} for bound nucleon momenta up 
to $800$~MeV/c.
Key result in considering two-nucleon system in the light-cone is 
the existence of a relationship between nonrelativistic~(NR) and LC equations for 
two-nucleon wave functions. 
If nonrelativistic  potential describes the phase shifts, the same is true for its LC analog. 
Hence there exists a simple approximate relation between  LC and  NR two nucleon waves  
functions and NN potentials. 
The proof (rather lengthy) is based on reconstruction of properties 
of NN potential from the Lorentz invariance of on-mass-shell NN amplitudes\cite{FS88,FS90}. 
One finds that the form of the LC potential which enters in 
the LC equation for scattering amplitude (Fig.\ref{FSeq})   is 
strongly constrained by the the angular momentum conservation.

After introducing irreducible part~$V$ (which does not contain two nucleon intermediate state) 
through the  lengthy algebra we obtain relativistic equation for deuteron wave function and 
related equation for two nucleon system with mass $M$ \cite{FS88,FS90} in the 
following form:
\begin{equation}
(4(m^2+p^2)-M_{2N}^2)\psi=\int
V(p,p') (1/(2\pi)^3)d^3p'/\sqrt{p^{'2}+m^2}\psi(p'),
\end{equation}
where $p$ and $p'$ are three dimensional "internal"  momenta of two nucleon system
which are related to the LC variables as follows:
\begin{equation}
\alpha=1+{p_3\over \sqrt{m^2+p^2}} \ \  \mbox{and}  \ \ k_t=p_t.
\label{lcmom}
\end{equation}
The relation between kernel $V$ as given by Feynman diagram and potential $U$
which enters in the nonrelativistic description of the NN interaction is given by:  
 \begin{equation} 
U(p_1,p_3)= {V(p_1,p_3) \over \sqrt{2E_1 \cdot 2E_2\cdot 2E_{3f}\cdot 2E_{4f}}}.
\end{equation}
This is the same relation that relates  relativistic
Feynman diagrams for QED to the  Coulomb potential. Therefore
\begin{equation}
[4(m^2+p^2)-M^2]/\sqrt{m^2+p^2}\psi=\int U(p,p') d^3p'/(2\pi)^3\psi(p').
\end{equation}
This equation has a unique  solution provided  the self consistency requirement is 
imposed   that the 
equation of Fig.\ref{FSeq} generates rotationally invariant $NN$ scattering 
amplitude which satisfies angular momentum conservation. As a result the deuteron wave 
function  in the two nucleon approximation  depends only on a single variable, $k$, which 
is defined as follows:
\begin{equation}
k = \sqrt{{m_N^2 + p_t^2\over \alpha(2-\alpha)}- m_N^2},
\label{klcdef}
\end{equation}
and relates in a straightforward way to the nonrelativistic deuteron wave function:
\begin{equation}
\Psi_d^{LC}(\alpha,p_t) = \Psi_d^{NR}(k)(m_N^2 + k^2)^{1\over 4}.
\label{dlc}
\end{equation}

To summarize, for two body system in two nucleon approximation the biggest difference 
between nonrelativistic or  virtual nucleon approximation and LC   is  in  the 
definition of ``internal'' momentum of two nucleons. For NR case the ``internal'' momentum 
corresponds to the relative momentum of two nucleons in the lab, while in LC it 
is defined according to Eq.(\ref{klcdef}).
This results in a qualitatively different relation between the wave function and the 
scattering amplitude for large nucleon momenta.

\subsection{Master equation for LC wave function  of a many nucleon system}

Similarly we 
can deduce many body equation for LC dynamics in terms of irreducible amplitudes 
of internucleon interactions\cite{FS90}:
\begin{equation}
[(\sum_{i=1}^{i=A} \epsilon_{i})^2-M_A^2]\psi_{A}(p_j)=\int V(p_i,p'_i)
\delta(\sum_{i=1}^{i=A}  p_{i})\frac{d^{3}p_{i}}{2\epsilon_{i} (2\pi)^{3}}(\sum_{i=1}^{i=A}  
\epsilon_{i})\psi_{A}, 
\end{equation}
where  $\epsilon_i=\sqrt{p_i^2+m^2}$  and $M_A$ is the invariant  mass of the eigenstate.

In the case of ground state for which binding  energy is small master equation 
for LC wave function obtains the same form as that given by relativistic theory 
for Schr\"{o}dinger wave function in c.m. in which particle production is neglected:
 \begin{equation}
[\sum \epsilon_{i}-M_A]\psi_{A}(p_i)=\int (V(p_i,p'_i)/2)\delta(\sum p_{i})\frac{d^{3}p_{i}}{2(2\pi)^{3}
\epsilon_{i}}\psi_{A} (p'_i).
\end{equation}
 
Master equation with potential $V$ reproducing Lorentz invariance 
of on-mass shell amplitude\cite{FS90}  allows to account 
for the angular momentum conservation  and to satisfy the requirement of 
separability.  Account of the angular momentum conservation for on-shell amplitudes 
within LC mechanics requires special many body forces described in  \cite{FS90}.
Another pattern how potential $V$ can be chosen consistent with the properties of 
on-mass-shell amplitudes is to explore 
similarity of LC equations to the center mass equation of relativistic 
noncovariant perturbation theory  in which antinucleon production is neglected.

Having master equation and fitting potentials to describe on mass shell amplitudes 
it should be feasible to calculate LC wave functions, spectral and decay functions.  
It is worth mentioning that due to factorization of LC momentum fraction $\alpha$ and 
recoil energy $p_{R+}$ (see Sec.3) the calculation of 
LC density matrix which enters in description of large $Q^2$ inclusive $A(e,e')X$ 
processes  does not require performing a more challenging calculation of 
the spectral function. The latter is the case in the nonrelativistic approach. 
Calculation of LC spectral function could be simplified by the angular condition  
which implies that the LC spectral function is a function of two variables only  
and by the sum rule relating LC density matrix to the  spectral function.

At present, for state of art analysis 
of many phenomena in high momentum transfer reactions 
nonrelativistic wave functions, spectral 
functions and  decay functions calculated within nonrelativistic theory of nuclei can 
be used as basis for building LC density matrix as well as LC spectral and 
decay functions\cite{eip3}.

 \section{Directions for the future studies}

The progress in studies of SRCs described above and challenging problems of QCD 
as well as understanding the implication of SRCs  in the dynamics of cold 
dense nuclear matter  calls for a systematic studies of reactions described above 
as well as including series of new hard processes. 

Here we briefly outline some possible directions for experimental research both for 
electron and hadron facilities.

\subsection{Inclusive $A(e,e')X$ reactions}

\subsubsection{Probing 2N and 3N SRCs}
We explained in Sec.3 that high $Q^2$ inclusive $A(e,e')X$ reactions at $x>1$ 
directly probe 
SRC probabilities. Further progress in this direction will be the study 
of $A(e,e')X$ reactions at  much broader range of $x$ and $Q^2$. 
In particular for  $x\sim 3$ it is highly desirable to reach at least 
$Q^2\sim 6\div 8$~GeV$^2$.
 The transition to DIS regime at $Q^2 \ge 10$~GeV$^2$ is very interesting. 
One expects that with an increase of $Q^2$ the ratios at fixed $x \ge 1$ should increase 
since DIS scattering at given $x$ probes LC fractions $\alpha \sim x +0.5$ \cite{FS81}. 
For example the carbon/deuteron ratio at x=1 should increase from $\sim 0.4$ to $\sim 5$.
Such regime will corresponds to deep inelastic scattering off superfast quarks in nuclei with 
$x\ge 1$\cite{FS88,hnm}. 
The first experimental signal of  DIS off superfast quarks will be the 
observation of the onset of Bjorken scaling in the region $x\geq 1$.

\begin{figure}[ht]
\centering\includegraphics[height=10cm,width=10cm]{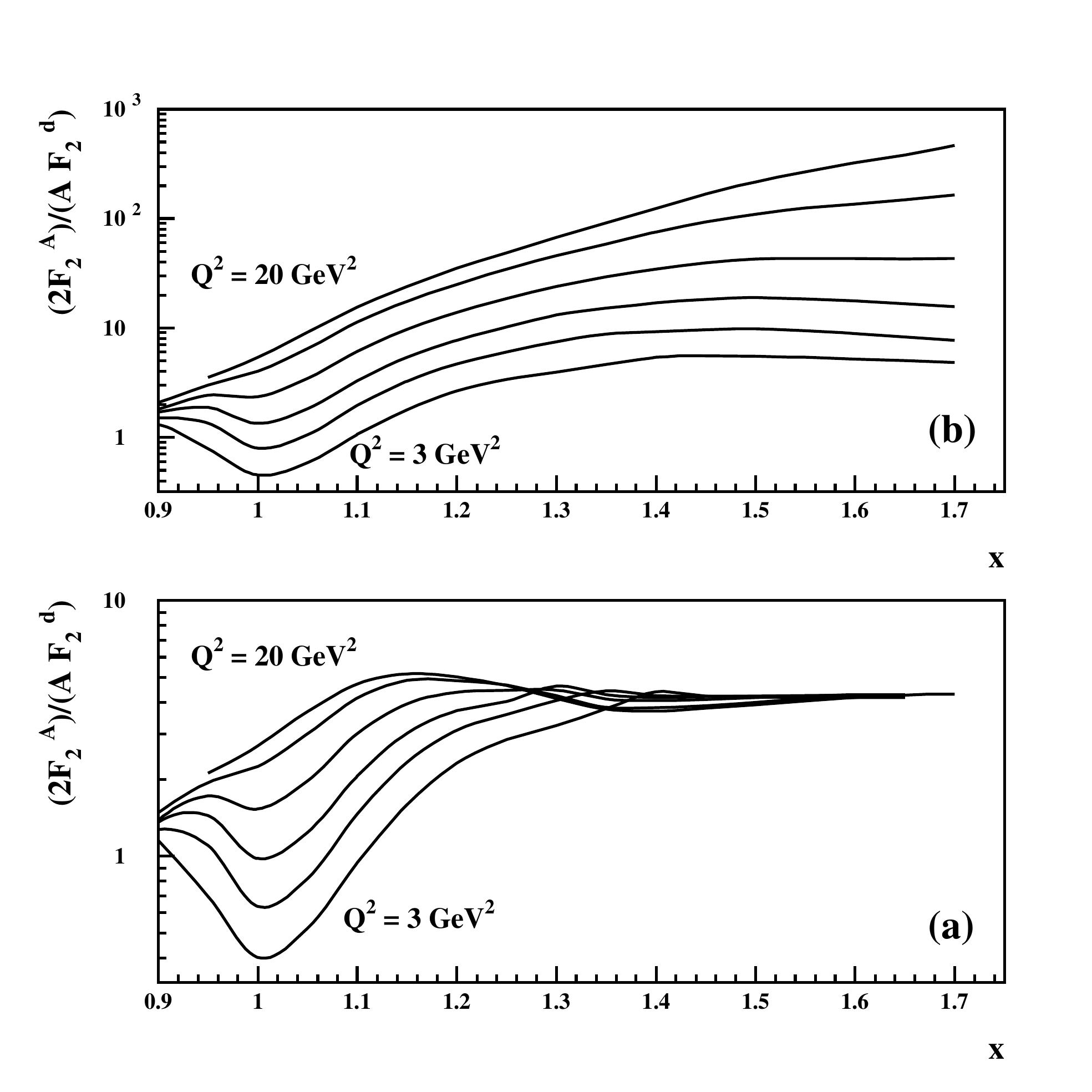}
\caption{Ratio of the per nucleon structure functions of carbon and deuteron for different 
$Q^2$ in the few nucleon and two nucleon correlation approximations including both 
quasielastic and inelastic contributions. Curves correspond to $Q^2=3,5,7,10,15$ and 
$20$~GeV$^2$ values.}
\label{ratioad}
\end{figure}

The leading twist contribution is expected to dominate at $Q^2\ge 12$~GeV$^2$ for x=1 and 
somewhat larger $Q^2$ for higher x (the interplay between leading twist and higher twist contributions 
(quasielastic scattering) depends on relative importance of the  2N and 3N correlations). 
The $x$ dependence of $F_{2A}(x\ge 1,Q^2), A\ge 4$ in the scaling limit  is expected 
in the few nucleon correlation model \cite{FS81,FS88} to be $\propto \exp(-bx)$ with $b\sim 8\div 9$, 
leading to a large increase of the $F_{2A}/F_{2D}$ ratio between $x=1$ and $x=1.5$, see Fig. \ref{ratioad}. 
Experimental  attempts to observe such "superfast" 
quarks were inconclusive: the BCDMS collaboration~\cite{BCDMS} has observed a very small $x > 1$ 
tail ($b\sim 16$), while the CCFR collaboration ~\cite{CCFR}   observed a tail consistent 
with the presence of very significant SRCs~($b\sim 8$). A possible explanation for the 
inconsistencies is that the resolution in x at  $x > 1$ 
of the high-energy muon and neutrino 
experiments is relatively poor, causing great difficulties in measuring $F_{2A}$ which strongly 
varies with x. The energy resolution, intensity and energy of Jefferson Lab at 
11GeV may allow one to study the inset of the scaling regime and thereby 
confirm the existence of superfast quarks.

\subsubsection{Study of isotopic structure of SRCs and the nuclear core 
in $A(e,e')X$ reactions}.

If $pn$ correlations dominate in high $Q^2$ $A(e,e')X$ reactions at  $x> 1$ one expects that,
\begin{equation}
{\sigma_{^3H}(1<x<2,Q^2)\over \sigma_{^3He}(1<x<2,Q^2)} \approx 1, \ \ \mbox{and} \ \ 
{\sigma_{^{40}Ca}(1<x<2,Q^2)\over 
\sigma_{^{48}Ca}(1<x<2,Q^2)}
\approx {N(^{48}Ca)\over N(^{40}Ca)}=1.4. 
\end{equation}
As we discussed  in the text inclusive $A(e,e')X$ reactions at $x> 2$ and 
production of fast backward nucleons in semi-exclusive reactions presently  
are  the only sources of information about 3N SRC. 

The $A(e,e')X$ reactions at $x>2$ allow one also to  probe the isospin structure of 3N SRCs. 
In this  process virtual photon is absorbed by a nucleon of 3N SRC with  
large momentum, $p_i$, which is balanced by two nucleons with a   
relative momentum  much smaller than momentum characteristic of  2N SRCs.

If 3N SRC emerges in the nuclear wave function predominantly 
through the iteration of two nucleon interactions one expects that 
$ppn$ and $nnp$ correlations to have about the same strength considerably 
exceeding the strengths of  $ppp$ and $nnn$ correlations.  The dominant contribution to the $x>2$ region comes from the configurations in which
recoiling mass is close to minimal.Our numerical studies with the realistic spectral 
function for $A=3$ nucleon system suggest that in this kinematics it is by factor 
$2 \div 3$ more likely (depending on the nucleon momentum)  for a  recoil pair to be 
in $I=0$, rather than $I=1$, state. As a result the ratio 
\begin{equation}
R_3={\sigma_{^3H}(2<x<3,Q^2)\over \sigma_{^3He}(2<x<3,Q^2)} \approx {\sigma_{el}(en)(Q^2)\over
\sigma_{el}(ep)(Q^2)} \ll 1,
\end{equation}
where $\sigma_{el}(eN)$ is the cross section of elastic electron-nucleon scattering in the 
kinematics of the $A(e,e')X$ reaction. This should be compared with the expectation of 
$R_3\sim 1$ for the scattering off two nucleon correlations.
Similarly a strong increase of the cross section with 
the number of neutrons can be expected for scattering off different isotopes like calcium. 

\subsection{Neutrino processes off nuclei}
 
Unique advantage of neutrino (antineutrino) initiated processes
is in the feasibility to probe density of antiquarks within a nucleon 
and nucleus. Therefore investigation of the correlation between $y$ dependence of 
processes: $\nu +A\to \mu +backward \ proton +X$ and momentum of the backward 
proton will help to unambiguously establish  dependence of meson currents on 
virtuality and nuclear density. 

\subsection{Spectral functions}

\subsubsection{ Probing 2N and 3N correlations}

Further studies of spectral functions at large initial momenta of 
removed nucleon~($p_i$) are necessary 
in kinematics in which rescattering effects are minimized.  
Generally this is the case when produced nucleon has a small transverse momentum 
and large energy relative to the residual system. 
In this kinematics one can probe both 2N and 3N correlations.
To enhance the contribution from 3N SRCs 
one can look for the ratio of cross sections of $^3He(e,e'p)X$, and  $^3He(e,e'n)X $ 
reactions in quasielastic kinematics in which the recoil invariant mass is 
sufficiently small to suppress the contributions 
of 2N correlations (such study will be complementary to the study of $A(e,e')X$ processes 
at $x>2$ discussed above).

\subsubsection{High excitation energies and possible presence of nonnucleonic components in nuclei}

The current meson exchange based models of NN interactions involve $N\Delta$ and  
$\Delta \Delta$ intermediate states. These states lead to a high momentum component 
of the nuclear wave function through the $N\Delta$ and  $\Delta \Delta$  correlations.  
The current calculations of spectral functions  with realistic NN interactions do not  
treat $\Delta$-isobar degrees of freedom explicitly. On the other hand,  if a nucleon 
is removed at large $Q^2$ from a $\Delta N$ correlation, the typical excitation energy 
will be of the order of $m_{\Delta}-m_N\sim 300$~ MeV. 
High values of recoil nucleus excitation energy will be characteristic also for 
scattering off more exotic configurations like six-quarks 
that are  close together - a kneaded quark state.  
In order to observe the evidence for nonnucleon components like $\Delta$-isobars in
$N\Delta$ and $\Delta\Delta$ SRCs
one needs large $Q^2 \ge 2\div 3 \mbox{GeV}^2$ to destroy instantaneously 
these correlations  and to suppress the contribution from  two step charge exchange 
processes. Note that  studying $x$ dependence  of these processes  
may help to estimate 
two step processes as they should be practically the same at $x=1$ and away from the 
quasielastic kinematics.
 
 One can also study a complementary process of knock out of a $\Delta$-isobar, preferably 
$\Delta^{++}$ which cannot be produced in the scattering off a single nucleon in the 
processes like $e+N\to e+ \Delta$ or  $p+p\to p+\Delta$. 
Similar to the processes discussed above we would 
need high enough energies of the produced $\Delta^{++}$ to suppress the charge exchange 
contribution.

Note that in this case too the study of $x$ dependence of $\Delta^{++}$ production rate 
relative to the nucleon rate will allow to separate mechanism of scattering off the preexisting  
$\Delta$-isobar like configurations from those events associated with $\Delta$ production due to  
charge exchange rescattering of nucleons.

\subsubsection{Spin structure of 2N correlations}
It is important to measure directly the ratio of the S- and D- wave contributions 
in $pn$ correlations in the momentum range where D-wave dominates.   This is possible 
in the scattering off the polarized deuteron in the reaction $e+^2\vec{H} \to e +p +n$ 
if one chooses parallel kinematics to minimize rescattering effects. 
In the case of tensor polarized deuteron the $T_{20}$ asymmetry is expressed through the ratio of 
the D- and S- wave momentum distributions, $w(k)/u(k)$. This reaction also provides a unique 
way to study relativistic effects which are predicted within light-cone approach 
to be strongly sensitive to  the angle between recoil  nucleon momentum, $\vec{p}_r$ and the 
reaction axis ($\vec{q}$)\cite{FS81,FS88}.
Similar investigations are possible  using  vector polarized deuteron target and studying  polarization 
of the interacting nucleon which  is also expressed through the $w(k)/u(k)$ ratio \cite{FSpold}. 
Such a measurement was performed in Ref.~\cite{passchier}  using  the reaction 
$\vec{e}+^2\vec{H} \to e +p +n$ 
at $Q^2 = .21$~GeV$^2$ for the recoil nucleon momenta $\vec{p}_s<$~350 MeV/c. 
It would be important to extend these measurements to much higher  $Q^2$ and sufficiently large $W$  
where dynamics of final state rescatterings is simplified and can be described within GEA. 
In this case it would be possible to cover much larger range of recoil nucleon momenta for 
a range of angles between $\vec{q}$ and $\vec{p}_r$ for which the FSI  is small and can be 
reliable calculated.

\subsection{Decay functions}

\subsubsection{Tests of factorization, mapping pp, pn correlations}

The first studies of decay function described in the review suggest 
several 
directions for further 
theoretical investigations.
First it is important to  find kinematics in which final state interaction is minimal. 

Secondly, to identify kinematic conditions  for which  factorization of
the cross section into a product of decay function and elementary electron-bound-nucleon 
cross section is justified. 
Such studies are desirable to perform  for  both electron scattering using range of
 $Q^2>1$~GeV$^2$ and for high momentum transfer (anti)proton-nucleus scattering.

Ultimately studies along these lines will allow us to study both $pp$ and $pn$ correlations
for larger range of  correlated nucleons momenta in which case one expects central forces
in NN potential to became dominant or comparable 
with respect to the tensor forces. The onset of this regime could be 
identified by   an increase of the pp/pn ratios with an increase of 
initial momentum of the nucleons in SRC.

It is also important to establish minimal momenta of the struck nucleon
for which correlation mechanism is still operational, namely how close it is to the Fermi momentum. 
Remember that for  carbon ($k_F(C) \sim 220 \, MeV/c$) the correlation is clearly seen at $k\ge 300 MeV/c$ 
while it may be setting in at somewhat smaller $k$.

\subsubsection{Looking for SRC involving $\Delta$-isobars}

As we mentioned above the $\Delta N$ like correlations may be present in nuclei. 
They may be manifested in the decay of correlations when a nucleon correlated with 
$\Delta$-isobar is removed. Hence one needs to look for production of backward isobars in electron 
and proton scattering in high momentum transfer kinematics we discussed.

Although the yield of $\Delta's$ comparable to that of nucleons is clearly excluded by 
near  saturation of decay function by $pn$ and $pp$ SRCs, a yield 
on the level of $\sim 10\%$ is possible. 
It is worth noting that this is a scale expected in quark exchange 
models of  NN interaction\cite{FS81}. In addition, there exist data on inclusive 
production  of $\Delta$'s in $e - "air" $ scattering at $E_e=5$~GeV
and at large $\alpha \ge 1$ which allowed to estimate
the ratio of  $\Delta^{++}$ and proton yields for same $\alpha$ of the order 
5\% \cite{Degtyarenko:1990qk}. The latter corresponds to the order of 10\%
high momentum component 
per nucleon
in the nuclear ground state wave function due to 
$\Delta$-isobars.

\subsubsection{Probing 3N correlations}

The structure of three nucleon SRCs can be explored  in the processes like 
$p+A\to p'+ p_1+N_2+N_3 + (A-3)^*$ in which  proton $p_1$ is  produced in 
large angle center of mass $pp$ scattering and two nucleons $N_1$ and $N_2$ 
represent recoiling particles,
or in analogous processes in $\bar p A$ scattering using the PANDA detector \cite{PANDA}.
Similar to the case of $A(p,2pn)X$  processes studied in 
Ref.\cite{eip1,eip2} the knock-out proton~($p_1$) in this case will have preferentially 
$\alpha_i< 1$, leading to recoiling nucleons with $\alpha_{2,3} >  1$ emitted backward. 
Since such reaction selects $\left<k_{ti}\right> \sim 0$  this will 
correspond to the production of  "$N_2$" and "$N_3$"  nucleons  
with back-to-back transverse momenta.
Such configurations will be significantly enhanced due to 3N SRCs.

It is worth emphasizing here that this reaction would allow to check the role of 
three-proton SRCs which cannot be easily generated from two nucleon SRCs, and which 
via isospin invariance are connected to three-neutron correlations.
The latter are important for understanding the equation of state  of neutron stars 
where the role of $nnn$ SRCs is expected to be enhanced due to much higher densities 
involved.

\subsection{Theoretical studies}

At several points in the text we explained that high energy processes require understanding 
of LC structure of nuclear wave functions as well as spectral and decay functions. 
A simple relation between nonrelativistic~(or virtual nucleon) and light-cone approximations 
exists only for the two nucleon system. Already for the case of the motion of NN pair in a  
mean field of the nucleus LC and virtual nucleon approximations yield 
significantly different results due to different treatment of the recoil system 
in these approximations.

It is important to develop further the  relativistic approaches to confront 
them with experimental data. In the case of LC approximation the most pressing task is to 
solve the light-cone three-nucleon bound state problem. 
The relevant equations which to a very high accuracy satisfy constrains imposed by 
rotational invariance of on-shell NN amplitudes have been derived in Ref.\cite{FS90}.
Solving numerically these equations would allow both to make predictions for 
various experiments discussed in the review and to address a delicate issue of 
matching LC and nonrelativistic spectral functions for three body systems.

In describing SRCs at very large internal momenta it is important also to 
develop theoretical approaches that describe NN interaction at very small separations.
In this respect it is important that recent approaches to derive nucleon-nucleon 
interaction based on Chiral QCD Lagrangian 
which are justified for small nucleon and pion momenta to be matched 
with theoretical approaches that account for asymptotic freedom and color screening of 
strong interaction. 

\medskip
\medskip

\noindent{\bf Acknowledgments}

We are thankful to Gordon Baym,  Claudio Ciofi degli Atti, Werner Boeglin, Shalev Gilad, Douglas Higinbotham, 
David Khmelnitskii, Konstantin Kikoin,
Gerald Miller, Shmuel Nussinov, Vijay Pandharipande, Eli Piasetzky, Larry  Weinstein,  Dmitri Yakovlev for many useful discussions. 
Special thanks to Ernest Henley for suggesting to write this review.
This work is supported  by  DOE grants under contract DE-FG02-01ER-41172 and DE-FG02-93ER40771 as 
well as by the Israel-USA Binational Science Foundation Grant.


\end{document}